\documentclass[12pt,a4paper]{article}
\pdfoutput=1
\usepackage{jcappub}

\hypersetup{colorlinks,bookmarksopen,bookmarksnumbered,citecolor=blue,
linkcolor=black,pdfstartview=FitH,urlcolor=blue}

\graphicspath{ {figures/} }

\usepackage{graphicx}
\usepackage{amsmath}
\usepackage{hyperref}
\usepackage{amssymb}
\usepackage{epstopdf}
\usepackage{placeins}

\def\bea{\begin{eqnarray}}
\def\eea{\end{eqnarray}}
\def\beq{\begin{equation}}
\def\eeq{\end{equation}}

\newcommand{\lsim}{\mathrel{\rlap{\lower4pt\hbox{\hskip1pt$\sim$}}
    \raise1pt\hbox{$<$}}}         
\newcommand{\gsim}{\mathrel{\rlap{\lower4pt\hbox{\hskip1pt$\sim$}}
    \raise1pt\hbox{$>$}}}         

\newcommand{\leftrightarrowraised}{\mathrel{\rlap{\lower-0pt\hbox{\hskip1pt$\partial$}}
    \raise6 pt\hbox{$\leftrightarrow$}}}

\newcommand{\vect}[1]{\boldsymbol{\rm #1}}
\newcommand{\Ho}{H_{1\textrm{DM}}}
\newcommand{\Ht}{H_{2\textrm{DM}}}
\renewcommand{\L}{\mathcal{L}}
\newcommand{\T}{\mathcal{T}}

\renewcommand{\L}{\mathcal{L}}
\newcommand{\boldtheta}{\boldsymbol{\theta}}

\newcount\hour \newcount\minute
\hour=\time \divide \hour by 60
\minute=\time
\title{On the direct detection of multi-component dark matter: sensitivity studies and parameter estimation}

\author[a,1]{Juan Herrero-Garcia,\note{\url{http://orcid.org/0000-0002-3300-0029}}}
\author[a,2]{Andre Scaffidi,\note{\url{http://orcid.org/0000-0002-1203-6452}}}
\author[a]{Martin White}
\author[a,3]{and Anthony G. Williams \note{\url{http://orcid.org/0000-0002-1472-1592}}}

\emailAdd{juan.herrero-garcia@adelaide.edu.au}
\emailAdd{andre.scaffidi@adelaide.edu.au}
\emailAdd{martin.white@adelaide.edu.au}
\emailAdd{anthony.williams@adelaide.edu.au}

\subheader{ADP-17-33/T1039}

\affiliation[a]{ARC Centre of Excellence for Particle Physics at the Terascale, Department of Physics, University of Adelaide, Adelaide, South Australia 5005, Australia}

\abstract{
We study the case of multi-component dark matter, in particular how direct detection signals are modified in the presence of several stable weakly-interacting-massive particles. Assuming a positive signal in a future direct detection experiment, stemming from two dark matter components, we study the region in parameter space where it is possible to distinguish a one from a two-component dark matter spectrum. First, we leave as free parameters the two dark matter masses and show that the two hypotheses can be significantly discriminated for a range of dark matter masses with their splitting being the critical factor. We then investigate how including the effects of different interaction strengths, local densities or velocity dispersions for the two components modifies these conclusions. We also consider the case of isospin-violating couplings. In all scenarios, we show results for various types of nuclei both for elastic spin-independent and spin-dependent interactions. Finally, assuming that the two-component hypothesis is confirmed, we quantify the accuracy with which the parameters can be extracted and discuss the different degeneracies that occur. This includes studying the case in which only a single experiment observes a signal, and also the scenario of having two signals from two different experiments, in which case the ratios of the couplings to neutrons and protons may also be extracted.}


\begin{document}
\maketitle

\section{Introduction} \label{sec:intro}
We know from gravitational effects that dark matter (DM) constitutes a significant
fraction of the energy density in the universe, but no confirmed detection in the laboratory has been made so far. Some of the most popular candidates are Weakly-Interacting-Massive-Particles (WIMPs), in particular those that have non-vanishing interactions with the standard model (SM) and therefore can be tested. In fact, they are actively being searched for in direct detection (DD) experiments, which look for their nuclear scatterings in underground detectors~\cite{Goodman:1984dc}. Interestingly, current and planned next-generation experiments are probing a very large portion of the parameter space of well-motivated theories of WIMPs. 

A plausible scenario is that DM is not made up of a single species, but that it has a multi-component nature. In this work we study direct detection signals in the presence of multi-component WIMP-like DM, i.e., several types of WIMPs (labelled by Greek sub indices $\alpha=1, 2\,... ,N$) with individual global energy density $\Omega_\alpha$ such that they constitute the observed total DM energy density of the Universe, $\Omega_{\rm DM} = \sum_\alpha^N \Omega_\alpha$. Purely on theoretical grounds, having the individual energy densities (the global $\Omega_\alpha$, or the local $\rho_\alpha$) exactly equal would seem to be a highly unnatural scenario, requiring a fine-tuning between masses and number densities (unless there is some underlying mechanism to equalise the densities). On the other hand, that the densities are similar up to order one factors seems rather plausible, that is to say, that there are several species contributing in a non-negligible way to the global and local energy densities. For instance, in the SM there are baryons forming different stable nuclei with a non-negligible density: H, He, Li..., and also electrons, photons and neutrinos. Therefore, it is not difficult to imagine that a similar situation could occur in the dark sector, which has an energy density five times larger than the visible one. 

There have been only a few works in the past regarding the direct detection of multi-component DM~\cite{Adulpravitchai:2011ei,Batell:2009vb,Dienes:2012cf,Profumo:2009tb,Bhattacharya:2017fid,Bhattacharya:2016ysw,Chialva:2012rq}. Let us discuss the main points studied there and the most relevant differences with our analysis. In Ref.~\cite{Batell:2009vb} the authors considered very small mass splittings ($<200$ keV) for the particles, such as arise in inelastic scenarios~\cite{Blennow:2015hzp,Bozorgnia:2013hsa,TuckerSmith:2004jv}. In Ref.~\cite{Dienes:2012cf} a continuum spectrum of closely spaced DM particles was considered. In this work we will focus mainly on the case of two DM states and consider splittings comparable to the DM masses. In Refs.~\cite{Profumo:2009tb,Dienes:2012cf} the possibility of testing multi-component DM using collider, indirect and direct searches was studied. In Ref.~\cite{Profumo:2009tb} the authors related the WIMP-nucleon scattering cross-section to the annihilation one, motivated by thermal freeze-out and supersymmetric scenarios. Therefore they were able to express the scattering cross-section as a function of the global (and local) density and the DM mass. In our study we will keep the analysis as phenomenological and model-independent as possible. In particular, we will not make any assumptions in the numerical analysis regarding the production mechanisms for the DM, i.e., we will keep the abundance and the scattering cross-sections as independent parameters. Furthermore, we will adopt a different statistical approach to that in Ref.~\cite{Profumo:2009tb} and we will study the effects of all relevant parameters entering the scattering rate. As pointed out in this last reference, indirect detection of one DM particle could mimic the effects of two components, as it may annihilate/decay not only to two gamma rays but also to Higgs/Z plus a photon. In addition these interactions are loop-suppressed. Regarding colliders, the authors studied the case of models with charged partners decaying into DM. In Ref.~\cite{Giudice:2011ib} the authors also considered the possibility of discriminating the number of DM particles generating missing energy distributions. In this work we will focus on how multi-component DM can be studied using only information stemming from direct detection signals. 

We will first discuss some general expectations of multi-component DM. In particular, the relevant quantity that enters in DD is the local DM number density, $n_{\rm loc}=\rho_{\rm loc}/m_{\rm DM}$, and we will discuss this in some detail. Afterwards, in order to draw quantitative conclusions, we will focus on the case of having a two-component DM signal in a direct detection experiment. The questions we would like to answer are two-fold: first, how significantly can we distinguish the one-component and the two-component hypotheses? Second, assuming that the two components can indeed be distinguished, with what accuracy can we extract the DM properties? In order to answer these questions, we will simulate a signal generated by two DM components in different target nuclei, assuming either spin-independent (SI) or spin-dependent (SD) interactions. We will adopt a frequentist framework in order to compare the one- and two-component hypotheses and to extract the preferred regions for the free parameters. We will also discuss the case of having two experimental signals, and how accurately one can extract the DM properties in this case, including the couplings to neutrons and protons.

This paper is structured as follows. In sec.~\ref{sec:dominance} we give some general remarks on multi-component DM. This section may be skipped for readers interested more in the main results of this work. In sec.~\ref{sec:direct} we present the framework relevant for direct detection of multi-component DM. In sec.~\ref{sec:hypothesis_test} we analyse in detail the case of a two-component DM event rate and how, from a given signal, the one- and two-DM hypothesis can be discriminated. Assuming that the 2DM hypothesis has been confirmed, in sec.~\ref{sec:par_estimation} we study the extraction of the DM parameters.  We conclude in sec.~\ref{sec:conc}. Finally we elaborate on some statistical methods used in App.~\ref{appendix:Tstat}, and show the results of the parameter estimation using Bayesian methods in App.~\ref{app:bayes}.

 \section{General remarks on multi-component dark matter densities}  \label{sec:dominance}

The relevant quantity for DD is the DM local energy density in the solar neighbourhood, which should obey $\rho_{\rm loc} = \sum_\alpha^N \rho_\alpha$. More precisely, it is the local \emph{number} density that is important in determining the total event rate in a DD experiment.  Given arbitrary local energy densities for two DM components, it is not entirely straightforward to determine which one dominates the event rate. For example, even though the heavier DM mass has a smaller number density, it probes a larger part of the velocity distribution, and therefore it is not straightforward which particle dominates the rate, the lighter or the heavier.

One feature to keep in mind is that the proportionality of the global and local densities is expected in the case of cold DM, such as WIMPs~\cite{Bertone:2010rv} (see also its effects in Ref.~\cite{Blennow:2015gta}). We will assume such a proportionality in the following discussion. Furthermore, for simplicity, we will focus on the case of two DM components, taken to have masses $m_2>m_1$ without loss of generality. 

The production mechanism of DM is unknown. Up to now, we have defined multi-component DM as the scenario in which several particles have similar energy densities. Another option is that they have similar number densities. The first one is relevant in the context of thermal freeze-out, while the second one could in principle be more natural in asymmetric scenarios (see Ref.~\cite{Petraki:2013wwa} for a review on the topic). In the latter case, however, the masses are typically of $\mathcal{O}$(GeV), and therefore DD is quite challenging, although detectors with low thresholds exist (typically with Germanium). Focusing on two components, in the case of similar local energy densities, we have that
\beq \label{eq:equalrho}
\rho_1 \approx \rho_2 \approx \frac{\rho_{\rm loc}}{2}\,,
\eeq
i.e., both species have a significant contribution to the energy density. Instead, for similar number densities $n_1 \approx n_2 \approx n_{\rm{loc}}$, we have
\beq \label{eq:equaln}
\rho_{\rm{loc}}= n_{\rm{loc}} (m_1+m_2) \approx n_{\rm{loc}}\, m_2\,,
\eeq
where in the last step we assumed $m_2 \gg m_1$. Therefore,
\beq\label{eq:R} \nonumber
n_{\rm{loc}}\approx\frac{\rho_{\rm{loc}}}{m_2}\,.
\eeq 
Of course, in the case of equal masses, i.e., $m_1=m_2$ (as roughly expected for asymmetric DM), both expressions in eqs.~\eqref{eq:equalrho} and \eqref{eq:equaln} are equivalent. This means that, in this case, direct detection signals for multi-component DM models are suppressed by the heaviest DM mass (really the total sum of the masses), and similarly with indirect detection signals.

\begin{figure}
	\centering
	\includegraphics[width=0.48\textwidth]{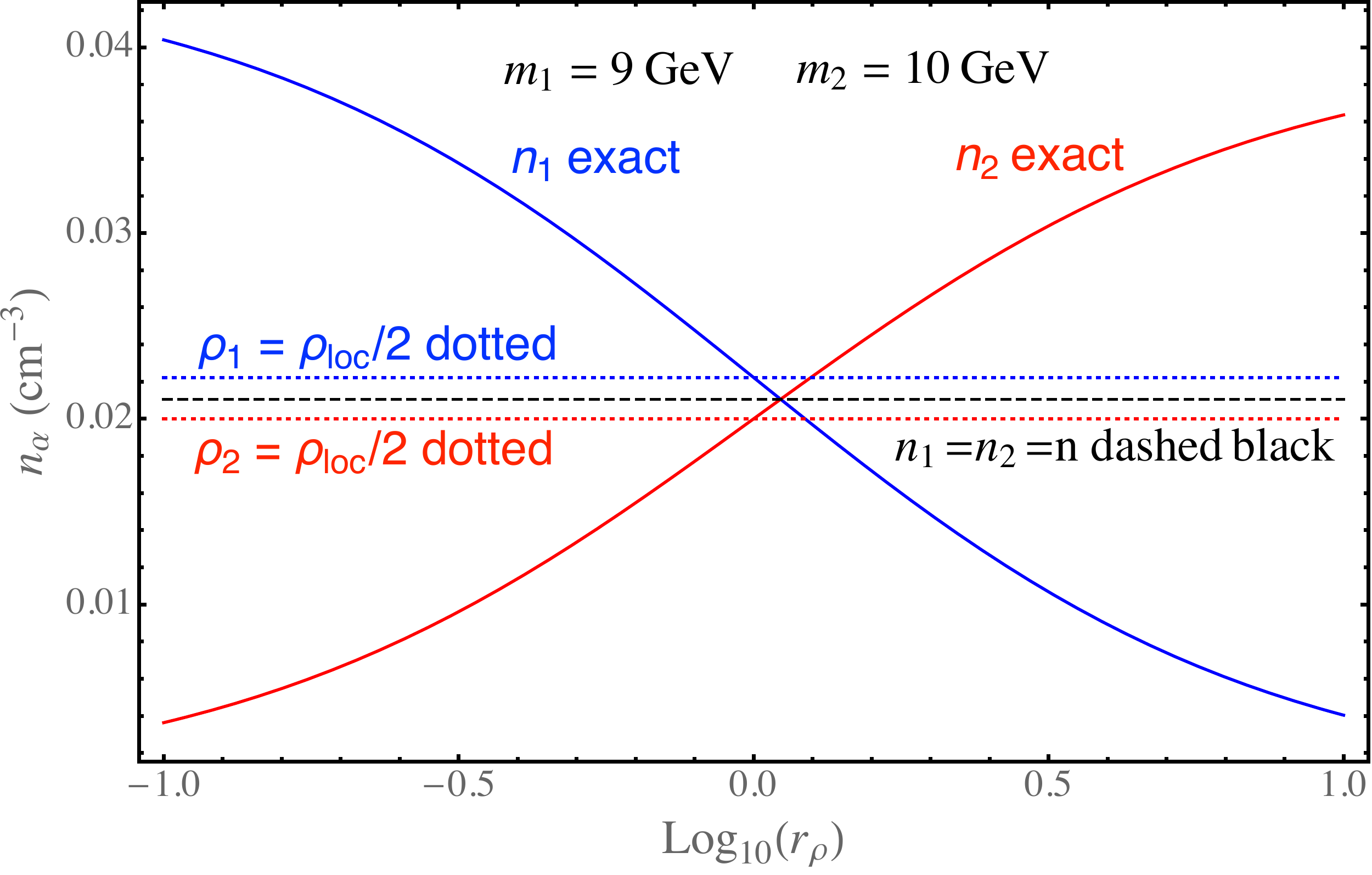}~~~\includegraphics[width=0.48\textwidth]{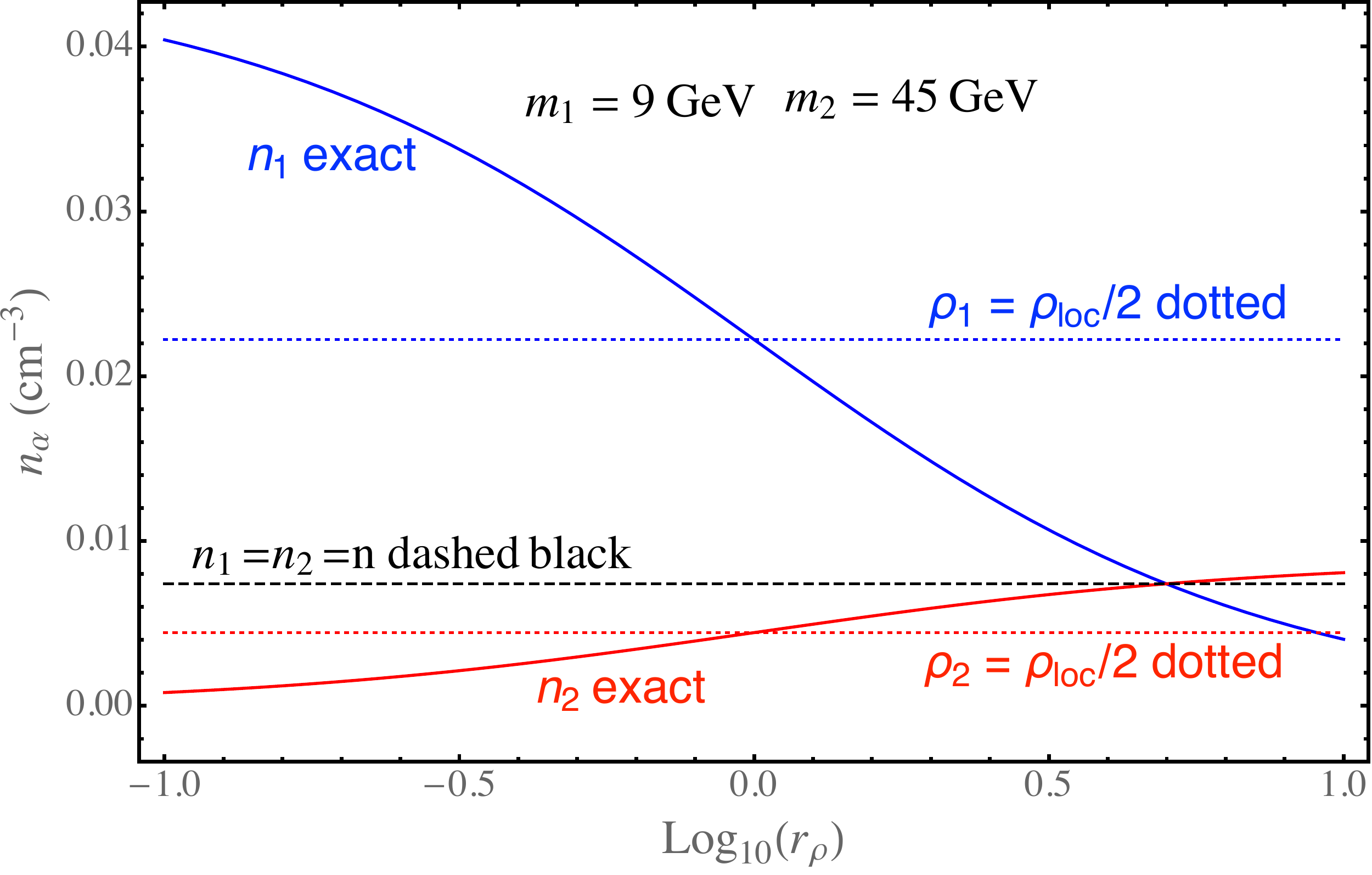}
	\caption{Number densities $n_\alpha$ with $\alpha=1,2$ versus $\log_{10}(r_\rho)$, where $r_\rho=\rho_2/\rho_1$. In the left panel we plot $m_1=9$ GeV and $m_2=10$ GeV. In the right panel, we plot $m_1=9$ GeV and $m_2=45$ GeV. We show the exact values of $n_\alpha$ in solid. Also shown are the different approximations: $n_1=n_2 =n$ (dashed black), $\rho_1=\rho_{\rm loc}/2$ (dotted blue) and $\rho_2=\rho_{\rm loc}/2$ (dotted red). For $m_1=m_2$ all three approximations collapse into a single horizontal line. A plot for $m_1=45$ GeV, $m_2=9$ is the mirror image of the right plot with respect to a vertical axis passing through $\log_{10}(r_\rho)=0$. } \label{fig:Fone}
\end{figure}

The number densities $n_1,\,n_2$ are plotted in Fig.~\ref{fig:Fone} versus $r_\rho=\rho_2/\rho_1$ for $m_1=9$ GeV and $m_2=10$ GeV in the left figure, and for $m_1=9$ GeV and $m_2=45$ GeV in the right one. We show the exact values in solid, and the different approximations $n_1=n_2$ in dashed black, $\rho_1=\rho_{\rm loc}/2$ in dotted blue and $\rho_2=\rho_{\rm loc}/2$ in dotted red. In the case of $\rho_1=\rho_2=\rho_{\rm loc}/2$, for $r_\rho=1$ it is of course exact. One can also see that taking $n_1 = n_2$ gives the average of the exact number densities. In the regime where the local energy density of particle 1 (2) is large, i.e. $r_\rho \ll 1 (\gg 1)$, the respective approximations are a (roughly constant) factor of two weaker, while in the opposite regime they are more discrepant.

When the masses are similar (left panel), both approximations are roughly the same. In fact all the approximations collapse for $m_1=m_2$. On the other hand, when the masses are very different (right panel), the approximations are quite different. In the case of $n_1=n_2$, it is an exact solution for $r_\rho =m_2/m_1= 1.1\,(5)$ in the left (right panel). This qualitative behaviour is similar for other DM masses: the assumption of equal number densities is an average of the real number densities for any true value of the local energy densities, while that of equal energy densities is a worse approximation at least for one of the two candidates whenever $r_\rho \neq 1$. 

In the rest of the paper, we will not make any approximation regarding energy densities, which we will treat as free parameters, just subject to the constraint that their sum gives the observed local energy density.

\section{Direct detection of multi-component DM} \label{sec:direct}

\subsection{The differential event rate} \label{sec:rate}
We present in the following the general notation for multi-component DM in detectors with different types of nuclei. For the time-being, we assume elastic spin-independent (SI) scattering of DM particles $\chi_\alpha$ ($\alpha=1, 2\,... ,N$) with masses $m_\alpha$ off nuclei with atomic and mass numbers $(Z_j,A_j)$ ($j=1, 2\,.. M$), depositing the nuclear recoil energy $E_{R}$. The total differential rate (usually measured in events/keV/kg/day) observed by a detector is given by the sum of the event rates of the individual DM particles on each of the nuclear elements:
\beq\label{eq:Rt} 
R (E_{R}, t) = \sum_{j=1}^M\sum_{\alpha=1}^N R_{j}^\alpha(E_{R}, t) \,,
\eeq 
where
\beq\label{eq:R0} 
R^\alpha_j(E_{R}, t) = x_j
\frac{\rho_\alpha \sigma_\alpha^p}{2 m_\alpha \mu_{\alpha p}^2} \, (A_{\alpha, j}^{\rm eff})^2 F_j^2(E_{R}) \,
\eta_{\alpha, j}(v_{m,j}^{(\alpha)}, t) \,, 
\eeq 
with $\rho_\alpha$ the individual local DM energy density (with the restriction $\rho_{\rm loc}=\sum_{\alpha=1}^N \rho_\alpha$), $\sigma_\alpha^p$ the individual DM--proton
scattering cross-section at zero momentum transfer,
$\mu_{\alpha p}$ the $\chi_\alpha$ particle--proton reduced mass and $F_j(E_{R})$
the nuclear form factor of element $j$. We also denoted the effective mass-number of the nucleus $j$ with DM $\alpha$ by 
\begin{align}
\label{AeeffDef}
A_{\alpha, j}^{\rm eff}= Z_j+(A_j-Z_j) \kappa_\alpha\,,
\end{align}
with 
\begin{align}
\kappa_\alpha\equiv f_\alpha^n/f_\alpha^p\;,
\end{align}
where $f_\alpha^{n,p}$ are the SI individual couplings of the DM particle $\alpha$ to neutrons and protons. $x_j$ is the mass fraction of element $j$ in the detector, i.e., $x_j=m_j/(\sum_j^M m_j)$.

We will discuss spin-independent (SI) and spin-dependent (SD) interactions in this paper; in the case of  SD interactions, Eq.~\eqref{eq:R0} can be used by substituting $A_{\alpha j}^{\rm eff} \rightarrow 1$ and the form factor $F_j^2(E_{R}) \rightarrow F_{\alpha,\, j}^{\rm SD}(E_{R}, \kappa_\alpha)$ now has a $\kappa_\alpha$ dependence. In the numerical analysis, for the SI form factors we will use the Helm parametrisation \cite{PhysRev.104.1466,LEWIN199687}, while for SD in xenon (and fluorine) we will use the results of Ref.~\cite{Klos:2013rwa}.

In addition to $\rho_{\alpha}$, the astrophysics enters in Eq.~\eqref{eq:R0} through the halo integral
\beq\label{eq:eta} 
\eta_{\alpha, j}(v_{m,j}^{(\alpha)}, t) \equiv \eta (f^{(\alpha)}_{\rm det},v^{(\alpha)}_{m,j}, t)=
\int_{v > v_{m,j}^{(\alpha)}} \negthickspace \negthickspace d^3 v 
\frac{f^{(\alpha)}_{\rm det}(\vect{v}, t)}{v} \,,
\eeq
with
\beq\label{eq:vm} 
v^{(\alpha)}_{m,j}\equiv v_m (m_\alpha,\,m_j)=\sqrt{ \frac{m_j E_{R}}{2 \mu_{\alpha j}^2}},
\eeq
where $v_{m,j}^{(\alpha)}$ is the minimal velocity of the particle $\alpha$ required to produce a recoil of energy $E_R$ in element $j$, and 
$f^{(\alpha)}_{\rm det}(\vect v, t)$
describes the distribution of DM particle velocities in the detector rest
frame, with $f^{(\alpha)}_{\rm det}(\vect v, t) \ge 0$ and $\int d^3 v f^{(\alpha)}_{\rm det}(\vect
v, t) = 1$.
 The velocity distributions in the rest frames of the detector and the galaxy are
related by a Galilean transformation, $f^{(\alpha)}_{\rm det}(\vect{v},t) = f^{(\alpha)}_{\rm gal}(\vect{v} + \vect{v}_e(t))$, where $\vect{v}_e(t)$ is the velocity vector of the Earth in the galaxy rest-frame.  Notice that $\eta_{\alpha j}(v_{m,j}^{(\alpha)})$ is a decreasing function of $v_{m,j}^{(\alpha)}$, which for large DM masses does not depend on $m_\alpha$. Throughout this paper, we will use the so-called Standard Halo Model (SHM), with   $\rho_{\rm loc}^{\rm exp} \simeq 0.4 \, {\rm GeV/cm}^3$, a Maxwellian velocity distribution $f_{\rm gal}(v)=\frac{1}{(2 \pi \sigma^2_{H_\alpha})^{3/2}} \exp{\Big(-\frac{3v^2}{2\sigma_{H_\alpha}^2}\Big)}$, and a cut-off at the escape velocity $v_{\rm esc} =550\,\rm km\, s^{-1}$. 

In principle, $\eta_{\alpha j}(v_{m,j}^{(\alpha)}, t)$ depends on the DM particle $\alpha$ in two different ways: directly, via its velocity distribution $f^{(\alpha)}(\vect v, t)$ (which, in addition may depend in a non-trivial way on the micro-physics of the DM, like its mass and interactions) and indirectly, through its mass $m_\alpha$ that enters into $v_{m,j}^{(\alpha)}$ (unless $m_\alpha \gg m_j$, in which case the dependence on the DM mass drops, $v_{m,j}^{(\alpha)} \rightarrow v_{m,j}$). In the following, we will assume that the functional form of the velocity distributions of the different DM components is equal, i.e., $f^{(\alpha)}(\vect v, t) \equiv f(\vect v, t)$. We will however consider the case of different velocity dispersions later on. Also, we will focus on constant rates, i.e., averaged over the year, so that
\begin{equation} \label{eq:aver}
\bar{R}(E_{R})=\frac{R(E_{R}, t_{\rm max})+ R(E_{R}, t_{\rm min})}{2}\,,
\end{equation}
where $t_{\rm max}\,(t_{\rm min})$ are the times of the year at which the rate reaches a maximum (minimum). \footnote{We will not discuss annual modulation signals, the reader is referred to refs.~\cite{Gelmini:2000dm,Savage:2006qr,Freese:2012xd,HerreroGarcia:2012fu,HerreroGarcia:2011aa,Lee:2013wza,Bozorgnia:2014dqa,DelNobile:2015nua} for studies on the topic.}

\subsection{The rate for two DM particles} \label{sec:framework}

We now fix the notation for the DD signals expected from 2 DM particles $\alpha=1,2$ with masses $m_1<m_2$, cross-sections with protons $\sigma_1^p,\,\sigma_2^p$, and densities $\rho_1,\,\rho_2$, such that $\rho_1+\rho_2=\rho_{\rm loc}$. We will also study their signals in two different detectors $j=A_1,\,A_2$, with mass and atomic numbers $(A_1,Z_1)\neq\,(A_2,Z_2)$, and taken to have mass fractions $x_1=x_2=1$ for simplicity (in the case of xenon, we will consider the different mass fractions of its isotopes). We define
\beq\label{eq:rhos} 
r_{\rho} \equiv \frac{\rho_2}{\rho_1}\,,\qquad  \text{such that}\qquad  \rho_2 = r_\rho\,\rho_1 =r_\rho\,\frac{\rho_{\rm loc}}{1+r_\rho}\,. 
\eeq 
From eqs.~\eqref{eq:Rt}, \eqref{eq:R0} and \eqref{eq:aver} (dropping the bar from the notation of $R(E_R)$), and using $\mu_{1p}=\mu_{2p}=m_p$, we can write the total rate as:
\begin{align}
\label{eq:rate_tot}
R_{1}(E_{R}) &=R_1^1(E_R) + R_1^2(E_R) \nonumber\\&=  C(r_\rho, \sigma_1^p) \,F_{A_1}^2(E_{R}) \,\left(\frac{(A^{\rm eff}_{1,1})^2}{m_1}\,\eta(v_{m,A1}^{(1)}) +\frac{(A^{\rm eff}_{2,1})^2}{m_2}\,r_{\rho}\,r_{\sigma}\,\eta(v_{m,A1}^{(2)})\right)\,,
\end{align}
where we defined
\beq\label{eq:Ci2} 
C(r_\rho,\sigma_1^p) \equiv \frac{\rho_{\rm loc}\,\sigma_1^p}{2 \,(1+r_\rho)\,m_p^2}\,,\,\qquad \text{and} \qquad r_{\sigma} \equiv \frac{\sigma_2^p}{\sigma_1^p} \,.
\eeq 
$R_{A_2}(E_{R}) $ is similar, after making the following substitutions: $A_1 \rightarrow A_2$, $A^{\rm eff}_{1,1} \rightarrow A^{\rm eff}_{1,\,2}$ and $A^{\rm eff}_{2,1} \rightarrow A^{\rm eff}_{2,2}$. In the case of a DM signal generated from two components, it is clear that the particle masses will determine the slope of their individual rates, while $r_\rho, r_\sigma$ and $\kappa_{1,\,2}$ will determine their relative normalisation. Therefore, a first conclusion is that using just information from a given direct detection signal, we can only distinguish if there are one or two components if the particles have different masses.

In the following numerical analysis, we will use different targets: fluorine, sodium, germanium and xenon. The strongest limits for the SI cross-section come from XENON1T~\cite{Aprile:2017iyp}, and are  $10^{-45}\,{\rm cm}^2$ for a 10 GeV DM particle and $10^{-46}\,{\rm cm}^2$ for $m_{\rm DM}=30$ GeV at 90\% C.L. For SD couplings with protons, PICO-2L~\cite{Amole:2016pye} (and PICO-60~\cite{Amole:2017dex} for heavier masses) set the strongest bounds, at the level of $10^{-40}\,{\rm cm}^2$ for a 10 GeV DM particle. The values of the energy threshold ($E_{\rm th}$), mass, time and exposures for the different nuclei based on future expected experimental sensitivities are given in table~\ref{tab:future_exps}. Notice that there are many different proposed experiments, and very large uncertainties are present in the literature regarding these values. For Ge, although smaller thresholds are possible ($\lesssim$ keV), we take a conservative value similar to the other detector ones. Therefore our analysis can be understood as a proof of concept, with more sophisticated experimental simulations needed once there is a signal. We provide some examples of illustrative proposed experiments in the last column of the table. \footnote{This is by no means an exhaustive list, and other elements and experiments are also very promising, for instance those using argon~\cite{Fatemighomi:2016ree,Calvo:2016hve}, which however typically have higher energy thresholds ($\sim \mathcal{O}(20)$ keV). Moreover, in addition to XENONnt~\cite{Aprile:2014zvw}, other very promising xenon experiments are DARWIN~\cite{Akerib:2015cja} and LZ~\cite{Aalbers:2016jon}. For further details of the current status of DD experiments, the interested reader is referred to Refs.~\cite{Undagoitia:2015gya,Liu:2017drf}.} We also provide in the third column the minimum DM mass that can be detected for each element assuming perfect energy resolution.

\begin{table}[tb!]
\centering
\begin{tabular}{ccccccc}
Element & $E_{\rm th}$ (keV) &$m^{\rm min}_{\rm DM}$ (GeV) & M (t) & T (y) & MT ($t \cdot y$)& Experiments  \\ \hline
F &  3 & 2& 0.5&2& 1&PICO-500~\cite{Amole:2017dex} \\
Na &  3 &2.5 & 0.25&10& 2.5&PICO-LON~\cite{Fushimi:2015sew} \\
Ge &  2 & 3.5& 0.4&10& 4&SuperCDMS~\cite{Agnese:2016cpb} \\
Xe &  1 & 3& 2&1.5& 3&XENONnT~\cite{Aprile:2014zvw}\
\end{tabular}
\caption{Experimental values used in the numerical analysis for future direct detection experiments. Columns 2 to 6 show respectively the recoil energy threshold $E_{\rm th}$ (keV), the minimum DM mass that can be detected in GeV, the mass of the detector $M$ in tonnes, the data collection time $T$ in years and the total exposure $MT$ ($t \cdot y$). The last column shows an illustrative experimental reference for the type of experiment considered. For xenon we consider both isotopes Xe$^{129}$ and Xe$^{131}$ with mass fractions $x_j$ equal to $0.264$ and $0.212$, respectively.}
\label{tab:future_exps}
\end{table}

\begin{figure}
	\centering
	\includegraphics[width=0.48\textwidth]{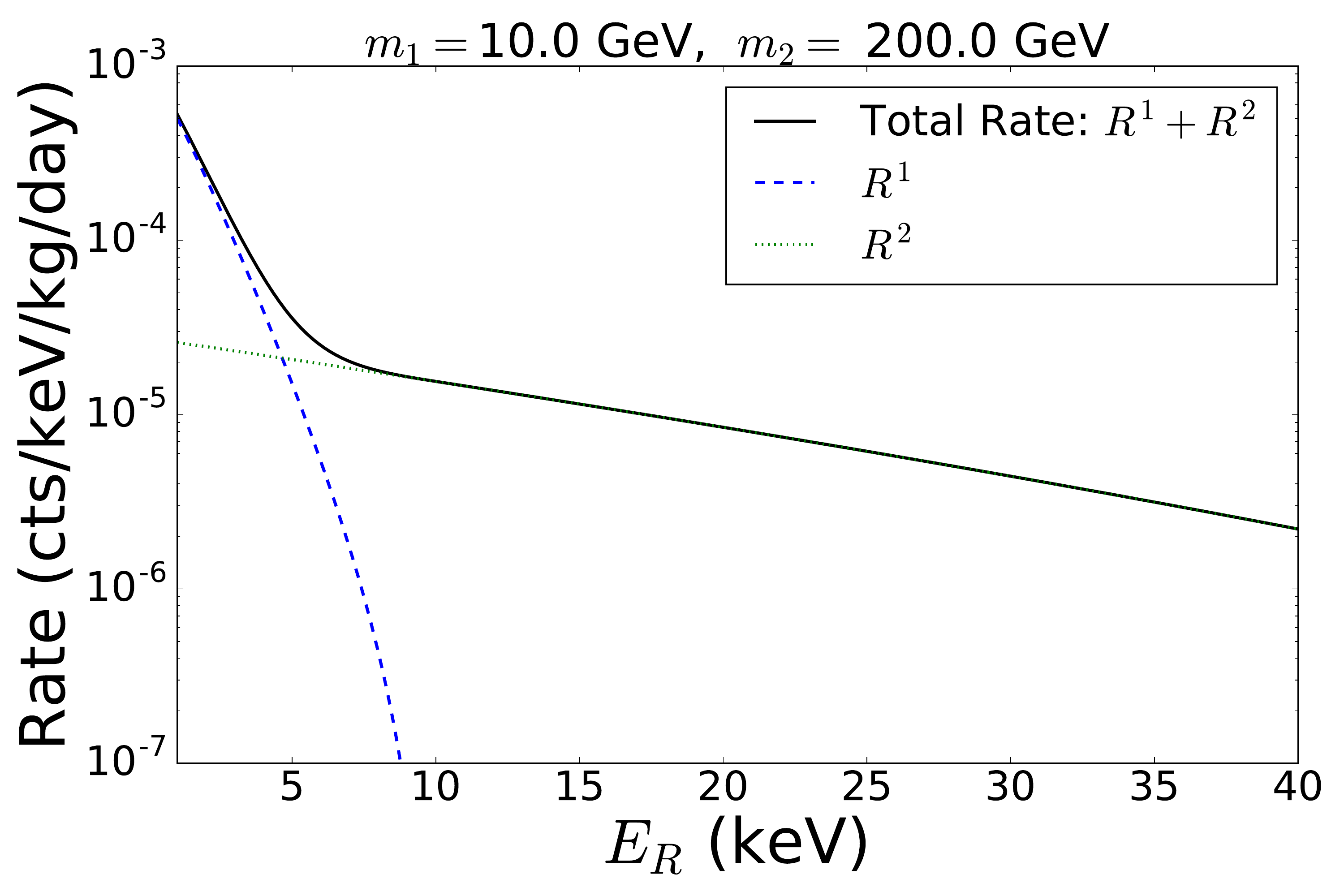}
	\includegraphics[width=0.48\textwidth]{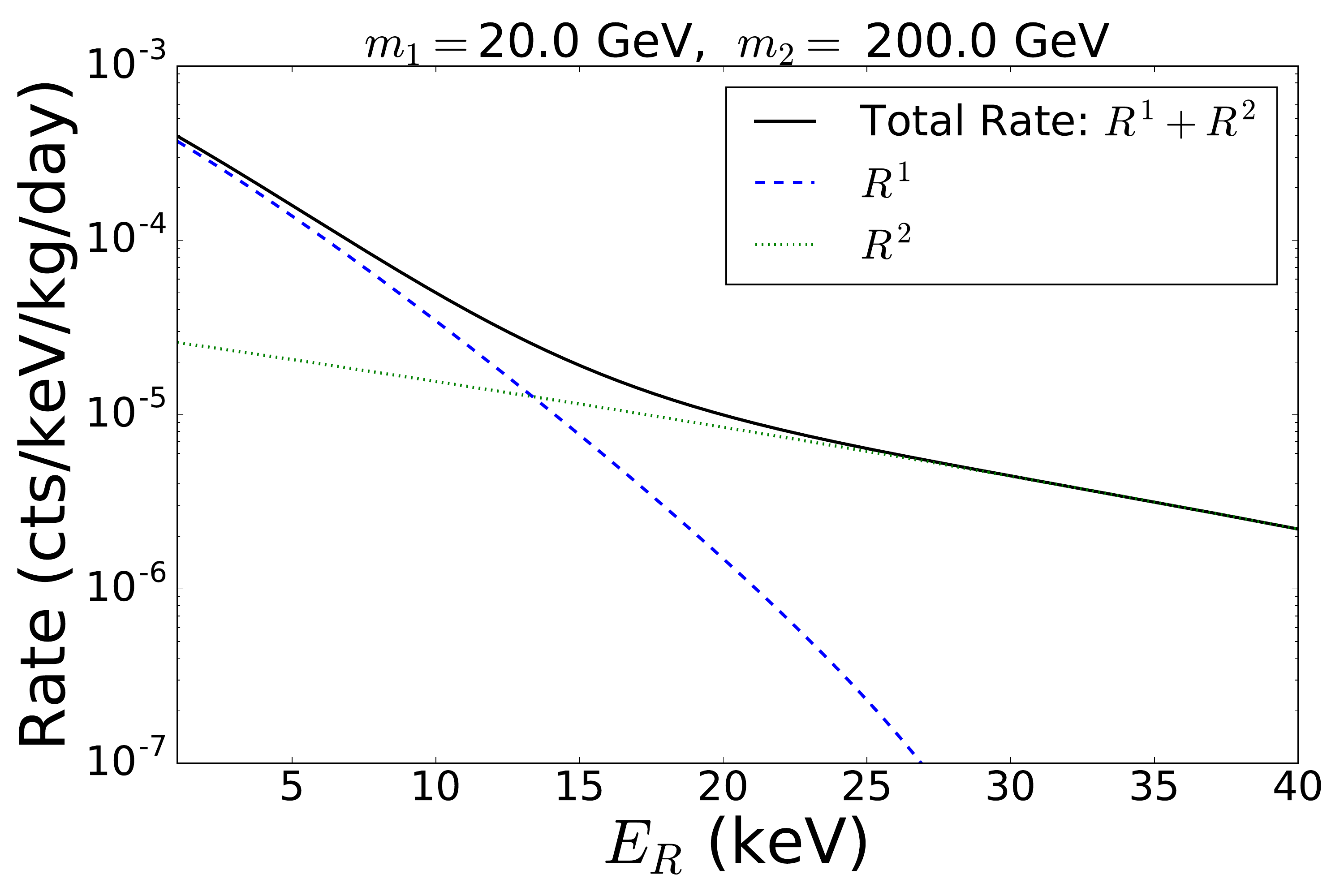}
	\includegraphics[width=0.48\textwidth]{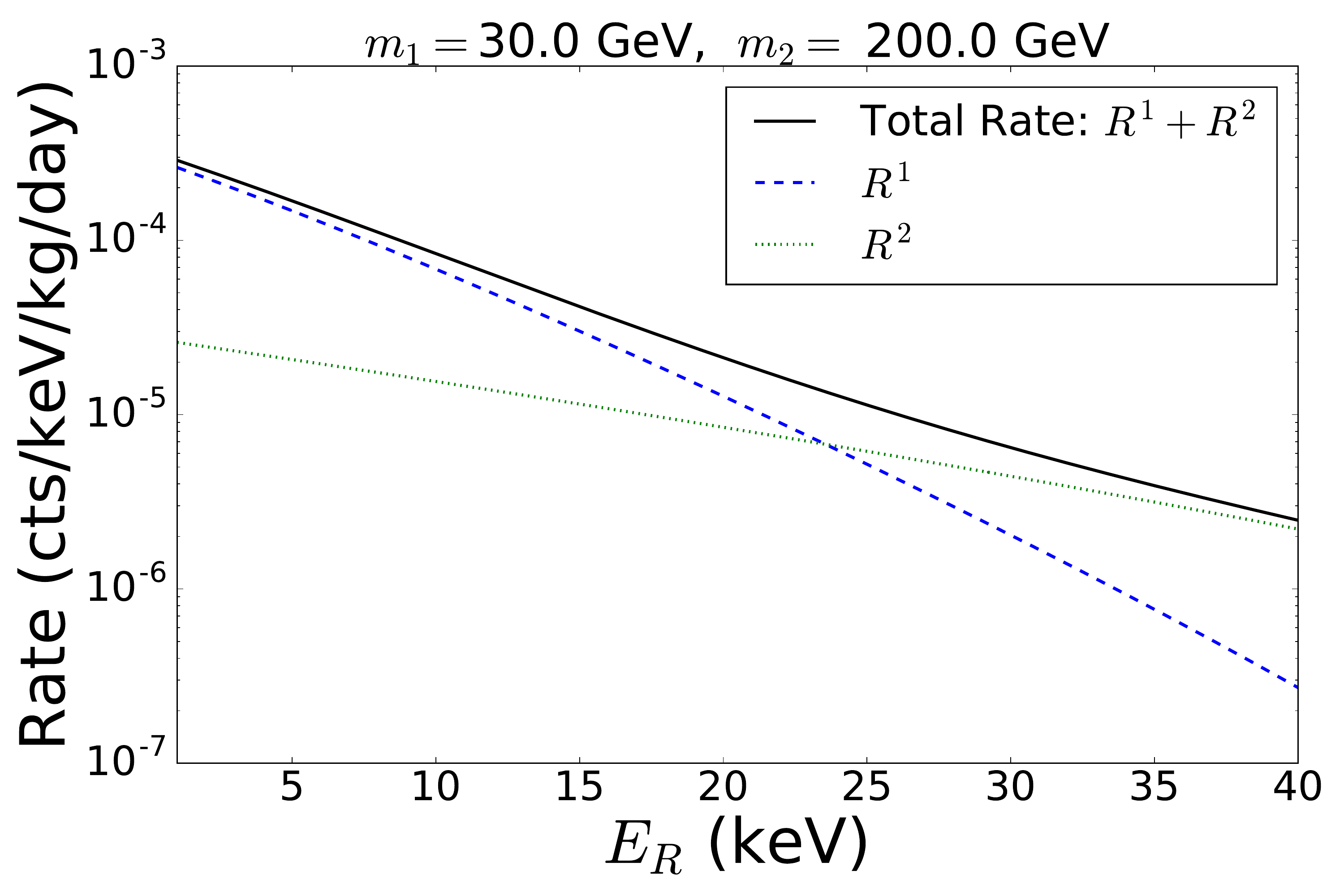}
	\includegraphics[width=0.48\textwidth]{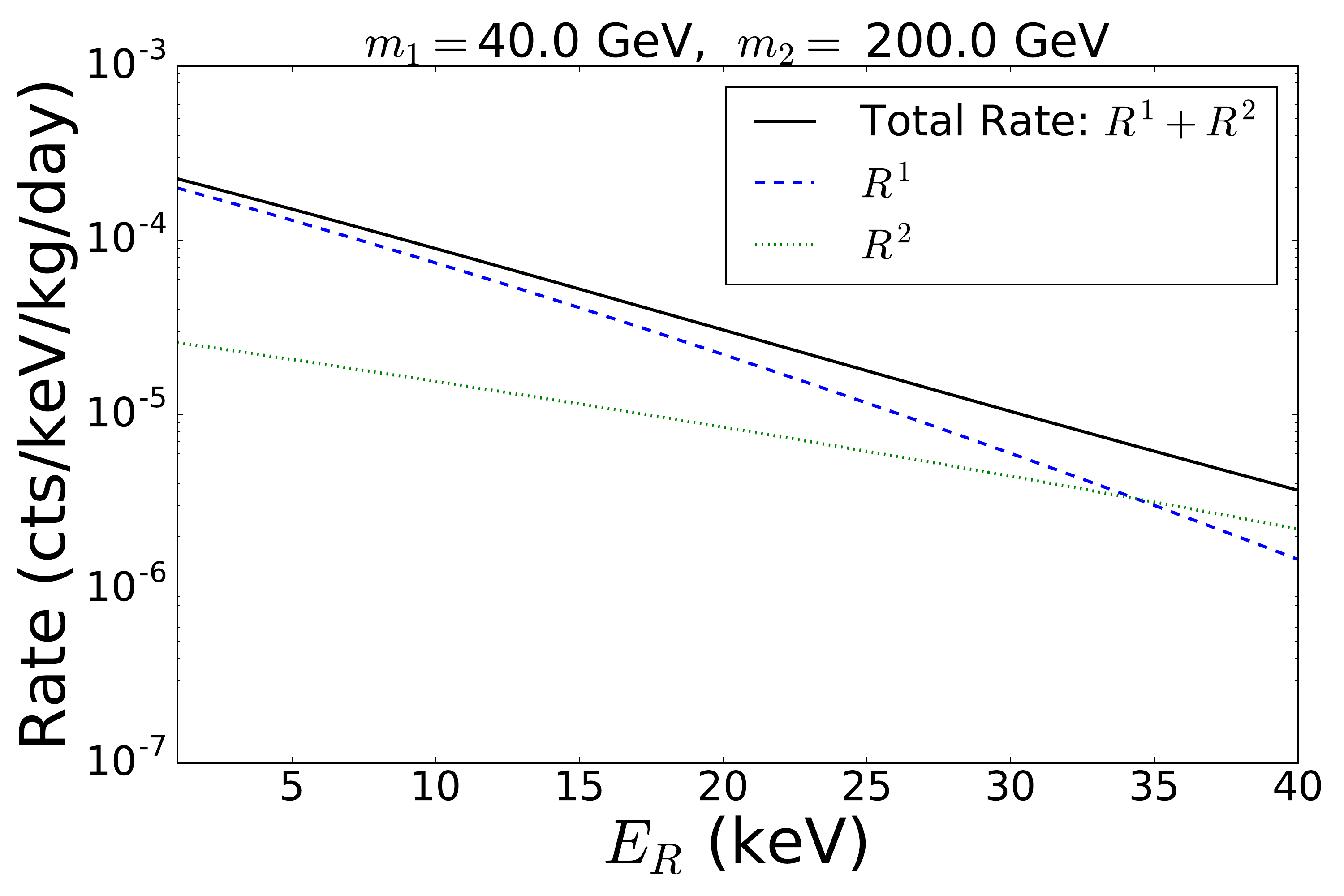}
	\caption{Total differential event rate for 2 DM particles (solid black), as well as their individual contributions (1 dashed blue, 2 dotted green) for a variety of DM mass splittings on the energy range $[2,30]$ keV. One should notice that the \emph{kink} feature in the combined spectrum rapidly vanishes with smaller mass splittings.  } \label{Rates_2comp}
\end{figure}

In Fig.~\ref{Rates_2comp} we show the differential spectrum of two component DM for a variety of DM mass splittings. One can see the different slopes of the two components and the presence of a €˜\emph{kink}€™ in the total rate, which rapidly vanishes for smaller mass splittings. This is the smoking gun of multi-component DM. We have checked that sensible energy resolutions do not significantly affect the spectra, and in the following we assume perfect energy resolution and efficiency. 

Something important to keep in mind in the case of multi-component DM is that, the heavier DM particle may contribute to the rates in different experiments. On the other hand, the lightest particle may only scatter in the lightest detector, as in the heaviest detector the recoils may be below its energy threshold. 

\begin{figure}
	\centering
	\includegraphics[width=0.48\textwidth]{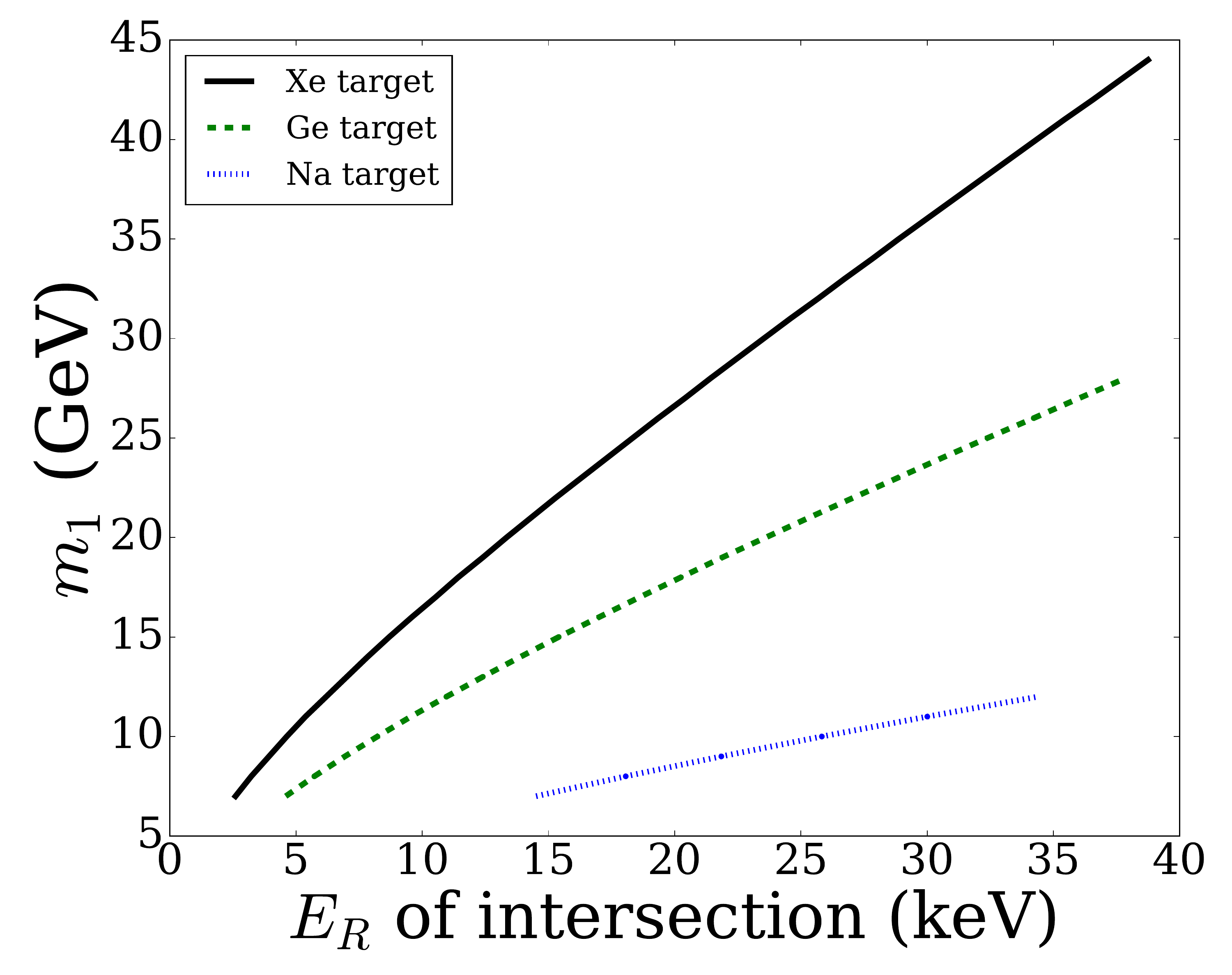}
	\includegraphics[width=0.48\textwidth]{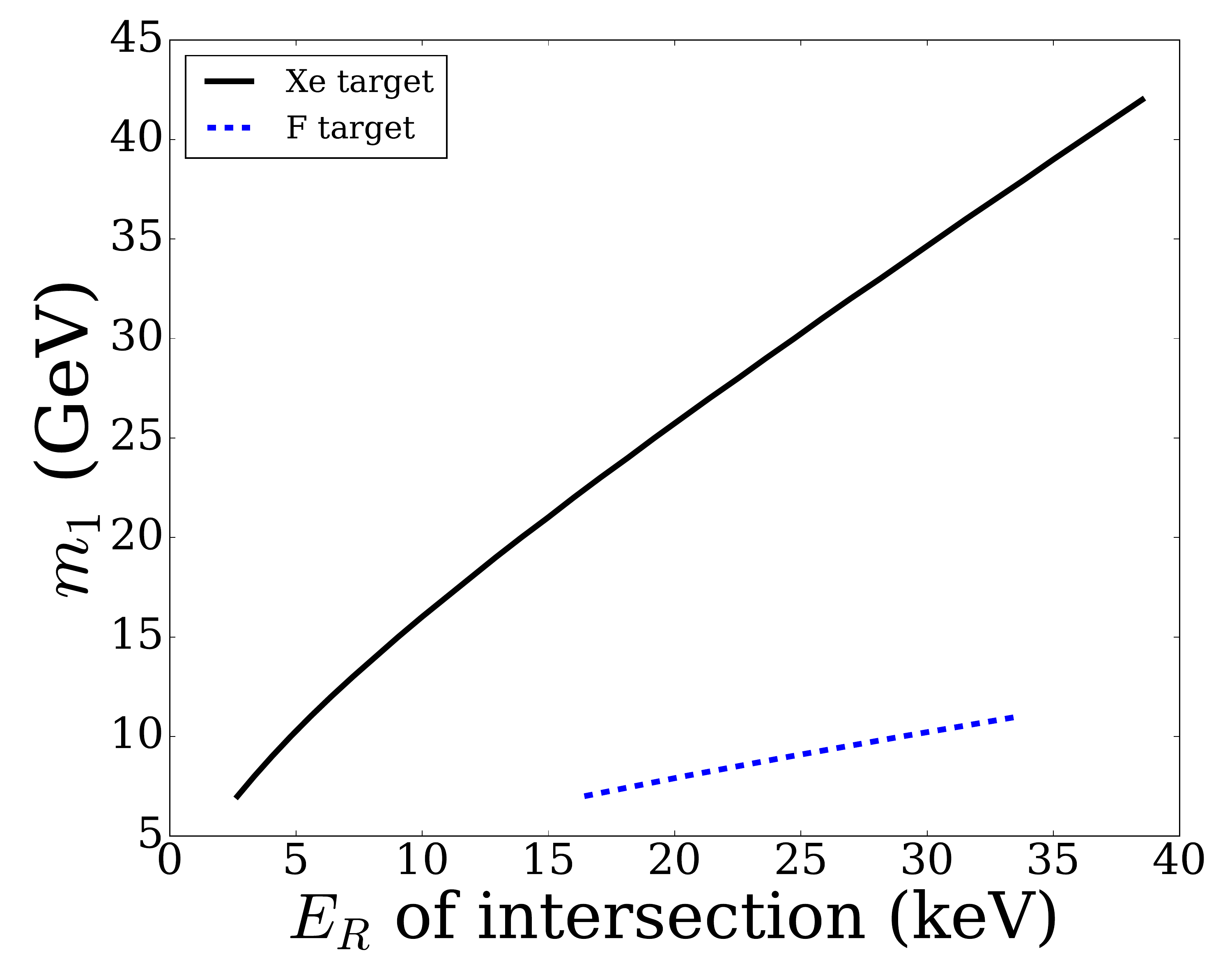}
	\caption{ The lightest DM mass $m_1$ (GeV) versus the recoil energy $E_R$ (keV) where the spectrum of the DM component 1 intercepts the spectrum of the DM component 2. We show different targets for both SI (left) and SD (right) interactions. For SI interactions we consider Xe, Ge and Na nuclei whereas for the SD case we only take Xe and F nuclei. We fix $m_2=200$ GeV and $r_\rho=r_\sigma=1$. Notice that the maximum recoil energy of the interception sets an upper limit on the maximum $m_1$ that can be discriminated. } \label{fig:Interception}
\end{figure}

We plot in Fig.~\ref{fig:Interception} the lightest DM mass $m_1$ (GeV) versus the recoil energy $E_R$ (keV) where the spectrum of the DM component 1 intercepts that of DM component 2, for both SI (left) and SD (right) interactions. The SI targets include xenon, germanium and sodium, whilst for SD interactions we only consider xenon and fluorine nuclei. We fix $m_2=200$ GeV and $r_\rho=r_\sigma=1$. The cut-off for the curves indicates the values of $m_1$ for which the spectra for the constituent components are parallel. One can see that this cut-off is highly dependent on the mass number of the nuclear target (i.e, the lower the mass number, the lower the cut-off). Notice that the maximum recoil energy that can be detected sets a conservative upper limit to the maximum $m_1$ that can be discriminated, i.e., such that the energy of the intersection is below its value. Similarly, the maximum $m_1$ sets the maximum energy of the intersection above which the rates cannot be discriminated. This means that there are both lower and upper bounds on the splitting between the DM masses $m_2$-$m_1$ for a discrimination between a two-component and one-component signal to be possible. This is due to the fact that, for fixed $m_2$, $m_1$ cannot be arbitrarily light, as it would not give recoils below threshold. Similarly, for a fixed $m_1$, $m_2$ cannot be arbitrarily large, as its number density (and therefore its rate) would be extremely suppressed.

\begin{figure}
	\centering
	\includegraphics[width=0.35\textwidth]{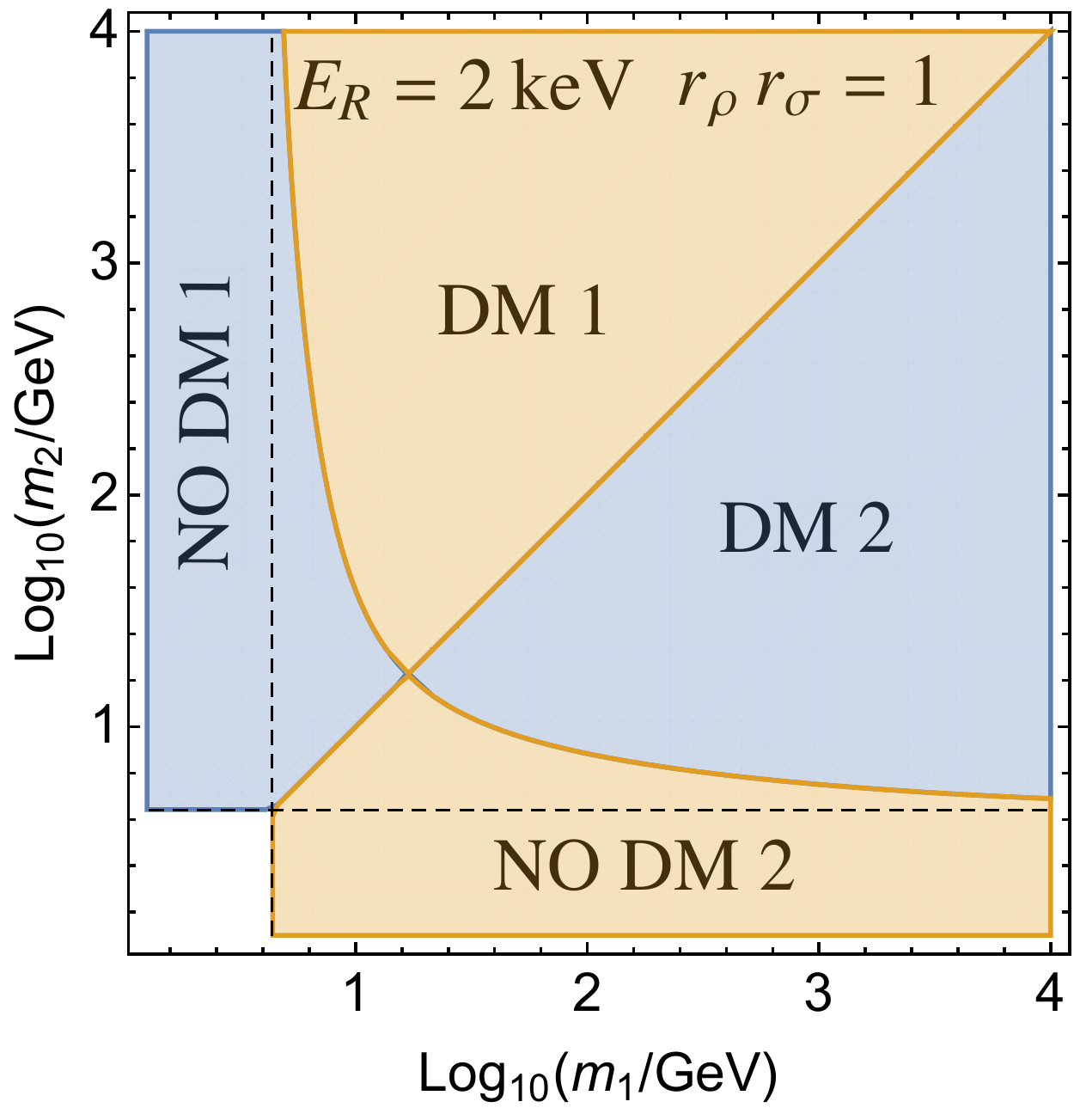}~~~\includegraphics[width=0.35\textwidth]{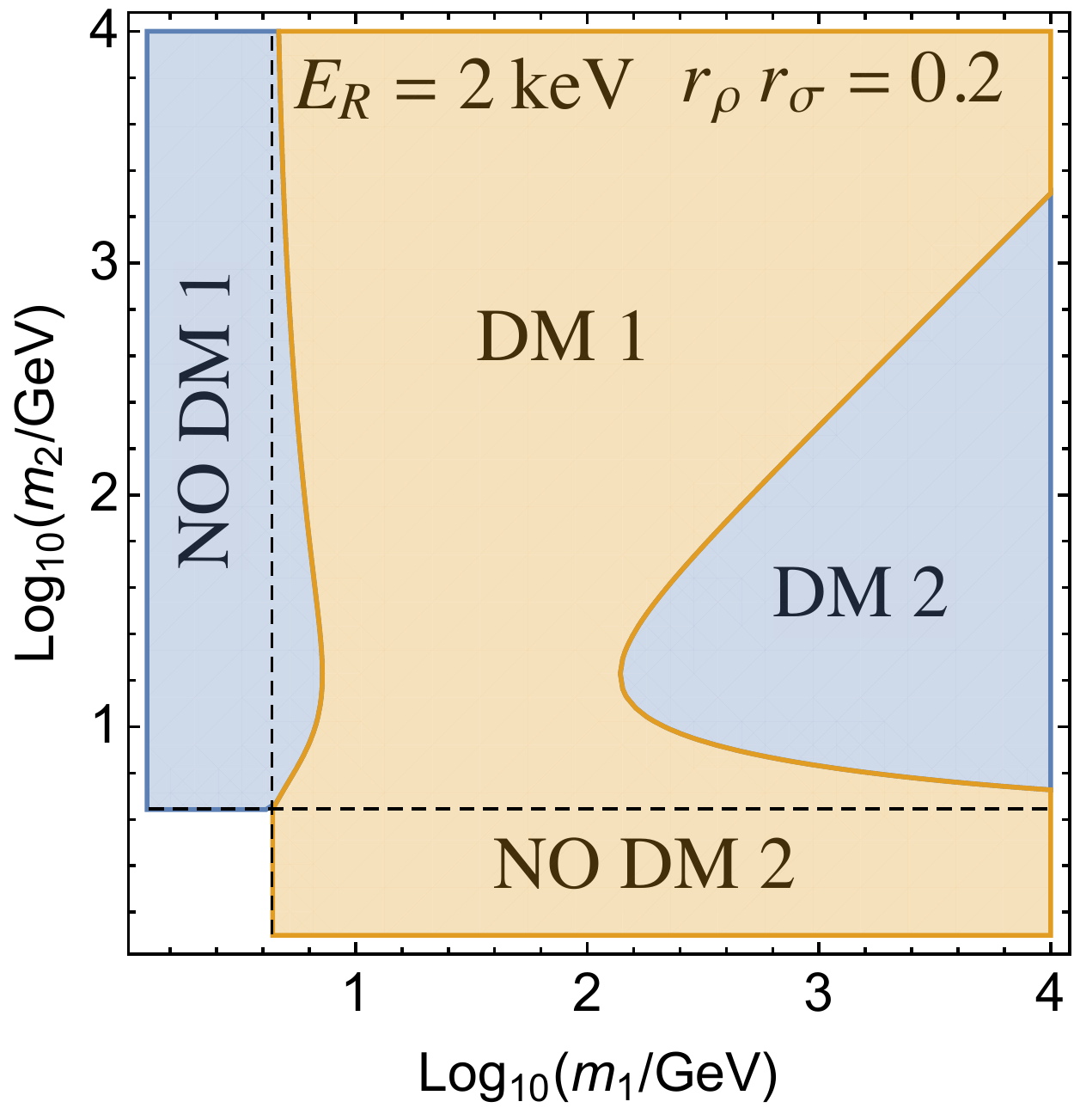}
	\includegraphics[width=0.35\textwidth]{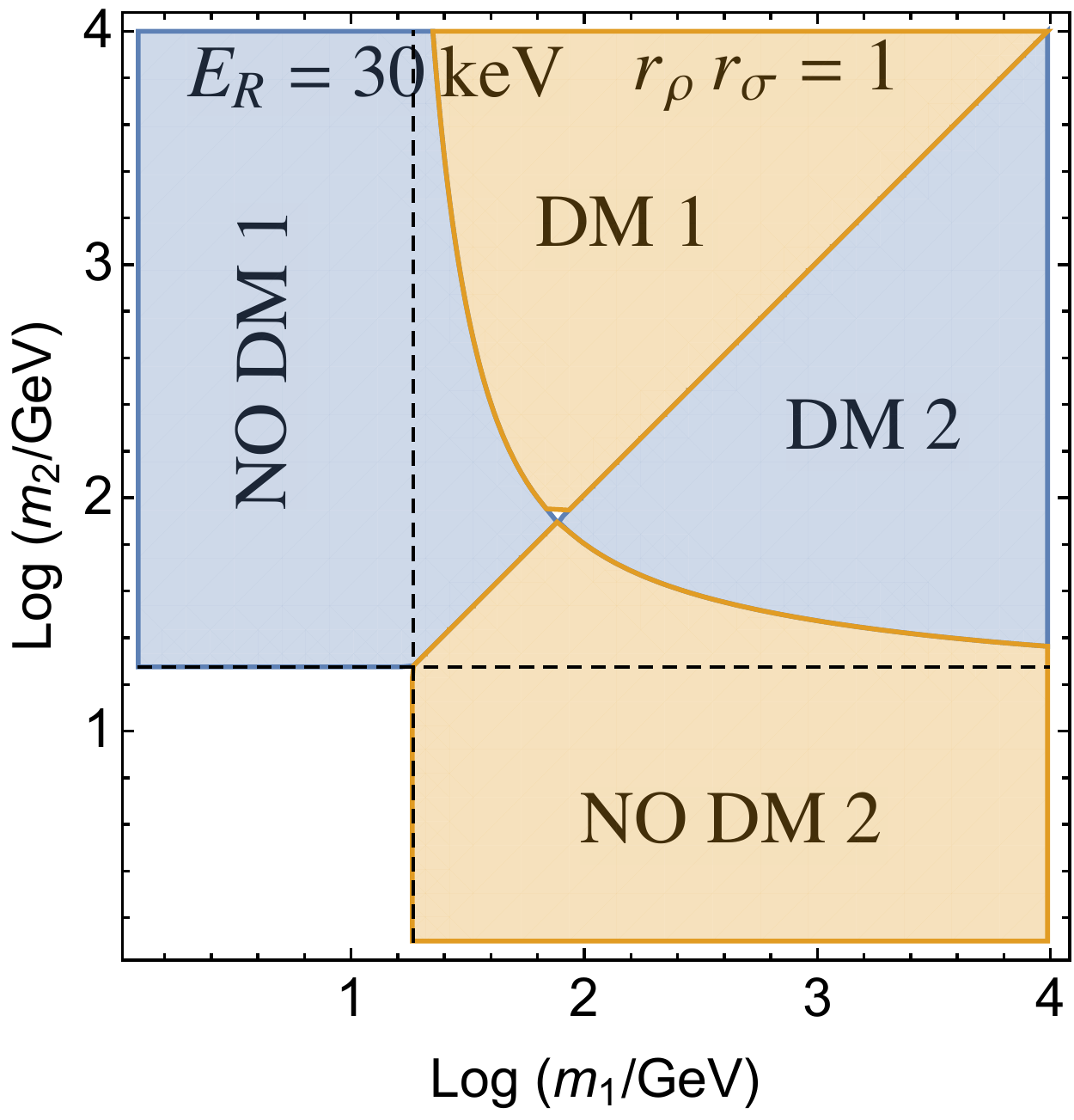}~~~\includegraphics[width=0.35\textwidth]{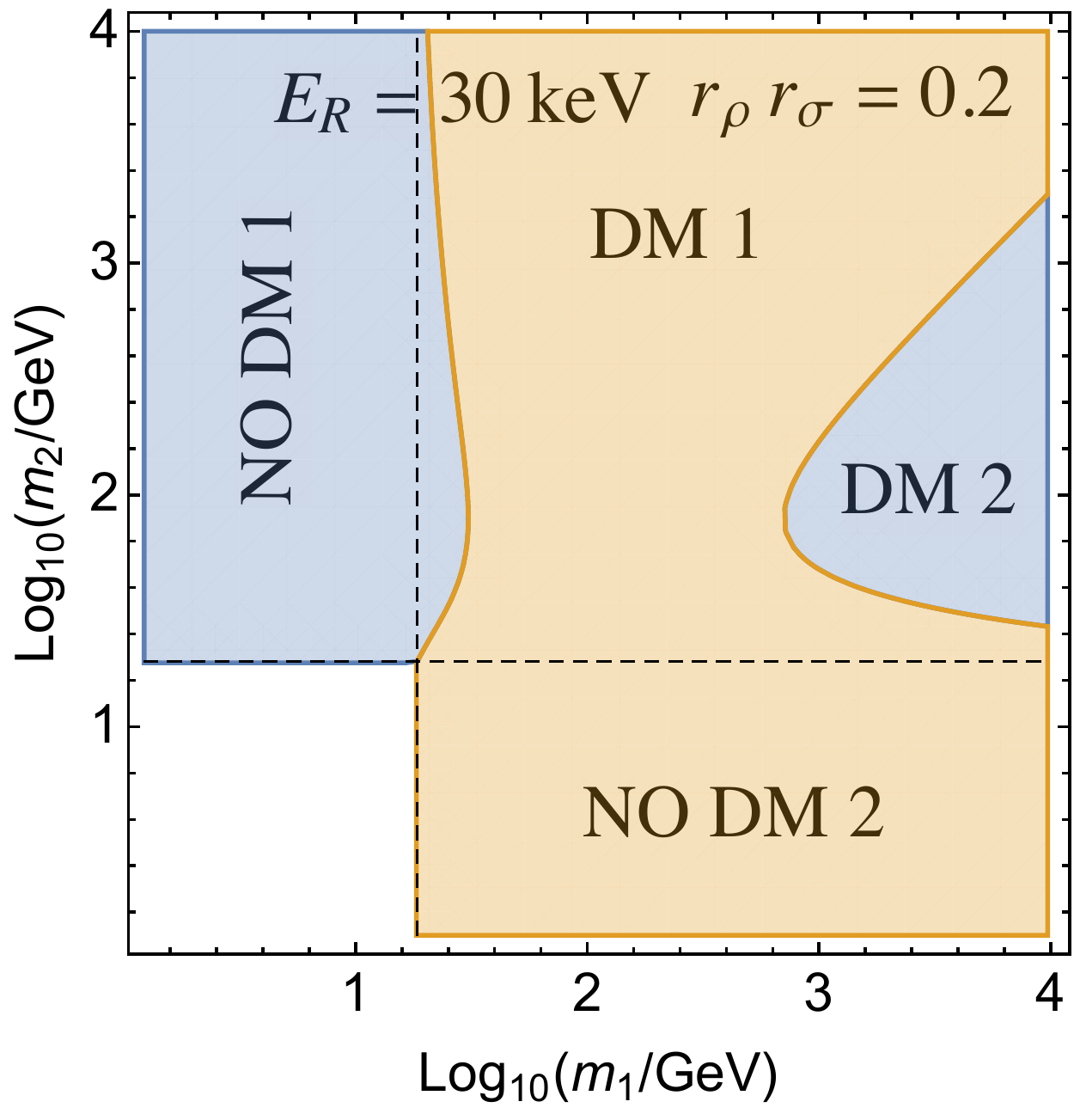}
	\caption{Regions in the plane $\log(m_1)$--$\log(m_2)$ where DM particle 2 (1) dominates as shaded light blue (shaded light brown) areas. The upper row is for $E_R=2$ keV and the lower row for $E_R=30$ keV. Left (right) column is for $r_\rho r_\sigma=1\,(0.2)$. The regions to the left (bottom) of the vertical (horizontal) dashed lines imply $v^{(1)}_m(E_R)>v_{\rm esc\,,det}$ ($v^{(2)}_m(E_R)>v_{\rm esc\,,det}$) and therefore there is no DM 1 (2) that can give recoils at that recoil energy.} \label{fig:F1}
\end{figure}

We can try to see which particle dominates the rate, the heaviest or the lightest. As we have seen, this depends on the recoil energy considered. For a fixed $E_R$, as $m_2>m_1$ (by definition), we have that $\eta(v_{m,A1}^{1})<\eta(v_{m,A1}^{2})$. However, it could well be that the lightest particle (number 1) dominates if $r_\rho \ll1$ or $r_\sigma \ll1$. Therefore which particle gives the largest contribution to the rate depends also on the product $r_\rho r_\sigma$. We plot in Fig.~\ref{fig:F1} the regions in the plane $\log(m_1)$--$\log(m_2)$ where DM particle 2 (1) dominates as shaded light blue (shaded light brown) areas for $E_R=2$ keV (upper panel) and for $E_R=30$ keV (lower row). We show results for two different values of $r_\rho r_\sigma=1,\,(0.2)$ in the left (right) panel. Of course, the areas are symmetric under interchange of particles 1 and 2 for $r_\rho = r_\sigma=1$. In this case, one can see that lightest DM species always dominates: DM 1 to the upper-left region of the diagonal, and DM2 to the lower-right, except in a small region where the lightest DM is so light that we are probing the tail of the velocity distribution, which is exponentially suppressed. For $r_\rho r_\sigma=0.2$, the regions where DM 1 dominates are somewhat larger than for $r_\rho r_\sigma=1$. A similar plot for $r_{\rho} r_{\sigma}=5$ can be obtained by the reflection of the $r_{\rho} r_{\sigma}=0.2$ plot by interchanging $m_1\leftrightarrow m_2$ everywhere, and therefore, in this case, the regions where DM 2 dominates are larger than for $r_\rho r_\sigma=1$. The regions to the left (bottom) of the vertical (horizontal) dashed lines imply $v^{(1)}_m(E_R)>v_{\rm esc\,,det}$ ($v^{(2)}_m(E_R)>v_{\rm esc\,,det}$) and therefore there is no DM 1 (2) that can give recoils at that recoil energy. They set the lowest DM mass that can be detected. For $E_R=30$ keV, the regions where the scattering is allowed shrink, as the minimum DM mass that can produce a recoil is larger than for $E_R=2$ keV.

We therefore conclude that in the general case, which particle dominates the rate is a rather model-dependent statement. If we take equal energy densities and cross-sections, i.e. $r_\rho r_\sigma=1$, again, which one dominates depends on the DM masses and cross-sections, see Eq.~\eqref{eq:rate_tot} and Fig.~\ref{fig:F1}. On the other hand, if we assume $n_1=n_2 \equiv n$, then $\rho_{\rm{loc}}=n\, m_{\rm tot}$, where $m_{\rm tot}=m_1+m_2$, and $r_\rho=m_2/m_1$. Therefore, in this scenario the total rate can also be computed using Eq.~\eqref{eq:rate_tot} for $r_\rho=m_2/m_1$, and it simplifies to:
\beq \label{eq:R_nsim}
R_{A_1}(E_{R}) = \frac{C_n(\sigma_1^p)}{m_{\rm tot}}\,\,F_{A_1}^2(E_{R}) \,\left((A^{\rm eff}_{1,1})^2\,\eta(v_{m,A1}^{(1)})+(A^{\rm eff}_{2,1})^2\,r_\sigma\,\eta(v_{m,A1}^{(2)})\right)\,,
\eeq
where we defined
\beq
 C_n (\sigma_1^p) \equiv \frac{\rho_{\rm{loc}}\,\sigma_1^p}{2 m_p^2}\,.
 \eeq
Interestingly, in this case the DM masses $m_1,\,m_2$ only enter in the rate through the dependence on $\eta(v_m)$. This means that the lower the DM mass, the larger the minimum velocity, and therefore the smaller the rate. Therefore, in this case, for $r_\sigma=1$ and $\kappa_1=\kappa_2$, the largest DM mass clearly dominates the rate. The implications of this will be further discussed below in sec.~\ref{sec:supp}.

\subsubsection{Suppression of the rate for equal number densities} \label{sec:supp}

In the case of equal number densities, if we assume that the DM particles have roughly the same kind of interactions with the SM ($r_\sigma=1$, $\kappa_1=\kappa_2$), calling $\sigma_1^p \approx \sigma_2^p \equiv \sigma_p$, with SI isospin-conserving interactions, the total rate for a particular nucleus can be written using Eq.~\eqref{eq:R_nsim} as:
\begin{eqnarray} \label{eq:Rtg} 
R(E_{R}) & \simeq& \frac{ \sigma_p}{2 m_p^2} \frac{\rho_{\rm{loc}}}{m_2} A^2 F^2(E_{R})\, \eta (v_m^{(2)})\,,
\end{eqnarray} 
where we used that $\eta (v_m^{(2)}) \gg \eta (v_m^{(1)})$ and $m_{\rm tot} \approx m_2$, which are valid when there is a significant hierarchy in masses, $m_1 \ll m_2$. Notice also that in the limit $m_2 \gg m_A$, $v_m^{(2)}$ becomes independent of the DM mass $m_2$, and so does $\eta (v_m^{(2)})$, so we could drop its superscript (we assume this in the following). From this expression, we can derive an upper bound on the heaviest DM mass to have a signal in a direct detection experiment for a given cross-section. For an experimental sensitivity of $R_{\text{exp}}(E_R)$ at a given recoil energy $E_R$, we get:
\beq\label{eq:RL} 
 \frac{\rho_{\rm loc}\,\sigma_p}{m_2} \simeq \frac{2\,m_p^2\,R_\text{exp}(E_R)}{A^2 F^2(E_{R})\,\eta [v_{m}(E_R)]} \,,
\eeq 
or similarly, for a fixed DM cross-section:
\beq\label{eq:RLb} 
m_2 \simeq \frac{1}{2\,m_p^2} \frac{\rho_{\rm loc}}{R_\text{exp}(E_R)}\sigma_p A^2 F^2(E_{R})  \,\eta [v_m(E_R)]\,.
\eeq 
Notice that one can use the inequality $\eta(v_m)< 1/v_m$ that is independent of the velocity distribution~\cite{Feldstein:2014ufa,Kavanagh:2012nr,Blennow:2015gta,Herrero-Garcia:2015kga}, and therefore derive an upper limit on the DM mass (or a lower limit on $\rho_{\rm loc}\,\sigma_p/m_2$) that is $f(v)$-independent as a function of the recoil energy:
\beq\label{eq:RLc} 
m_2 < \frac{1}{2\,m_p^2\,v_m(E_R)} \frac{\rho_{\rm loc}}{R_\text{exp}(E_R)}\sigma_p A^2 F^2(E_{R})\,.
\eeq 
Therefore one can check the $E_R$ at which the upper bound is weakest. As an example, using $\sigma_p=10^{-45} \, {\rm cm}^{2}$, and taking the the form factor equal to one, in Germanium, for an expected sensitivity of $R_{\rm exp}\geq 10^{-5}$ cts/kg/day at a typical recoil energy $E_{\rm R}=2$ KeV, we get that $m_2 \lesssim 200$ GeV, i.e., unless the heaviest mass is lighter than $200$ GeV, the rate will be too low to be detected. For the same conditions in Iodine (DAMA) and Xenon, we obtain upper bounds of $\sim600$ and $\sim700$ GeV respectively. Similarly, if the DM properties were known from other probes (colliders, indirect detection), eq.~\eqref{eq:RLc} could be used to derive an upper limit on the event rate as a function of recoil energy. The bottom line is that, for equal number densities, if the DM components have similar interaction strength, the heaviest mass of the DM group dictates the sensitivity of DD.  

\section{Hypothesis testing: one versus two DM components} \label{sec:hypothesis_test}

In the following (sec.~\ref{sec::stats}) we detail the statistical methods used to calculate the sensitivity of future experiments to discriminate a two-component DM scenario from the one-component one. We formulate this analysis as a frequentist hypothesis test, with the null-hypothesis being one-component, and the alternative hypothesis being two-component. The test should be able to select the two-component interpretation when the energy recoil spectrum has a significant \emph{kink} feature, as shown in Fig.~\ref{Rates_2comp}, since this feature cannot be explained with only one DM particle. The statistical techniques used in this study have been previously applied and thoroughly detailed in the context of neutrino mass ordering analyses~\cite{Blennow:2013oma,Ciuffoli:2017ayi,ciuffoli2013sensitivity}.
We then show in subsequent subsections that the best model discriminator is the mass splitting between the two DM components. We first show in sec.~\ref{sec:equal} results taking as free parameters $m_1$, $m_2$ and the overall normalisation $\sigma_p^1$, keeping $\kappa_{1,\,2}=r_\rho=r_\sigma=1$ and assuming the SHM for both DM particles with $\sigma_{\rm H1} = \sigma_{\rm H2} = 270$ km/s. Afterwards,  in sec.~\ref{sec:different} we proceed to see how the discrimination is worsened by allowing $r_\rho \neq r_\sigma\neq1$, the dispersion velocities $
\sigma_{\rm H1}\neq\sigma_{\rm H2}\neq \sigma_{\rm SHM}$, or the ratios of couplings $\kappa_1\neq\kappa_2\neq1$.

\subsection{Test statistic for hypothesis testing}
\label{sec::stats} 
We construct the hypothesis test in a frequentist framework, defining the null hypothesis to be the one-component scenario which we denote $H_{1\text{DM}}$. Similarly we denote the two-component alternative hypothesis $\Ht$.   Notice that the one DM hypothesis is a subset of the two DM hypothesis, as can be seen easily by taking $r_\rho = 0$ in Eq.~\eqref{eq:rate_tot}. $\Ho$ is said to be a \textit{nested} hypothesis of $\Ht$. We will see later that this has important ramifications. 

Suppose that we have a detector that has observed a set of  binned count measurements $\mathbf{x} = {x_1,x_2...,x_N}$ over $N$ bins with uncertainties $\sigma_i=\sqrt{x_i}$. We  parameterise the likelihood of observing this data given a hypotheses $\Ho/\Ht$ with a binned Gaussian distribution:
\begin{align}
\label{Likelihood:Defn}
\L(\mathbf{x}\,|\,H_\alpha) =\prod\limits^N_i\, \frac{1}{\sqrt{2\pi}\sigma_i}e^{-\frac{\left[x_i-\mu_i (\theta_{\alpha})\right]^2}{2\sigma_i^2} }\qquad \;\alpha=[\Ho,\Ht],
\end{align}
where $\mu_i (\theta_{\alpha})$ is the expected number of counts in bin $i$ as a function of the model parameters $\theta_{\Ho}/\theta_{\Ht}$ under the hypothesis $\Ho/\Ht$. Maximising the likelihood in Eq.~\eqref{Likelihood:Defn} with respect to the hypothesis parameters $\theta_\alpha$ is equivalent to minimising -2 times the log-likelihood as follows:
\begin{align}
\label{chisquare:Defn}
	\min_{\theta_\alpha}\big(-2\ln\mathcal{L}\big) =\min_{\theta_\alpha} \sum_i^N\frac{\left[x_i-\mu_i (\theta_{\alpha})\right]^2}{\sigma_i^2}  \equiv \min_{\theta_\alpha} \chi^2(\theta_\alpha)\;.
\end{align}
This is the familiar `chi-square' statistic which, as showed by Pearson \cite{doi:10.1080/14786440009463897} in the limit of large $x_i$ follows a $\chi^2$ distribution with $N-n(\theta_\alpha)$ degrees of freedom, where $n(\theta_\alpha)$ is the number of parameters $\theta_\alpha$. 

One should immediately notice that Eq.~\eqref{chisquare:Defn} only provides a  `goodness of fit' for the hypothesis $\alpha$, and does not explicitly reject one in favour of the other. For this task, we require a test-statistic $\mathcal{T}$ that explicitly discriminates between $\Ho$ and $\Ht$. A commonly used test statistic is:
\begin{equation}
\label{TestStat}
\mathcal{T}= \min_{\theta_{\Ho}} \chi^2(\theta_{\Ho}) -\min_{\theta_{\Ht} }  \chi^2(\theta_{\Ht})\,.
\end{equation}
 Notice that the definition of $\mathcal{T}$ is such that the larger its value, the larger the preference for $\Ht$, and the smaller its value, the more $\Ho$ is preferred. 
 
 In order to quantify how much the data supports either hypothesis, we require the limiting probability distribution of the $\mathcal{T}$ statistic in the case either $\Ho$ and $\Ht$ is true. Under some general assumptions, the theorem from Wilk \cite{wilks1938large} states that in the case of nested hypotheses (i.e when $\Ho$ is true), $\mathcal{T}$ will follow a $\chi^2$ distribution with $k \equiv n(\theta_{\Ht}) - n(\theta_{\Ho})$ degrees of freedom, where $n$ is the number of parameters that parameterise a given hypothesis. We denote this by $\mathcal{T}^{\rm1DM}$.  
 
However, in the case that $\Ht$ is true, we show in Appendix \ref{appendix:Tstat} that the $\mathcal{T}$ statistic, which we denote $\T^{\rm 2DM}$, will follow a Gaussian distribution that is solely dependent on the true parameter values $\mu(\theta_{\Ht}^{\rm true})$  with mean given by $\mathcal{T}^{\rm 2DM}_0$ and standard deviation $2\sqrt{\mathcal{T}^{\rm 2DM}_0}$, where
\begin{equation}
\label{T0}
\mathcal{T}^{\rm 2DM}_0 \equiv \mathcal{T} (x_i=\mu_i (\theta_{\Ht}^{\rm true}))= \min_{\theta_{\Ho} } \sum^n_i \,\left(\frac{\mu_i (\theta_{\Ht}^{\rm true})-\mu_i (\theta_{\Ho})}{\sqrt{\mu_i (\theta_{\Ht}^{\rm true})}} \right)^2\,.
\end{equation}
That is, $\mathcal{T}^{\rm 2DM}_0$ has no statistical fluctuations from the data and is often called the \textit{`Asimov likelihood}'. The Asimov likelihood can be qualitatively thought of as an approximation of the median value of a test statistic. See  Ref.~\cite{cowan2011asymptotic} for details. Note that large values of $\mathcal{T}^{\rm 2DM}_0$ disfavour the null hypothesis $\Ho$. 

\begin{figure}
	\centering
	\includegraphics[width=0.6\textwidth]{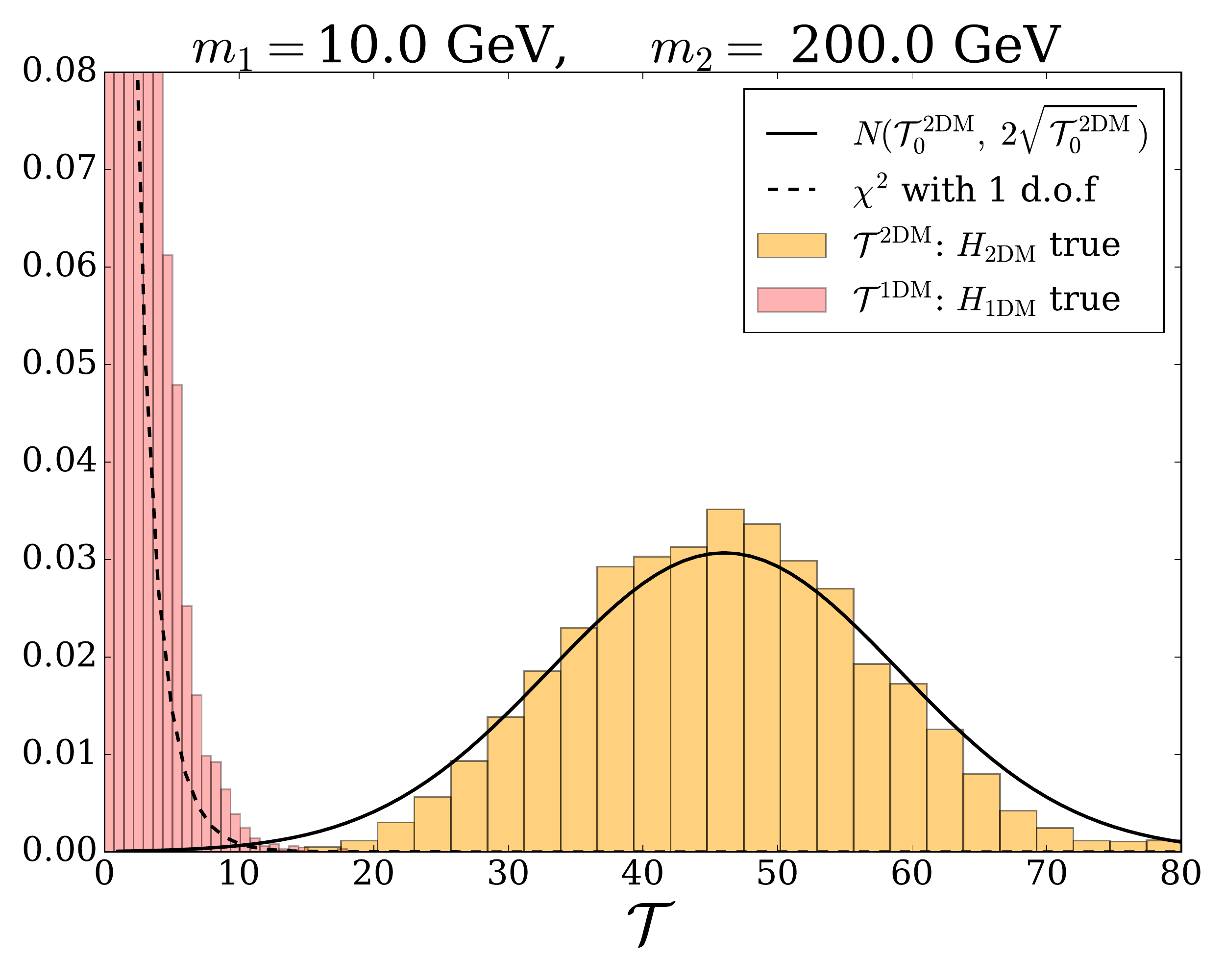}
	\caption{Normalised probability distribution of the test statistic $\mathcal{T}$ in a Xe target for true one-component $\mathcal{T}^{\rm1DM}$ (light red) and two-component $\mathcal{T}^{\rm2DM}$ (light yellow) DM hypothesis. We fixed $m_1=10$ GeV and $m_2=200$ GeV. We used $8500$ samples. We see that the one-component DM follows a $\chi^2$ distribution with one degree of freedom, while the two-component DM hypothesis is well-approximated by a Gaussian centered at $\mathcal{T}^{\rm 2DM}_0$, with standard deviation $2 \sqrt{\mathcal{T}^{\rm 2DM}_0}$. } \label{Ttest}
\end{figure}

In Fig.~\ref{Ttest} we show the probability distribution of $\mathcal{T}$ for the two hypotheses in a Xe target: $\Ho$ in light red and $\Ht$ in light yellow. We fix $m_1=10$ GeV and $m_2=200$ GeV and use $8500$ Monte Carlo samples. Since the only free parameters we are considering are the two DM masses, the difference in degrees of freedom of the two hypotheses is $k = 2-1 = 1$ degrees of freedom. Hence, from Wilk's theorem, we expect that under $\Ho$, $\T$ will follow a $\chi^2_{1\rm  d.o.f}$ distribution.  One can see that $\mathcal{T}^{\rm 2DM}$ is  approximately Gaussian distributed with median $\mathcal{T}^{\rm 2DM}_0$ and standard deviation $2 \sqrt{\mathcal{T}^{\rm 2DM}_0}$, while $\mathcal{T}^{\rm 1DM}$ is approximately estimated by a $\chi^2$-distribution with one degree of freedom as expected.\footnote{One will also notice that the width of the Gaussian approximation in Fig.~\ref{Ttest} slightly overestimates $\T^{\rm 2DM}$.   This of course could be primarily because we have assumed Monte Carlo realisations of the recoil spectrum are Gaussian distributed instead of Poisson. We have exploited the assumption that for large number of events the bin by bin distribution starts looking Gaussian. In practice, however, we are only interested in the median of the $\Ht$ distribution, since its variance  does not enter in the calculation of the median sensitivity, as will be discussed in sec.~\ref{medSens}.}

\subsubsection{Median sensitivity}
\label{medSens}
The aim of this analysis is to quantify how sensitive an ``average experiment" operating at future benchmarks is to rejecting $\Ho$ in favour of $\Ht$. The quantity that allows us to do this is called the \textit{median sensitivity} and can be easily visualised as follows: Under $\Ht$ an experiment will generate a $\T^{2DM}$ statistic that is approximately Gaussian as shown in Fig.~\ref{Ttest}. In this vein, the median (mean) of this Gaussian $\T^{\rm2DM}_0$ quantifies the experiments' `average' capability.

Since we know that the distribution of $\T$ under $\Ho$ is a $\chi^2$ with $k$ degrees of freedom, we can then calculate the probability of having a $\T^{\rm2DM}_0$ at least as extreme as the one we observed. This is called the \textit{p-value} and is given by
\begin{align}
\label{p-val}
p = \int_{\T^{\rm2DM}_0}^{\infty}\,f(\T|\Ho) \,d\T = 1-\rm CDF^{\chi^2}_{k\;\rm d.o.f}(\T^{\rm2DM}_0)\;,
\end{align}
where $\rm CDF^{\chi^2}_{k \;\rm d.o.f}$ is the cumulative density function for the $\chi^2_{k\;\rm d.o.f}$. As a result, a larger $\T^{\rm2DM}_0$ which favours $\Ht$ will produce a smaller p-value, and vice-versa. We define $\alpha$ to be the probability of making an error of the 1st kind, i.e, rejecting $\Ho$ if its true. If $p<\alpha$, then the data supports $\Ht$ over $\Ho$. In more conventional words, we have defined our critical region to be where there is a low probability to observe $\T$ if $\Ho$ is true but a high probability if the alternative hypothesis $\Ht$ is true. 
The p-value in Eq.~\eqref{p-val} can be converted to a two-sided number of unit Gaussian standard deviations, which we will denote throughout the rest of this paper as $Z$, and call it the \textit{median sensitivity} using
\footnote{An alternate definition of the median sensitivity as seen in Ref.~\cite{Blennow:2013oma} is given by the CL at which an experiment will reject the wrong hypothesis with a probability of 50\%, that is, with a rate for an error of the second kind of 0.5. We will use the definition in Eq.~\eqref{nsigma} for this study.}
\begin{align}
\label{nsigma}
Z(p) = \sqrt{2}\,\rm erfc^{-1}(p)\;,
\end{align}
where $\rm erfc(p) \equiv 1-\rm erf(p)$ is the complimentary error function. If $Z\geq Z(\alpha)\equiv5$ then the median experiment can reject $\Ho$ in favour of $\Ht$ at the 5-sigma confidence level (CL). In the special case that the difference in number of degrees of freedom is $k = 1$, the sensitivity is just given by~\cite{cowan2011asymptotic} 
\begin{align}
\label{sqrtT}
Z|_{k=1} = \sqrt{\T^{\rm2DM}_0}\;.
\end{align}

\subsection{Fixed parameters except for the DM masses} \label{sec:equal}
\begin{figure}[t!]
	\centering
	\includegraphics[width=0.49\textwidth]{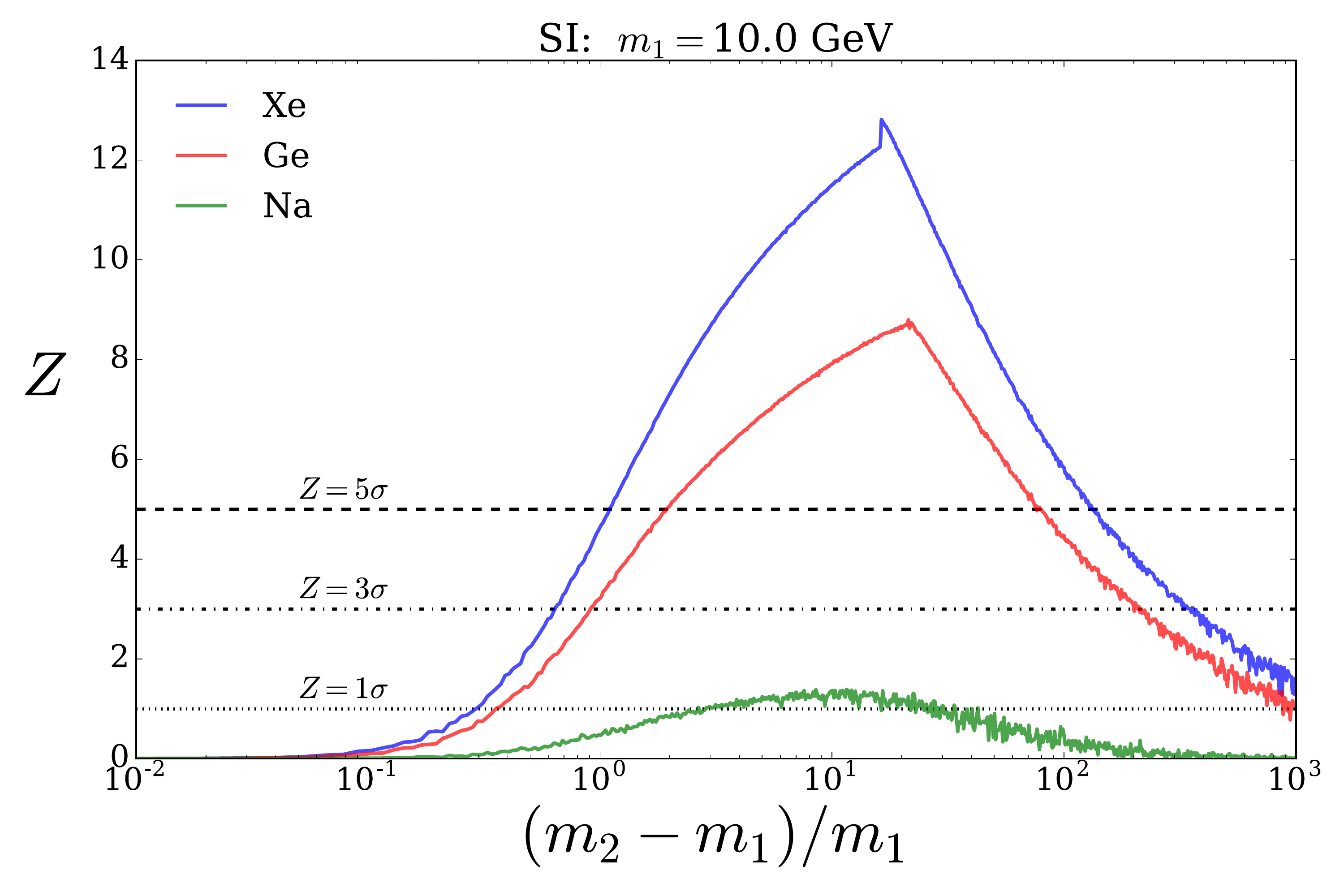}
	\includegraphics[width=0.49\textwidth]{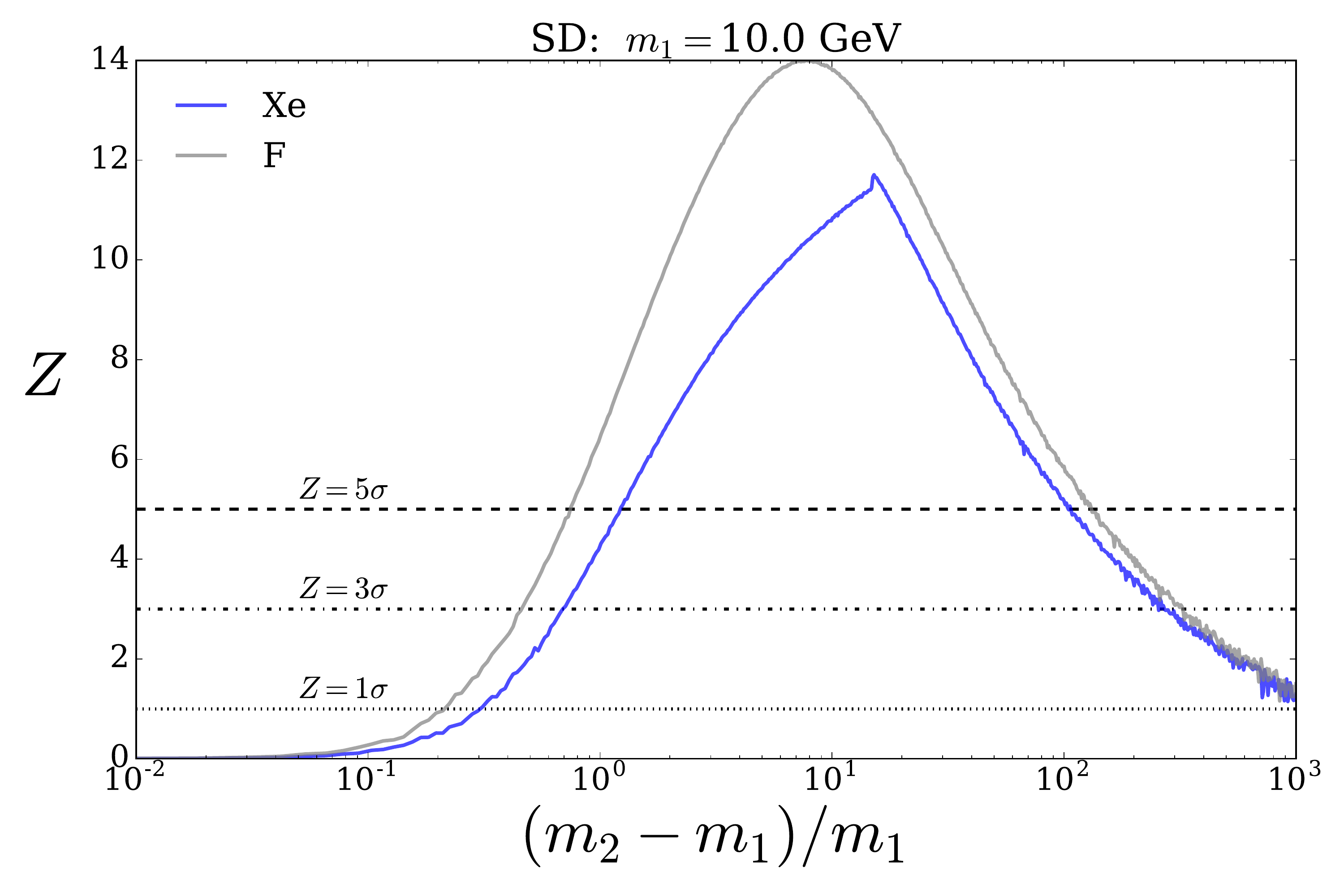}
	\caption{Significance $Z$ with which the median experiment can reject the one-DM hypothesis in favour of the two-DM hypothesis as a function of the mass splitting $(m_2-m_1)/m_1$, for fixed $m_1$ equal to 10 GeV. We show as dotted, dash-dotted and dashed (black) horizontal lines the 1, 3 and 5 $\sigma$ C.L. The left panel shows the SI targets xenon (in blue), germanium (in red) and sodium (in green). The right panel shows the SD targets fluorine (in grey) and xenon (in blue). } \label{split}
\end{figure}
\begin{figure}[ht]
	\centering
	\includegraphics[width=1\textwidth]{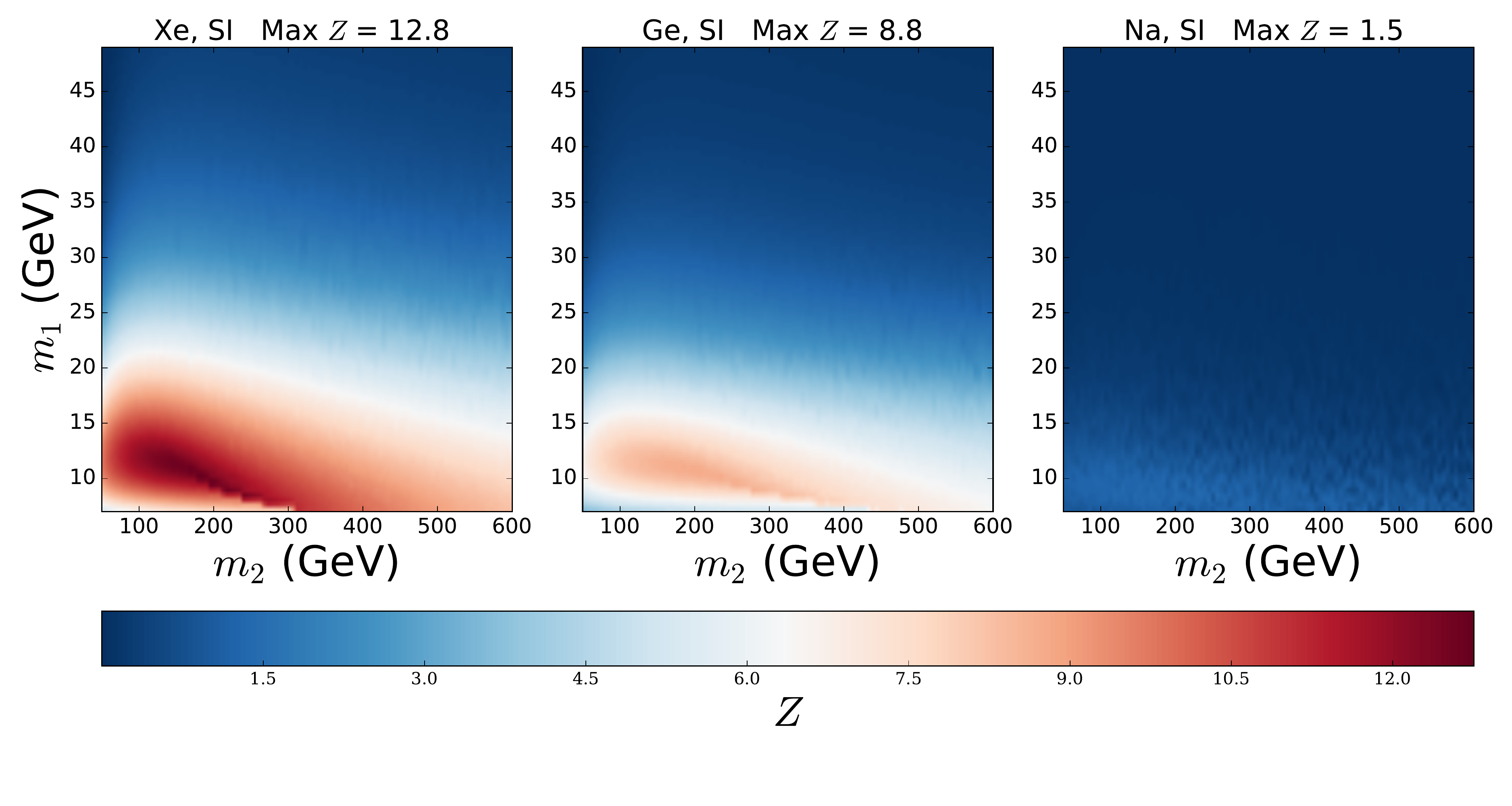}
	\includegraphics[width=1\textwidth]{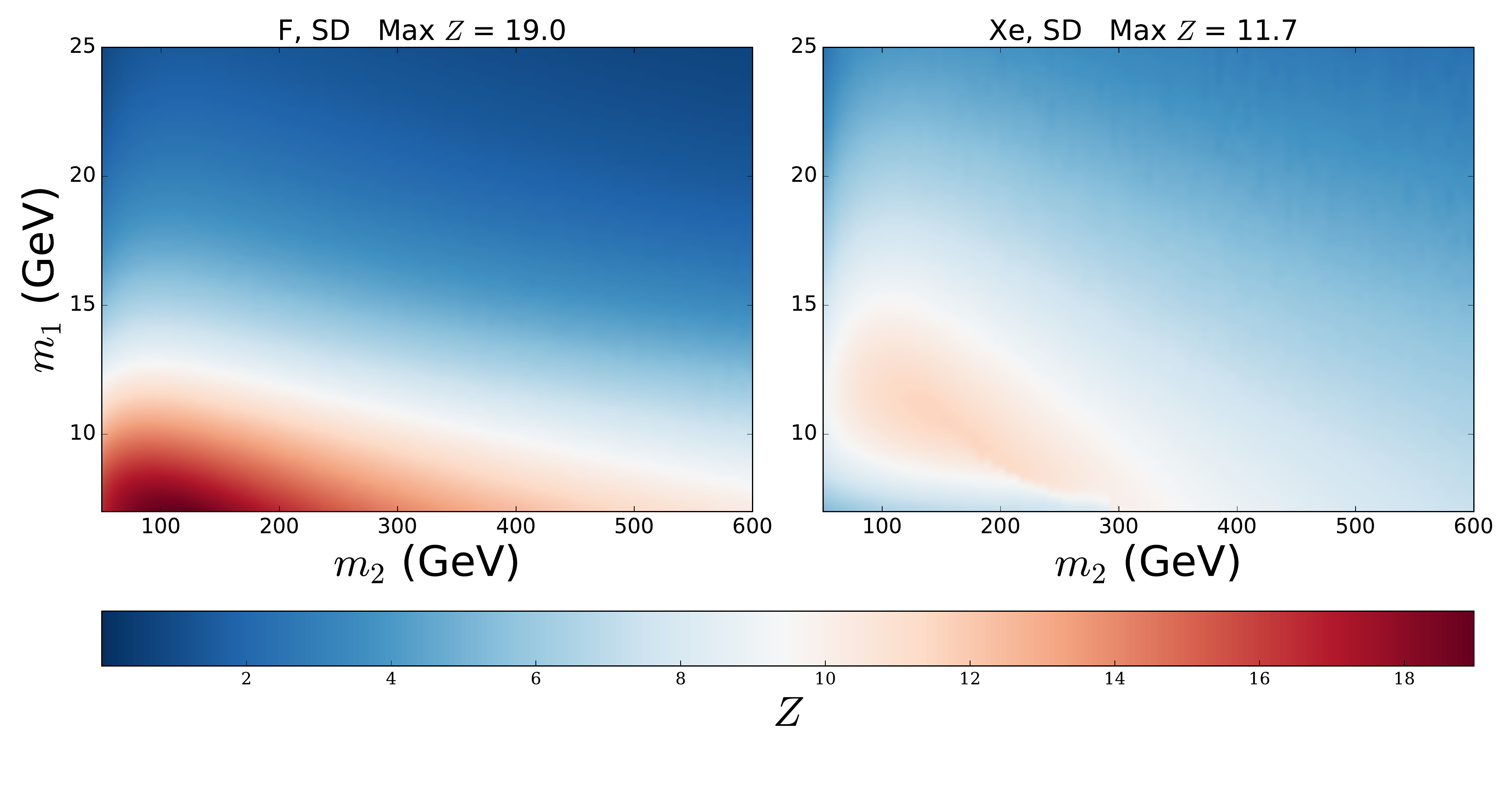}
	\caption{Significance $Z$ with which the median experiment can reject the one-DM hypothesis in favour of the two-DM hypothesis in the $m_2-m_1$ plane for different target nuclei. All other model parameters are fixed when generating the Asimov data: $r_\rho = r_\sigma= 1$, $\kappa_{1,\,2} = 1$ and $\sigma_{H_{1,\,2}} = 270$ km/s. We do not plot the symmetric region around the axis $m_1=m_2$ for clarity. The top panel is for SI interactions for Xe, Ge and Na, while the bottom panel is for SD interactions for F and Xe.} \label{scan_m1_m2}
\end{figure}
We perform the analysis discussed in sec.  \ref{sec::stats} firstly for the case that all hypothesis parameters are fixed except for $m_1$ and $m_2$. As a result, the difference in number of degrees of freedom between $\Ho$ and $\Ht$ is $k=1$. Thus, Eq.~\eqref{sqrtT} is used in this case to calculate the median sensitivity.  Later on in sec.~\ref{sec:different}, we will study how our results are changed in more general cases by relaxing the restrictions on other parameters. We do the following analysis both for SI and SD WIMP-nucleon couplings. 

For SI interactions we will consider future sodium, germanium and xenon experiments, with a true SI cross-section with protons of $\sigma_{1\rm(SI)}^p=10^{-45}\,{\rm cm^2}$. For SD interactions we will consider future fluorine and xenon experiments with true SD cross-section with protons of $\sigma^p_{1\rm(SD)}=10^{-40}\,{\rm cm^2}$. We adopt the experimental configurations shown in tab.~\ref{tab:future_exps}. The choice of the nuclei for our assumed SI interaction experiments is such that there is a large range in masses, while for SD interactions there is an additional feature: fluorine (xenon) is most sensitive to DM couplings to protons (neutrons). We also assume a true (known) $r_\rho, \,r_\sigma=1$, i.e., equal energy densities and cross-sections for the two DM particles, as well as equal couplings to protons and neutrons ($\kappa_{1,\,2}=1$). 

In order to illustrate how the mass splitting controls the hypothesis discrimination, we show in Fig.~\ref{split} the median significance $Z$ as defined in Eq.~\eqref{sqrtT} versus the normalised mass splitting $(m_2-m_1)/m_1$, for a fixed lightest DM mass $m_1$ equal to $10$ GeV. The left panel shows SI targets: xenon (in blue), germanium (in red) and sodium (in green). The right panel shows the SD targets fluorine (in grey) and xenon (in blue). We also show for reference the 1, 3, and 5 sigma contours as horizontal dotted, dash-dotted and dashed (black) lines. In the case of sodium, the significance is always very poor. This is due to two reasons. Firstly, the rate goes for spin-independent interactions as $A^2$. As a result, one obtains suppressed statistics in a Na based experiment as opposed to Xe and Ge when similar exposures are used. This dominates the height of the curve in figure 6, left panel (c.f. figure 6 right panel for SD). Secondly, in general one is always more sensitive to recoils when roughly $m_\text{DM}\sim m_\text{Nucleus}$. Hence since one will only ever observe a significant `kink' in the recoil spectrum for large enough mass splittings, a Na experiment will have trouble significantly rejecting the one-component hypotheses because it won't see the heavier particle. This effect is reflected by the relative shift of the peaks of the three curves (red, green and blue) in the left panel of figure 6. In all other cases, one should notice that the median significance is globally maximised at a mass splitting that is approximately equal to $\sim 10$ and drops to zero for very small $\lesssim 0.1$ or very large $\gtrsim 10^3$ mass splittings. This illustrates that there exists only a finite window in masses, roughly $m_2\sim \mathcal{O}(1-100)\,m_1$, where there can be a significant ($\gtrsim 5\,\sigma$) discrimination.

In Fig.~\ref{scan_m1_m2} we show the median sensitivity in the full mass plane $m_2-m_1$. The top panel is for SI and the bottom panel for SD. One can see that the DM mass regions of large significance increase with the mass of the nucleus. Indeed, as discussed above, the significance for sodium is always negligible. Notice also that the regions shrink for very large masses of the heaviest DM, where its number density is so suppressed that it gives no signal in the detector. The exposure for such heavy masses is a critical factor to obtain a signal and achieve discrimination. For xenon, the largest sensitivity occurs when the lightest DM mass is around 10-20 GeV, and the heavy mass is in the range 50-400 GeV. For germanium, the largest significance occurs when the heaviest DM is lighter than roughly 300 GeV. For the SD case the main features are preserved for xenon, however the maximum significance that can be achieved is slightly lower. Fluorine however achieves maximum significance for $m_2$ in the range 50-200 GeV. Another feature of the SD fluorine result is that the median significance also drops above $m_1\gtrsim 15$ GeV.  

An important note to make is that the median (statistical) sensitivity scales as usual with the square root of the exposure 
\begin{align}
Z\propto \sqrt{MT}\;.
\end{align}
This is true for all detector types for both SI and SD interactions. This is important, since the results we show in this section can always be scaled accordingly for different experimental exposures to those given in tab.~\ref{tab:future_exps}.  

\subsection{General scenarios} \label{sec:different}

In the following we drop some of our simplifying assumptions to see how our results are affected. We use values of $m_1$-$m_2$ that yield high median significance: we fix $m_2$ to have a mass of 150 (200) GeV for SD (SI) respectively. We then consider three different generalised scenarios (we do not show the results for sodium since in all cases the maximum median significance achieved is negligible):
\newcommand{\set}{1}

\begin{figure}[t!]
	\centering
	\includegraphics[width=\set\textwidth]{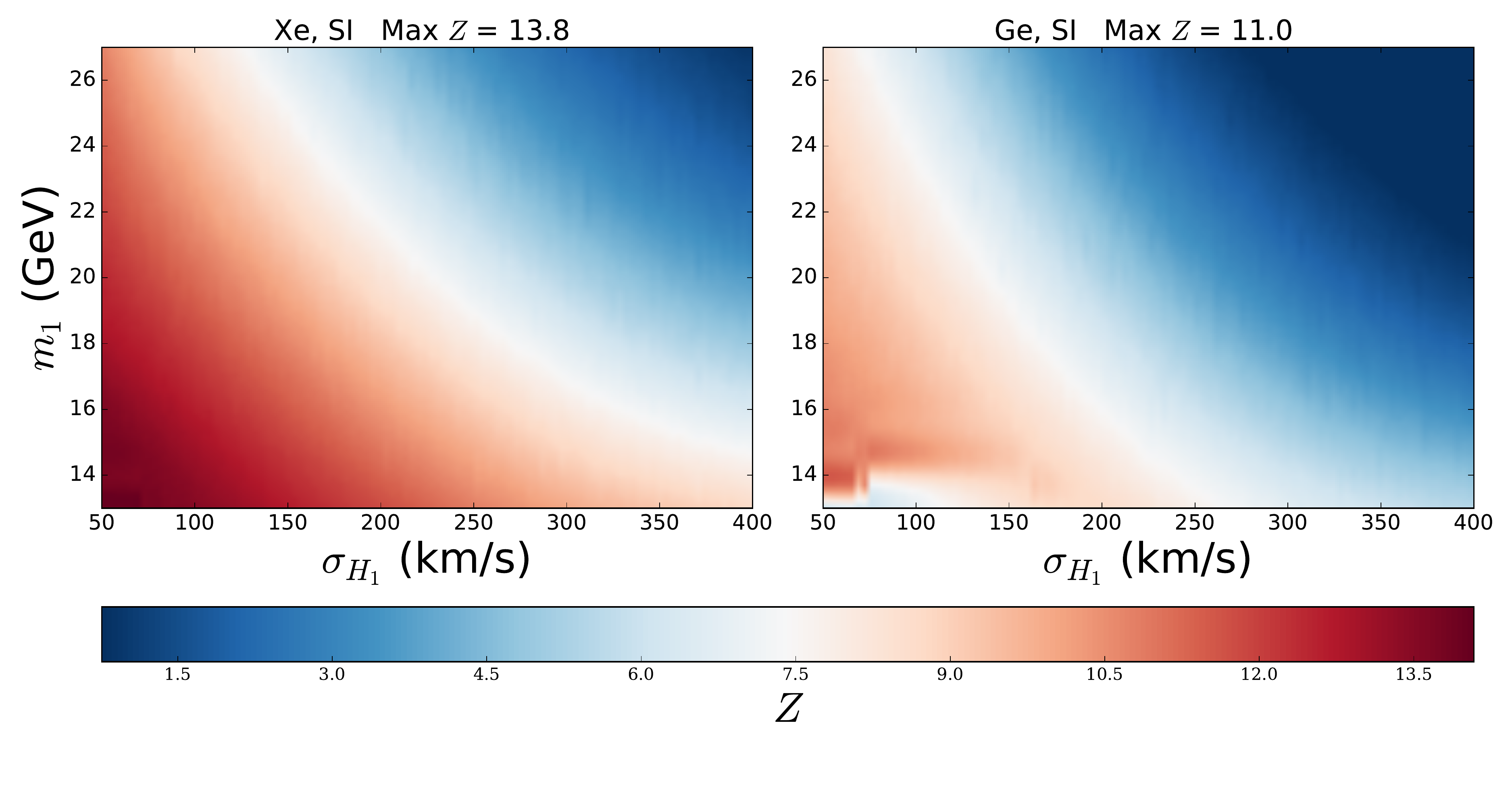}\\
	\includegraphics[width=\set\textwidth]{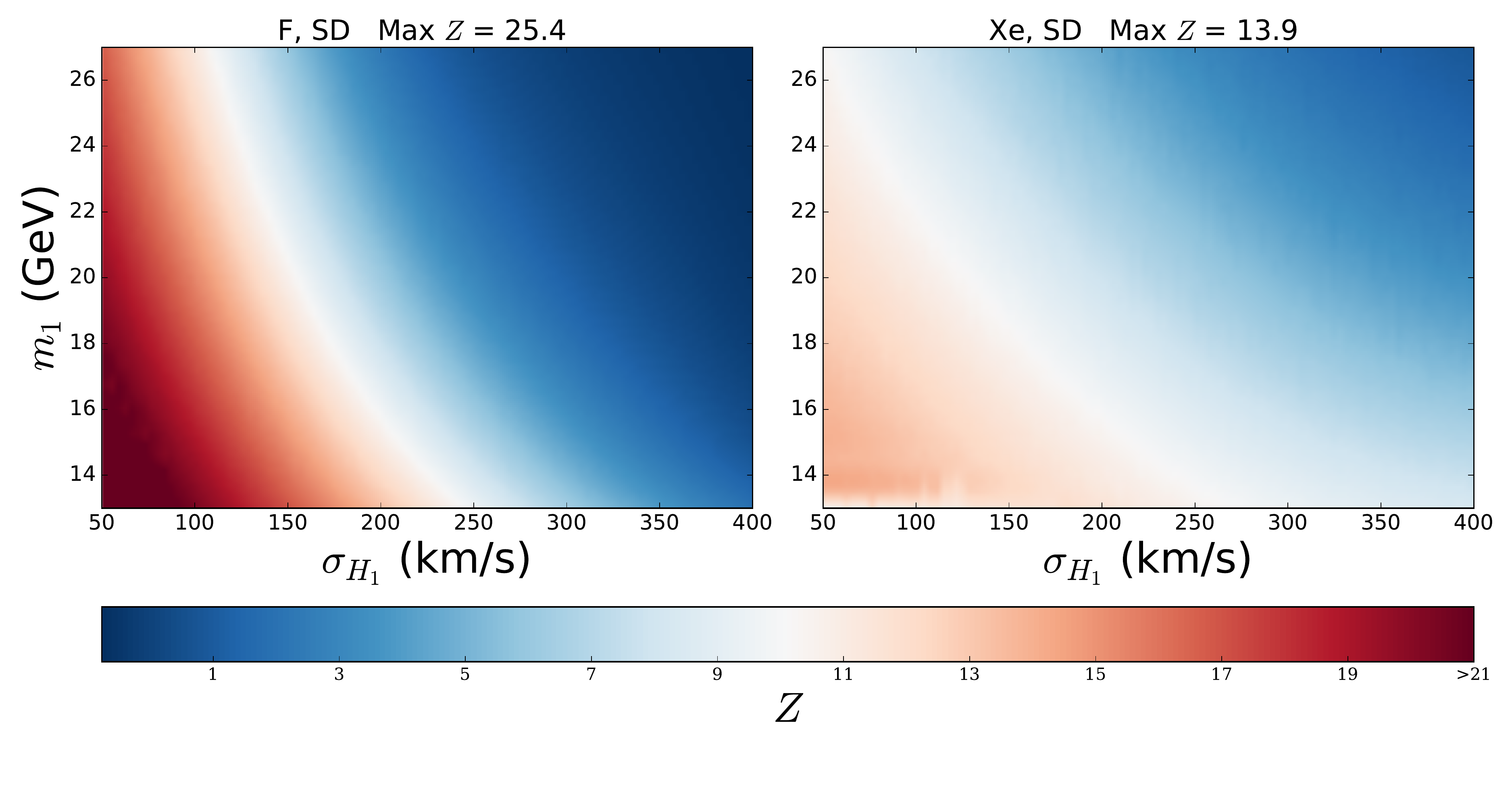}
	\caption{Significance $Z$ with which the median experiment can reject the one-DM hypothesis in favour of the two-DM hypothesis in the $\sigma_{H_1}$--$m_1$ plane. The fixed parameters are $m_2 = 150\, (200) $ GeV for SD (SI), $r_\rho=r_\sigma = 1$, $\kappa_{1,\,2} = 1$ and $\sigma_{H_{2}} = 270$ km/s. The top panel shows SI targets: Xe, Ge and the bottom is for SD targets: F and Xe. } \label{m1vd}
\end{figure}

\begin{figure}
	\centering
	\includegraphics[width=\set\textwidth]{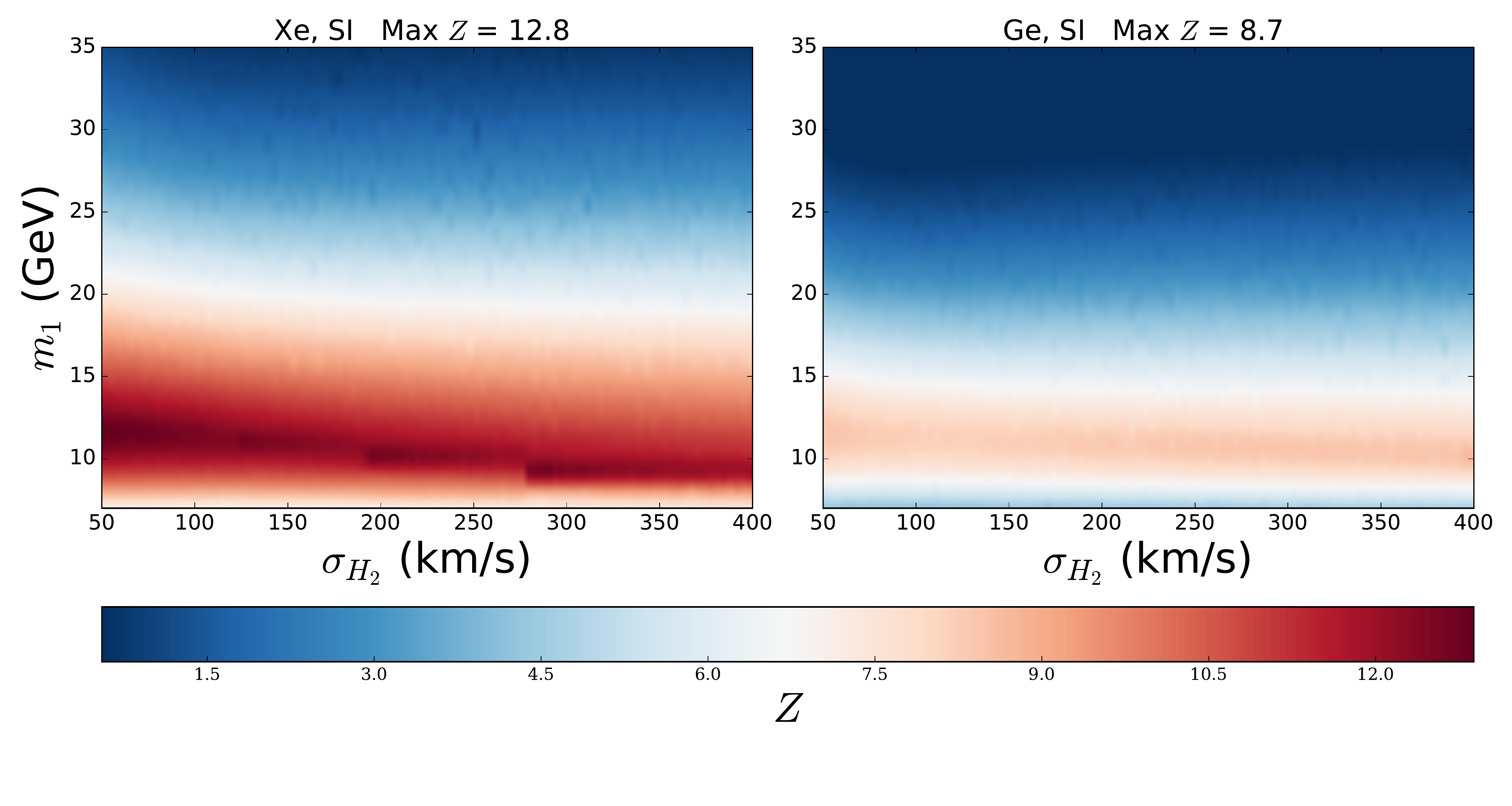}
		\includegraphics[width=\set\textwidth]{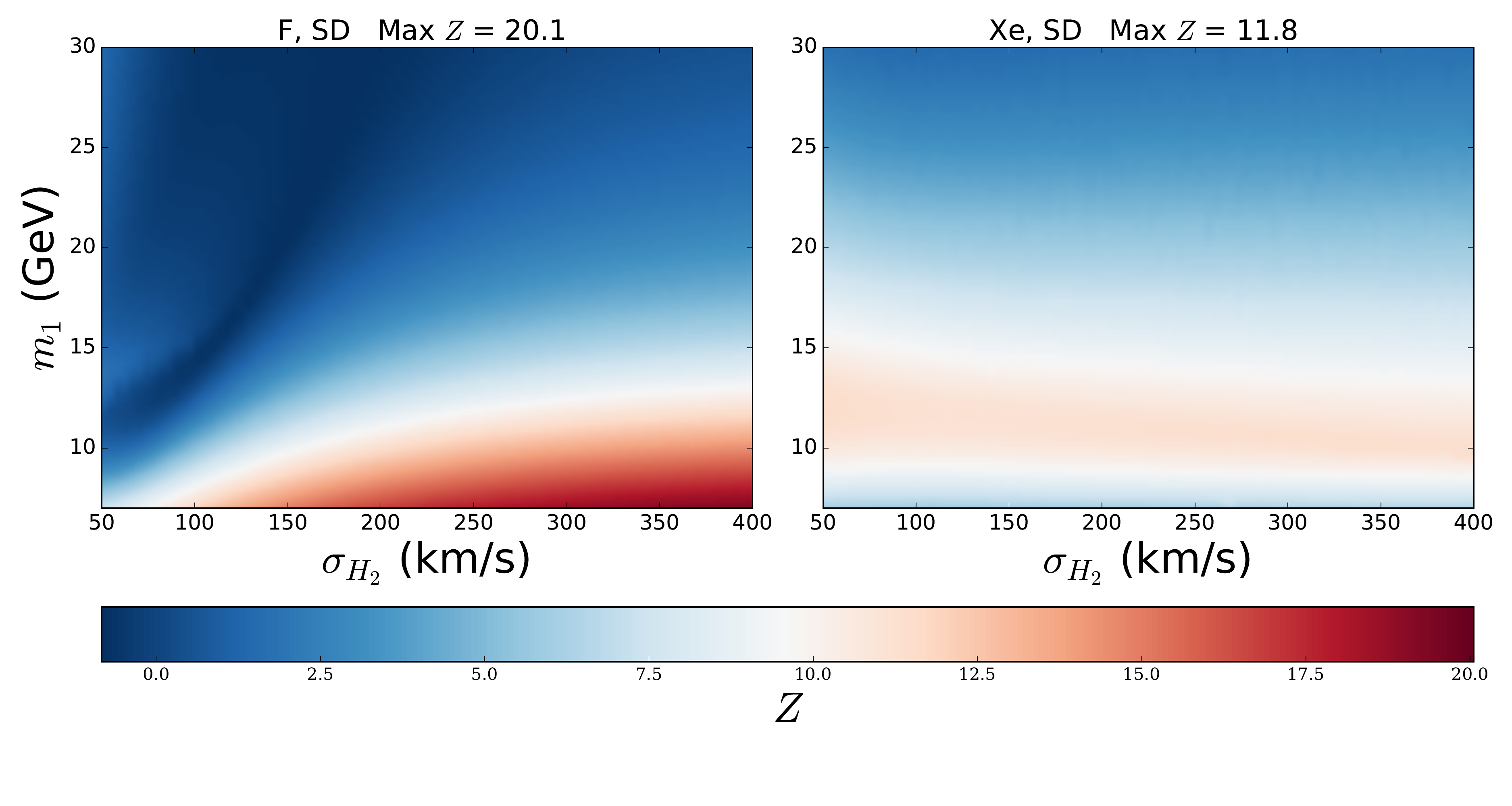}
	\caption{The same as Fig.~\ref{m1vd} but in the $\sigma_{H_2}$--$m_1$ plane, where at each point $\sigma_{H_1}$ is fixed to 270  km/s. } \label{m1v2}
\end{figure}

\begin{figure}
	\centering
\includegraphics[width=\textwidth]{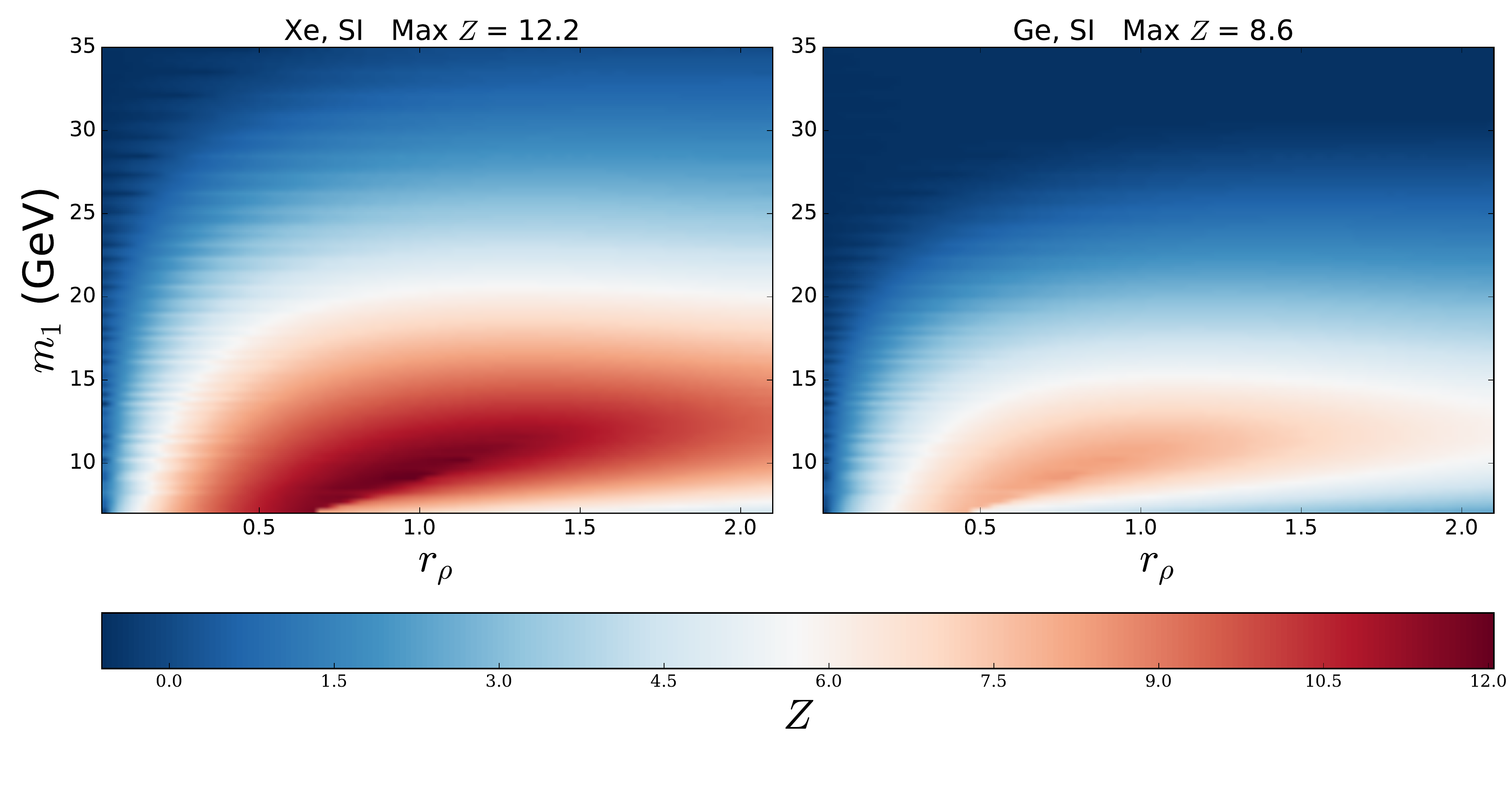}
	\includegraphics[width=\textwidth]{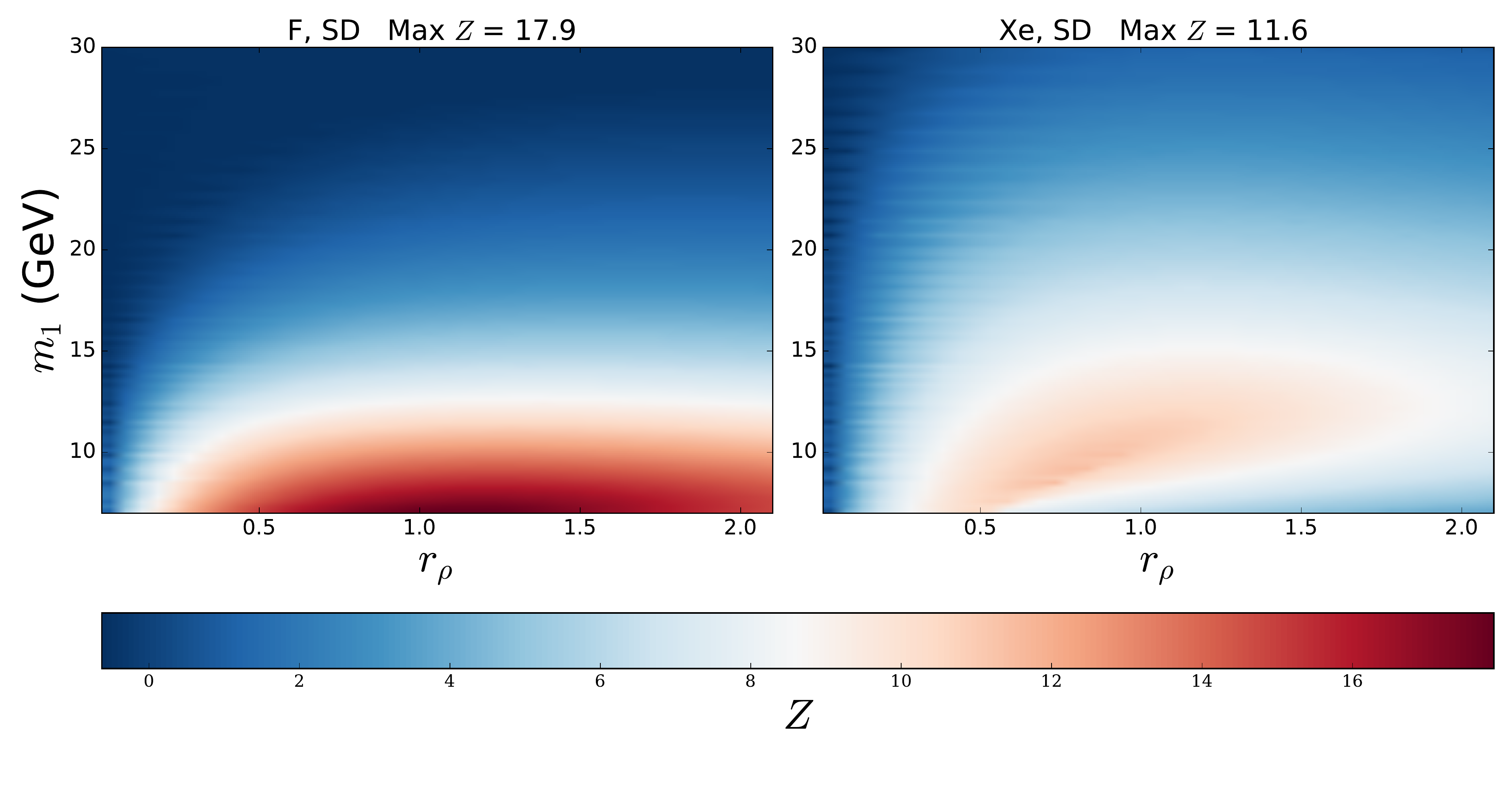}
	\caption{Significance $Z$ with which the median experiment can reject the one-DM hypothesis in favour of the two-DM hypothesis in the $r_\rho$--$m_1$ plane. The fixed parameters are $m_2 = 150\, (200) $ GeV for SD (SI), $\kappa_{1,\,2} = 1$ and $\sigma_{H_{1,\,2}} = 270$ km/s. The top panel shows SI targets: Xe and Ge, while the bottom is for SD targets: F and Xe.  } \label{rrho}
\end{figure}

\begin{figure}
	\centering
	\includegraphics[width=\textwidth]{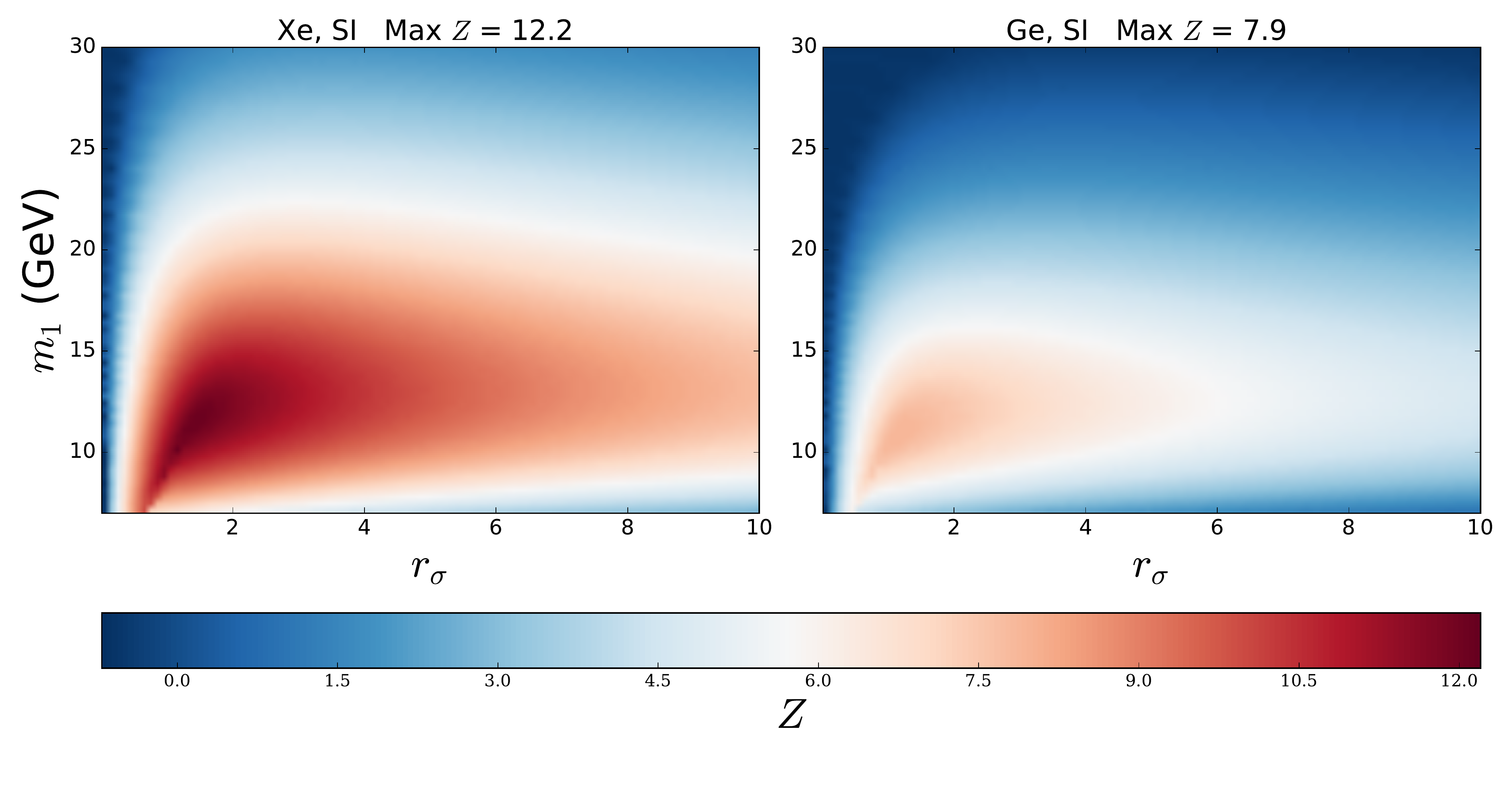}
	\includegraphics[width=\textwidth]{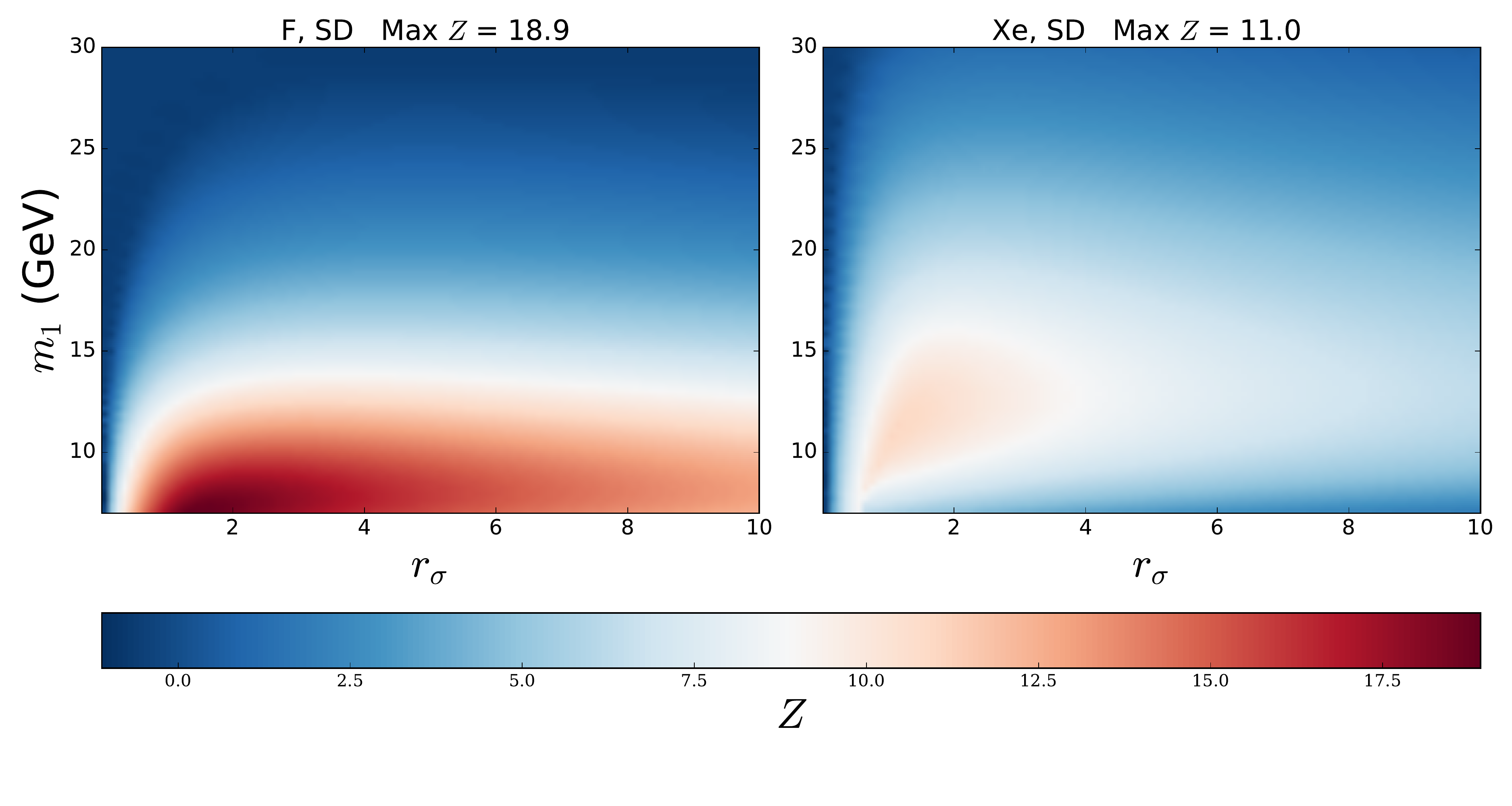}
	\caption{The same as Fig.~\ref{m1vd} but in the $r_\sigma$--$m_1$ plane.} \label{rsigma}
\end{figure}

\begin{figure}
	\centering
	\includegraphics[width=\textwidth]{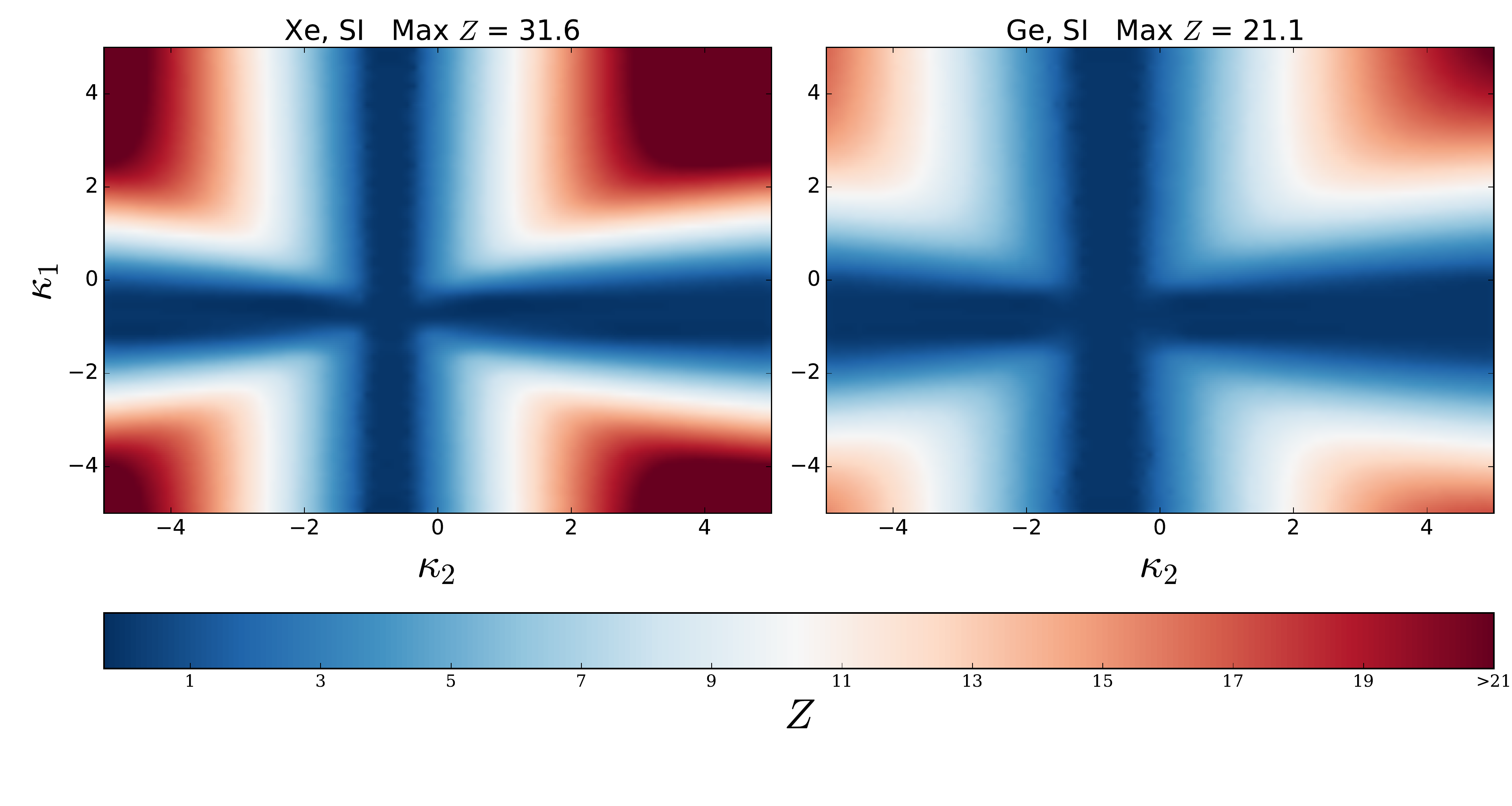}
	\includegraphics[width=0.65\textwidth]{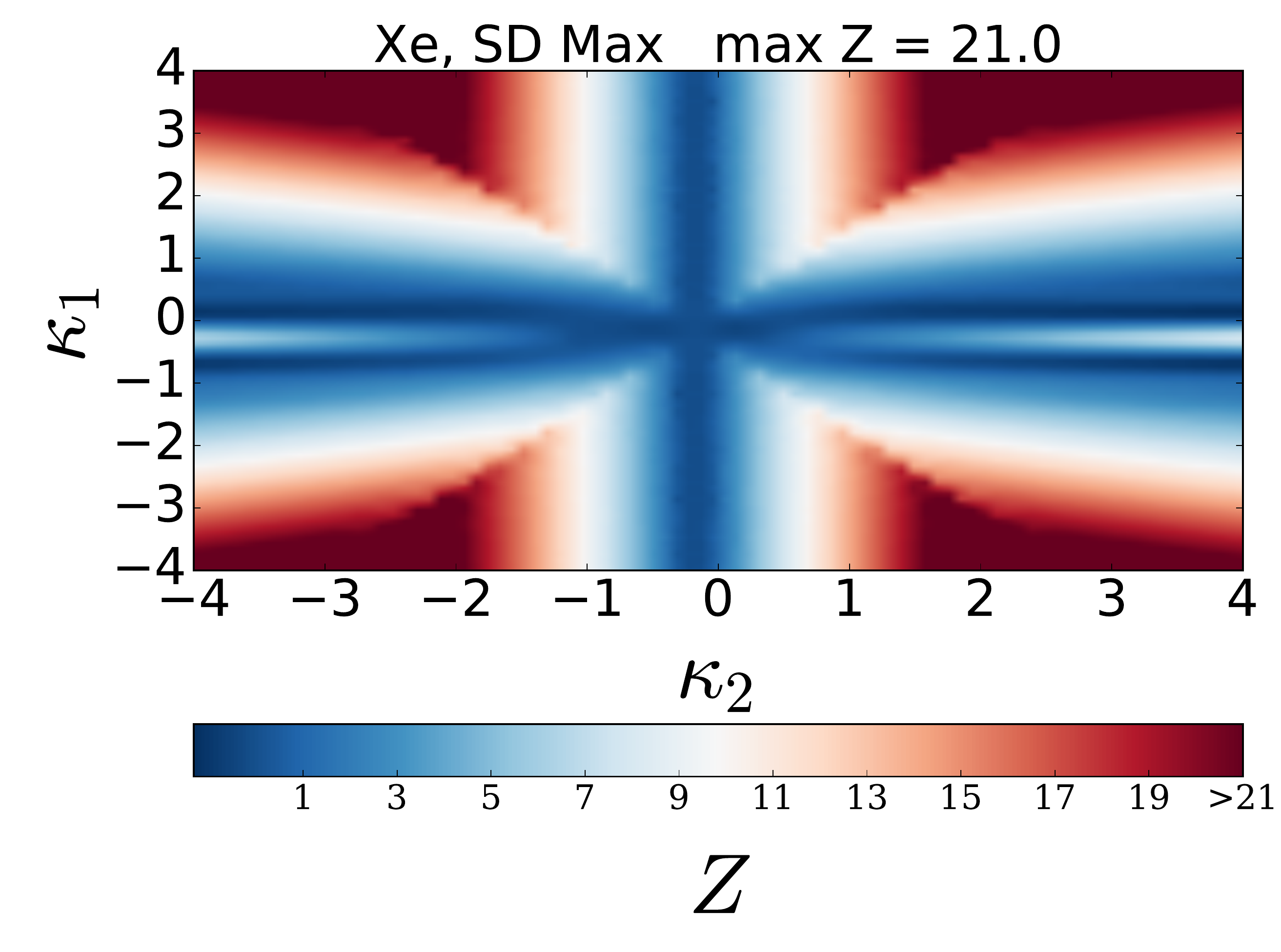}~
	\caption{Significance $Z$ with which the median experiment can reject the one-DM hypothesis in favour of the two-DM hypothesis in the $\kappa_{2}$--$\kappa_1$ plane. The fixed parameters are $m_2 = 150\, (200) $ GeV for SD (SI), $r_\rho = 1$ and $\sigma_{H_{1,\,2}} = 270$ km/s. The top panel is for SI targets: Xe and Ge. The bottom is for a SD Xe target. We do not show the results for F since the median significance does not change with varying $\kappa$, and hence is not illustrative.} \label{kappaScan}
\end{figure}

\begin{enumerate}
\item In the first general scenario we consider we include the velocity dispersions of the two components' velocity distributions into the parameter space of $\Ho$ and $\Ht$: $\Ho = [\sigma_1^p,m_1,\sigma_{H_1}]$ and $\Ht = [\sigma_1^p,m_2,m_2,\sigma_{H_1},\sigma_{H_2}]$.    
In Fig.~\ref{m1vd} we show the median sensitivity in the plane of $m_1$ versus $\sigma_{H_1}$ for the SI targets (top panel) and SD targets (bottom panel). We fix the velocity dispersion of the heavier particle to $\sigma_{H_2} = 270$ km/s, while the lightest (DM one) has a free velocity dispersion in the range $50<\sigma_{H_{1}}<400\,{\rm km/s}$. The low velocity dispersion case corresponds to a stream.  The median sensitivity is calculated realising that $\Ht$ has two more degrees of freedom than $\Ho$. Hence, $k=2$ and the median sensitivity is calculated using Eq.~\eqref{nsigma}. In the case of SI xenon (and similarly for germanium), one can see that there is a correlation between the lightest DM mass and its velocity dispersion, which is expected. This is because the median significance is large, whether for very low DM masses with large dispersions such that events are above threshold, or for medium-range masses with small dispersions. With respect to the SHM case, for $m_1$ in the region of $\sim15$ GeV, the hypotheses can be somewhat better resolved by a smaller velocity dispersion. Sodium (not shown) can only discriminate between $\Ho/\Ht$ at the level of 1.6-sigma.

In Fig.~\ref{m1v2} we show $m_1$ versus $\sigma_{H_2}$ for the SI targets (top panel) and SD targets (bottom panel). In a similar way to Fig.~\ref{m1vd}, here we fix $\sigma_{H_1} = 270$ km/s and let $\sigma_{H_2}$ run over $50<\sigma_{H_{2}}<400\,{\rm km/s}$. We see that the median significance is almost insensitive to $\sigma_{H_2}$ (at $m_1\sim12$ GeV) for all targets except for fluorine in the SD case, where $Z$ starts to decrease for smaller $\sigma_{H_2}$ ($\lesssim 100$ km/s). The lack of dependence on $\sigma_{H_2}$ is due to the fact that for heavy WIMPs the whole velocity distribution is probed. 

\item In the second scenario we drop the assumption of equal energy densities and cross-sections, i.e, we take $r_\rho \neq\,r_\sigma $. We now allow $r_\rho$ and $r_\sigma$ to enter the parameter space of $\Ht$ in the range  $0.01<r_{\rho}<2.5$ and $0.01 <r_\sigma < 10$:  $\Ho = [\sigma_1^p,m_1]$ and $\Ht = [\sigma_1^p,m_2,m_2,r_\rho,r_\sigma]$.  Since $r_\rho$ and $r_\sigma$ only exist within the parameter space of $\Ht$ then in this case we have $k=3$ and hence in order to calculate the median significance we need to resort to Eq.~\eqref{nsigma}.

In Fig.~\ref{rrho} we show the $r_\rho$--$m_1$ plane for SI (top panel) and SD (bottom panel) interactions, whilst keeping $r_\sigma = 1$. One can see that, for any fixed DM mass, $Z$ drops once we move away from the $r_\rho=1$ case. Conversely, for the SD case in a fluorine target the significance is approximately constant in the range $0.25<r_\rho<2.5$.  These results are similar to that of Fig.~\ref{rsigma} which shows the $r_\sigma$--$m_1$ plane for SI (top panel) and SD (bottom panel) interactions for fixed $r_\rho=1$. However, the region of high significance is slightly more elongated along the $r_\sigma$ axis for both the SI and SD cases. The SD fluorine target significance now drops lower at $r_\sigma<0.1$. We can conclude that if $r_\rho$ [$r_\sigma$] are different from one, but not more than a factor $\mathcal{O} (1\, [10])$ from it, the sensitivity is not significantly worsened.
 
\item The final scenario we consider involves isospin-violating DM (see also~\cite{Feng:2011vu,Yaguna:2016bga}). We perform a scan over the WIMP-nucleon coupling ratios $\kappa_1$ and $\kappa_2$. We let the coupling ratios enter the hypotheses' parameter space in the ranges $\kappa_1,\kappa_2\in[-5,5]$, keeping in mind that these parameters enter the rate differently for SI and SD interactions: $\Ho = [\sigma_1^p,m_1,\kappa_1]$ and $\Ht = [\sigma_1^p,m_2,m_2,\kappa_1,\kappa_2]$. The total difference in number of degrees of freedom is $k=2$ and hence the median sensitivity is again calculated using Eq.~\eqref{nsigma}.

In figure \ref{kappaScan} we show the $\kappa_2$--$\kappa_1$ plane for SI (top panel) and SD (bottom panel) interactions. In the SI case, one can easily see from inspection of Eq.~\eqref{AeeffDef} that $A^{\rm eff}\rightarrow 0$ as $\kappa\rightarrow -0.7$ for Xe and $-0.8$ for Ge. Hence if either one of $\kappa_1$ or $\kappa_2$ are close to zero, the rate due to this component will be vanishing and hence there will be no good hypothesis discrimination. One will also notice that the median sensitivity grows for larger $\kappa_{1,\,2}$, where $A^2_{\rm eff}\propto |\kappa_{1,\,2}|^2$, becoming independent of the signs of the $\kappa_i$. In the SD case, since Xe targets are more sensitive to WIMP-neutron couplings, we see that the rate diminishes for $\kappa\rightarrow 0$. In the case of a fluorine target however, since the nuclear mass is much lower than Xe, for a given $E_R$, the SD form factor is approximately 1, and hence varying $\kappa$ does not significantly effect the median sensitivity. For this reason we do not show the result for the fluorine target. \footnote{The median sensitivity is $Z > 15\;\; \forall\; \kappa_{1,\,2}\in[-5,5]$ and hence not much is gained from inspection of this result.}

\end{enumerate}


\section{Parameter estimation: extracting the DM properties from 2DM signals} \label{sec:par_estimation}
Previous studies have conducted parameter estimation analyses for ensembles of WIMP parameters in current generation detectors under the assumption of only one component DM~\cite{Pato:2010zk,Peter:2013aha}. In the following we want to study the case where the 2 DM hypothesis has been confirmed, so that one would try to extract the DM parameters. We again adopt a frequentist framework to conduct our parameter estimation by  making use of the method of maximum likelihood. We use the \texttt{MultiNest} \cite{feroz2009multinest} package which obtains likelihood samples within a user-specified prior boundary via an importance nested sampling algorithm. We also do the analysis under a Bayesian framework for completion, however we present these results in appendix~\ref{app:bayes}. We choose this prior range to span over parameter values that give good discrimination between $\Ho$ and $\Ht$ as observed from the results of sec.~\ref{sec:hypothesis_test}.  In the following sections we briefly discuss the statistical methods and then present our results.

\subsection{Statistical methods}

\subsubsection{Method of maximum likelihood}
The method of maximum likelihood (ML) allows us to make estimations about model parameters by finding the parameter configuration $\hat{\boldtheta}$ that maximises the likelihood function $\L(\mathbf{x}|\boldsymbol{\theta})$. The values $\hat{\boldtheta}$ are called the ML estimators of the model. For this study we wish to examine the parameter space of only two of the 8 model parameters at a time. One may explicitly `profile out' all other parameters by writing the likelihood function in terms of the ML estimators of these parameters. The result is called the profile likelihood 
\begin{align}
\label{profLike}
\L(\mathbf{x}\,|\,\theta_1,\theta_2) = \L(\mathbf{x}\,|\,\theta_1,\theta_2,\hat{\hat{\theta_3}}...\hat{\hat{\theta_{n}}})\;,
\end{align}
where the double hat notation denotes the ML estimators of $\theta_3...\theta_n$ for the given values of $\theta_1$ and $\theta_2$. Maximising the likelihood in Eq.~\eqref{profLike} with respect to $\theta_{1,\,2}$ is equivalent to maximising the profile likelihood ratio (PLR)
\begin{align}
\label{PLR}
\lambda (\theta_1,\theta_2) = \frac{\L(\mathbf{x}\,|\,\theta_1,\theta_2,\hat{\hat{\theta_3}}...\hat{\hat{\theta_{n}}})}{{\L(\mathbf{x}\,|\,\hat{\boldtheta})}} \equiv\frac{\L(\mathbf{x}\,|\,\theta_1,\theta_2,\hat{\hat{\theta_3}}...\hat{\hat{\theta_{n}}})}{{\L_{\rm max}}} \;.
\end{align}
In our analysis we will call the point in the $\theta_1$-$\theta_2$ plane that maximises the profile likelihood ratio the `best-fit' point.  In the limit of large statistics, the distribution of $-2\ln\lambda(\theta_1,\theta_2)$ will tend towards a $\chi^2$ with $k=2$ degrees of freedom as given by Wilk's theorem. As a result, this leads to a critical region that is defined by a cut on $\lambda$ that we will choose to represent contours of the standard 2$\sigma$ (95.45\%) frequentist confidence level (C.L).  

 \subsection{Results}
 The likelihood function we use for our analysis is a Gaussian as defined in Eq.~\eqref{Likelihood:Defn} but we now require that the likelihood is a function of all eight model parameters shown in tab. \ref{tab:benchmarks}:
 \begin{align}
 \label{glike}
 \L(\mathbf{x}\,|\,\boldtheta) =\prod\limits^N_i\, \frac{1}{\sqrt{2\pi}\sigma_i}e^{-\frac{\left[x_i-\mu_i (\boldtheta)\right]^2}{2\sigma_i^2} }\;\quad \boldtheta = [m_1,m_2,\kappa_1,\kappa_2,r_\rho,r_\sigma,\sigma_{H1},\sigma_{H2}]\;,
 \end{align} 
 where again, $x_i$, $\sigma_i$ and $\mu_i(\boldtheta)$ are the data, error and predicted number of counts in the $i$th energy bin.
 In a similar way to the approach in sec.~\ref{sec::stats}, we use the Asimov likelihood in our parameter estimation in order to obtain a measure of the median experiments ability to determine the model parameters. The likelihood that enters the profile likelihood ratio is then 
\begin{align}
\label{AmimovLikePS}
\L(\mathbf{x} = \mu (\boldtheta^{\rm true})\,|\,\boldtheta) =\prod\limits^N_i\, \frac{1}{\sqrt{2\pi\mu_i (\boldtheta^{\rm true})}}e^{-\frac{\left[\mu_i (\boldtheta^{\rm true})-\mu_i (\boldtheta)\right]^2}{2\mu_i (\boldtheta^{\rm true})} }\;.
\end{align}

\begin{table}[t!]
	\centering
	\begin{tabular}{cccccccccc}
		Benchmark & $m_1$ &$m_2$ & $\kappa_1$ & $\kappa_2$ & $r_\rho$ & $r_\sigma$ & $\sigma_{H_1}$& $\sigma_{H_2}$&Targets  \\ \hline
		SI &  15 & 200& 2&2.5& 0.7&1.1&220&230& Xe, Ge, Xe+Ge \\\hline
		SD &  7 & 150& 2.5&1.5& 0.9&0.9&220&230& Xe, F, Xe+F \\\hline
	\end{tabular}
	\caption{Parameter benchmarks used to generate the Asimov data in Eq.~\eqref{AmimovLikePS}. These benchmarks are chosen taken into account the results of sec.~\ref{sec:hypothesis_test}, i.e., choosing illustrative regions of parameter space that give a large discrimination between the one and two-component hypotheses. DM masses $m_{1,\,2}$ are measured in GeV. }
	\label{tab:benchmarks}
\end{table}

 As for generating the Asimov data $\boldtheta^{\rm true}$, we use a  few benchmark scenarios as given in tab.~\ref{tab:benchmarks} that contain parameter values that ensure a good discrimination between the one and two-component hypotheses, as deduced from the results of sec.~\ref{sec:hypothesis_test}. We provide benchmarks for SI and SD, that generate data for both one signal and two signals in detectors with different nuclei. We exclude sodium since the results from sec.~\ref{sec:hypothesis_test} show that no significant discrimination between $\Ho$ and $\Ht$ is possible. Another thing to note is that we don't assume that the DM particles have equal couplings to protons and neutrons, i.e., $\kappa_1 \neq \kappa_2 \neq 1$, or equivalently $A_{\rm eff, 1} \neq A_{\rm eff, 2}$. For the case of just one experimental signal from one detector, this could be absorbed into $\sigma^p_1$ and $r_\sigma$. That is, from one signal one cannot extract information about couplings to neutrons and protons, while one can do this using two signals (also using information on the coupling to protons from neutrinos from the sun, see Refs.~\cite{Kavanagh:2014rya,Blennow:2015oea,Blennow:2015hzp,Ferrer:2015bta,Ibarra:2017mzt}). However, in the case of two experimental signals in two different types of nuclei (we use Ge and Xe for SI, and Xe and F for SD) information about the interactions with protons and neutrons can in principal be obtained. Our chosen ranges for the scan are shown in table~\ref{tab:priors}.
  \begin{table}[!t]
  	\centering
  	\begin{tabular}{ccc}
  		Parameter & Range & Type \\\hline
  		$m_{1,\,2}$ (GeV) & $[10,1000]$   &Flat \\\hline
  		$\sigma_{H_1,\, H_2}\,({\rm km/s})$& $[20,400]$&Flat\\\hline
  		$\sigma_{p}\,({\rm cm^2})$ & $[10^{-48}, 10^{-38}]$ &Log \\\hline
  		$r_\rho$ & $[10^{-2},10^2]$ &Log\\ \hline
  		$\kappa_{1,\,2}$ & [-5,5]&Flat\\\hline
  	\end{tabular}
  	\caption{Ranges for the parameter estimation.}
  	\label{tab:priors}
  \end{table}

\begin{figure}[t!]
	\centering
	\includegraphics[width=0.42\textwidth]{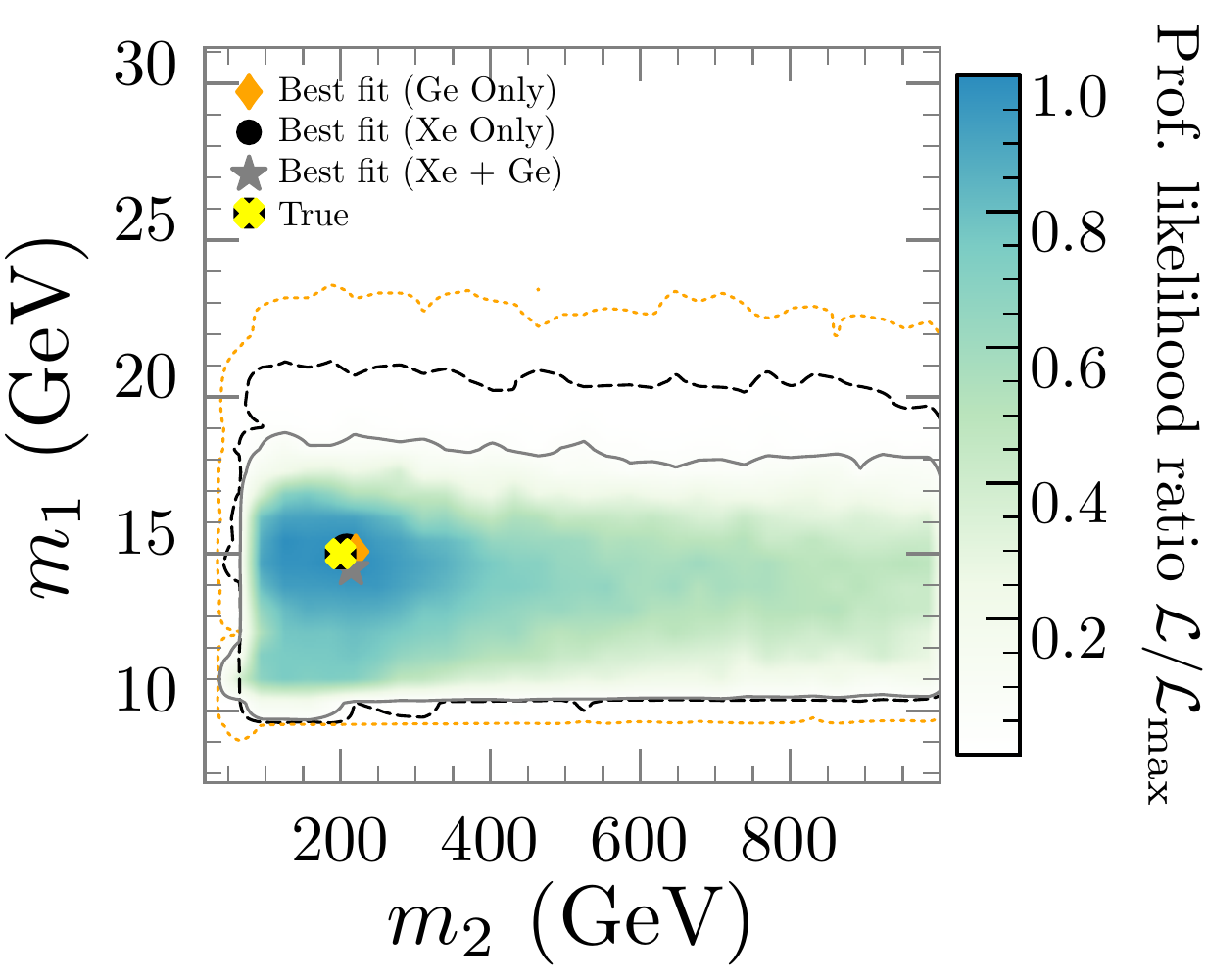}\quad
	\includegraphics[width=0.42\textwidth]{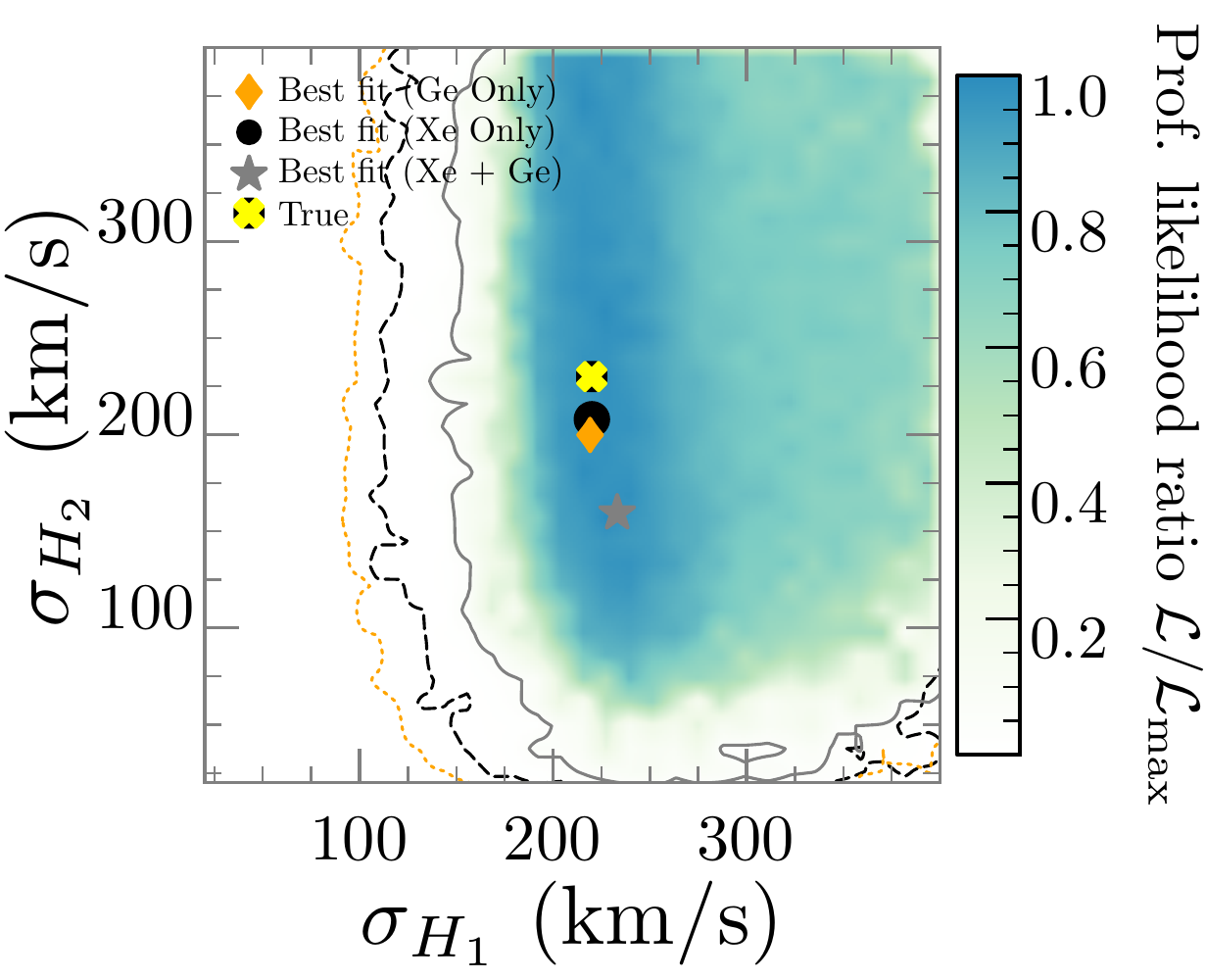}\quad
	\includegraphics[width=0.42\textwidth]{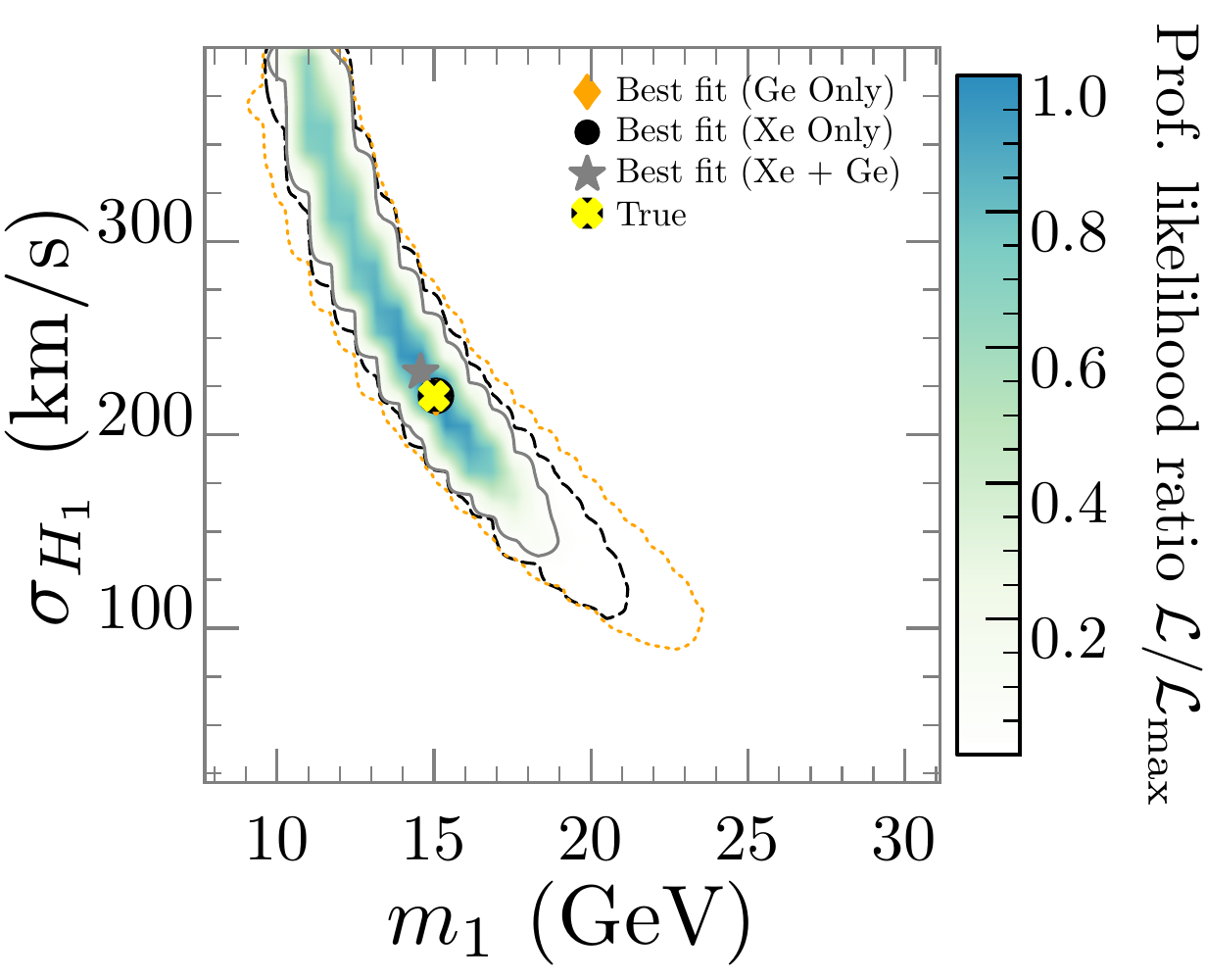}\quad
	\includegraphics[width=0.42\textwidth]{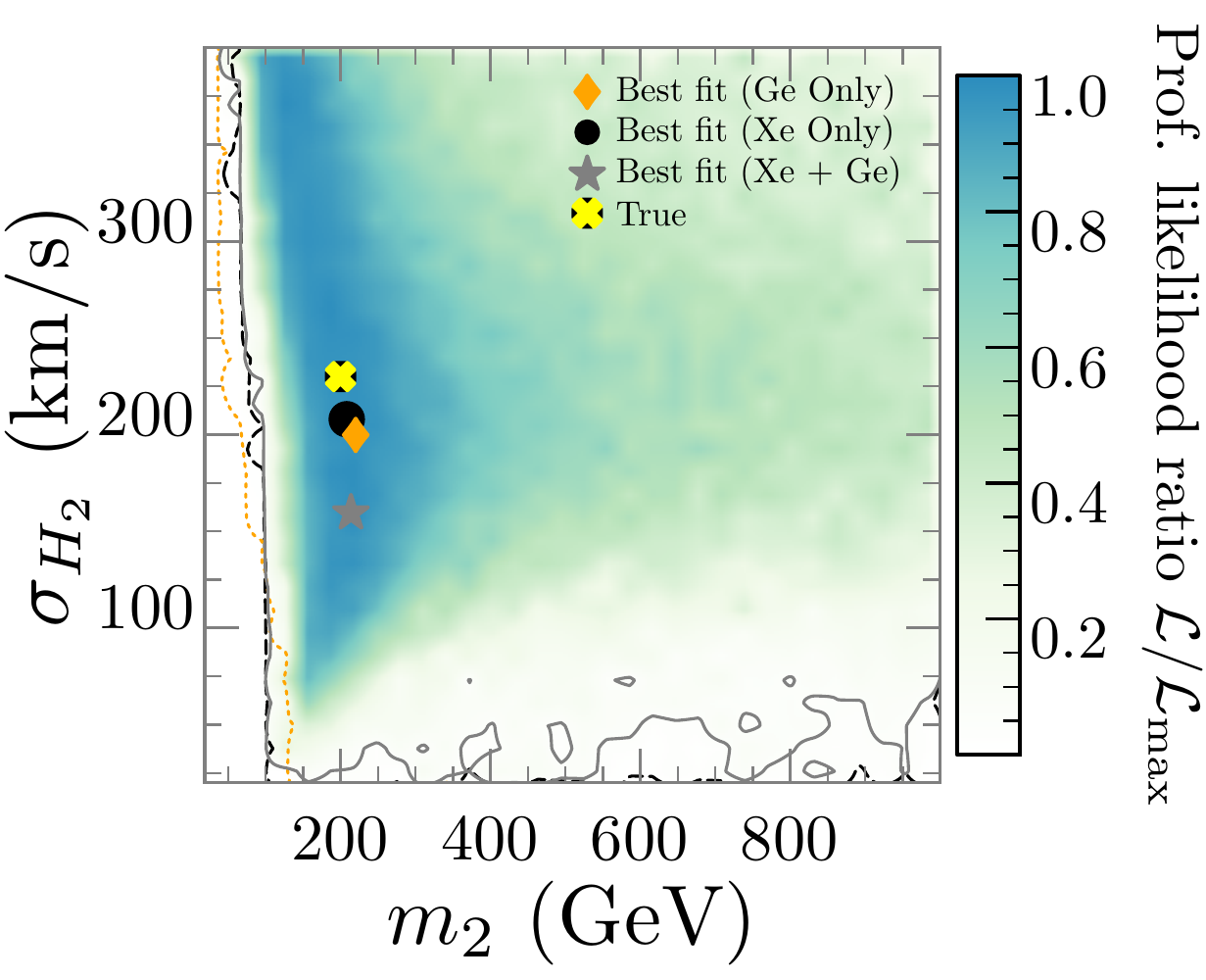}\quad
	\includegraphics[width=0.42\textwidth]{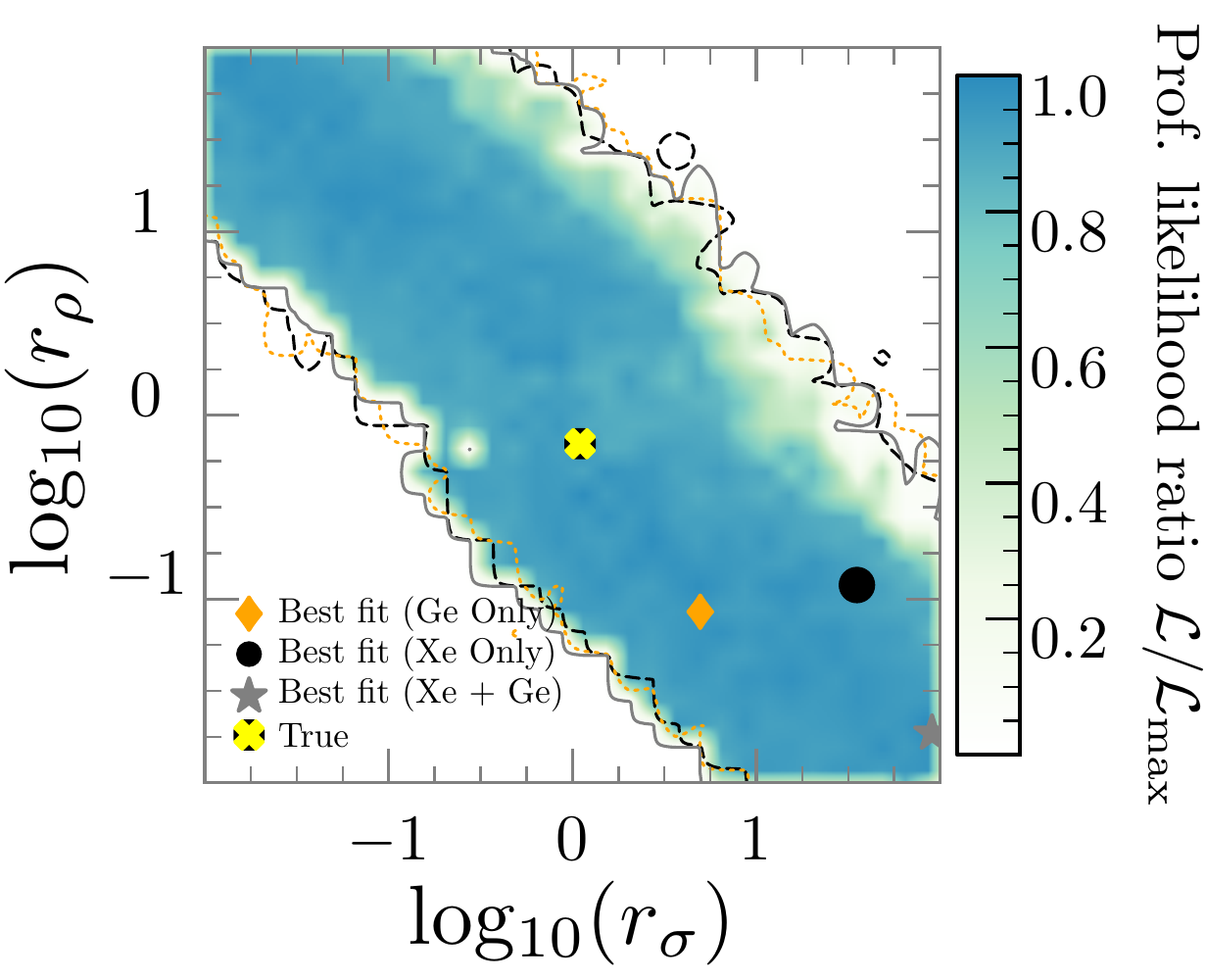}
	\includegraphics[width=0.42\textwidth]{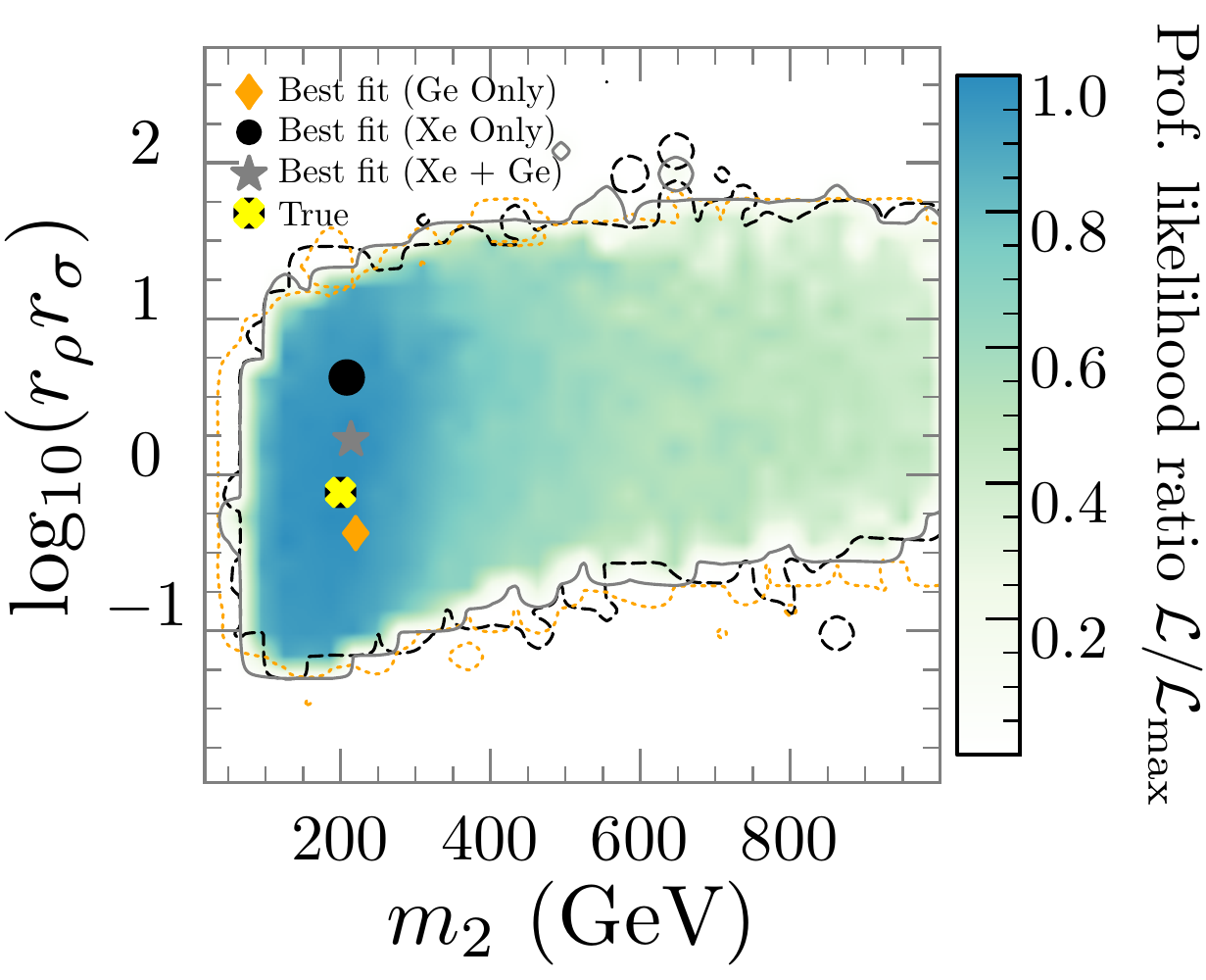}\quad
	\includegraphics[width=0.42\textwidth]{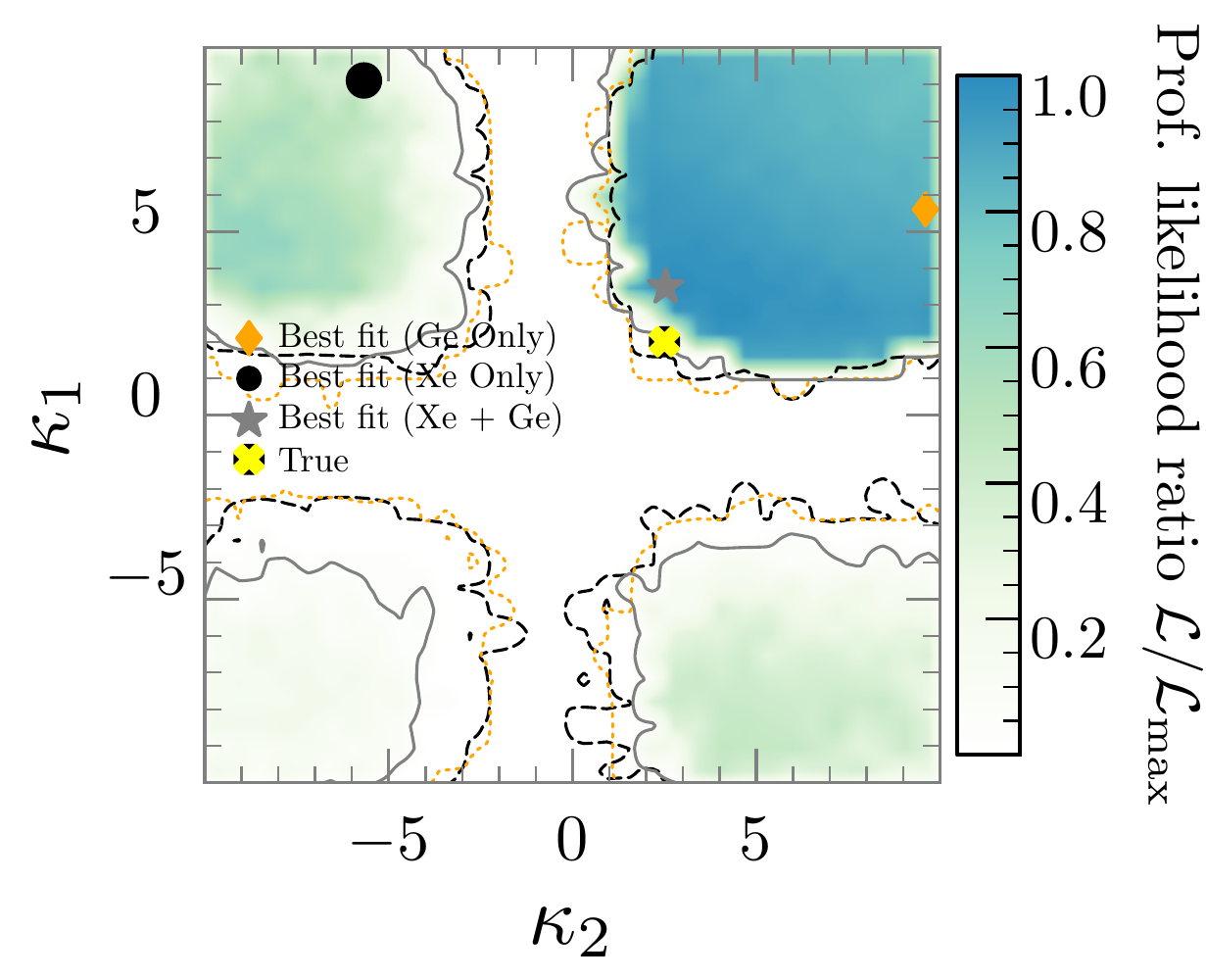}\quad
	\includegraphics[width=0.42\textwidth]{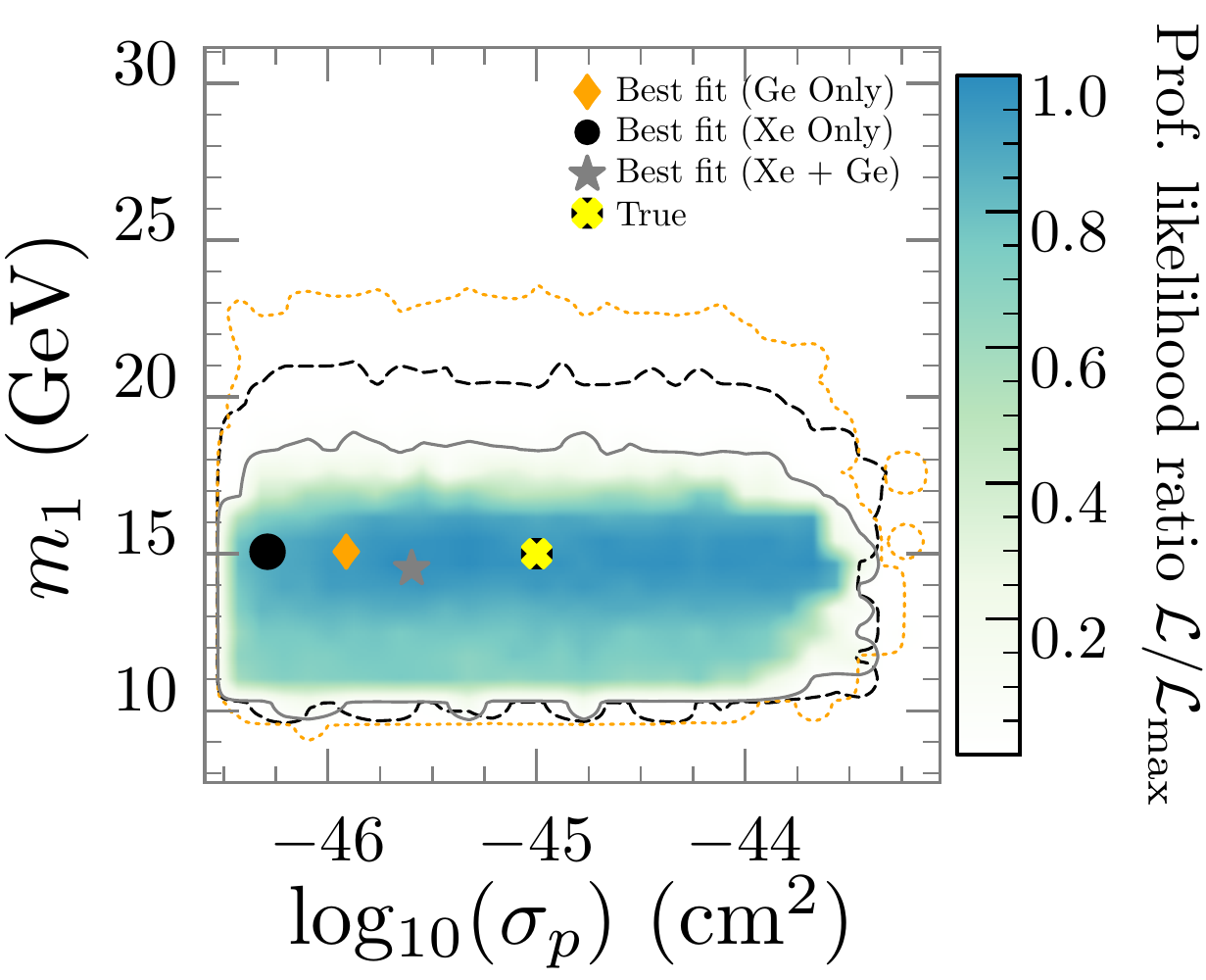}
	
	\caption{SI results of the parameter estimation for one and two signals in Xe and Ge.  The colour scale shows PLR while the solid (grey), dashed (black) and dotted (orange) lines show $2\sigma$ C.L contours for the combined Xe+Ge, Xe-only and Ge-only signals respectively. The best-fit PLR point for the Ge/Xe/combined Xe+Ge signal is shown with a diamond/dot/star. The true parameter values (see tab.~\ref{tab:benchmarks}) are depicted by the yellow cross.} 
		
	 \label{2exp_fit}
\end{figure}

\begin{figure}
	\centering
	\includegraphics[width=0.42\textwidth]{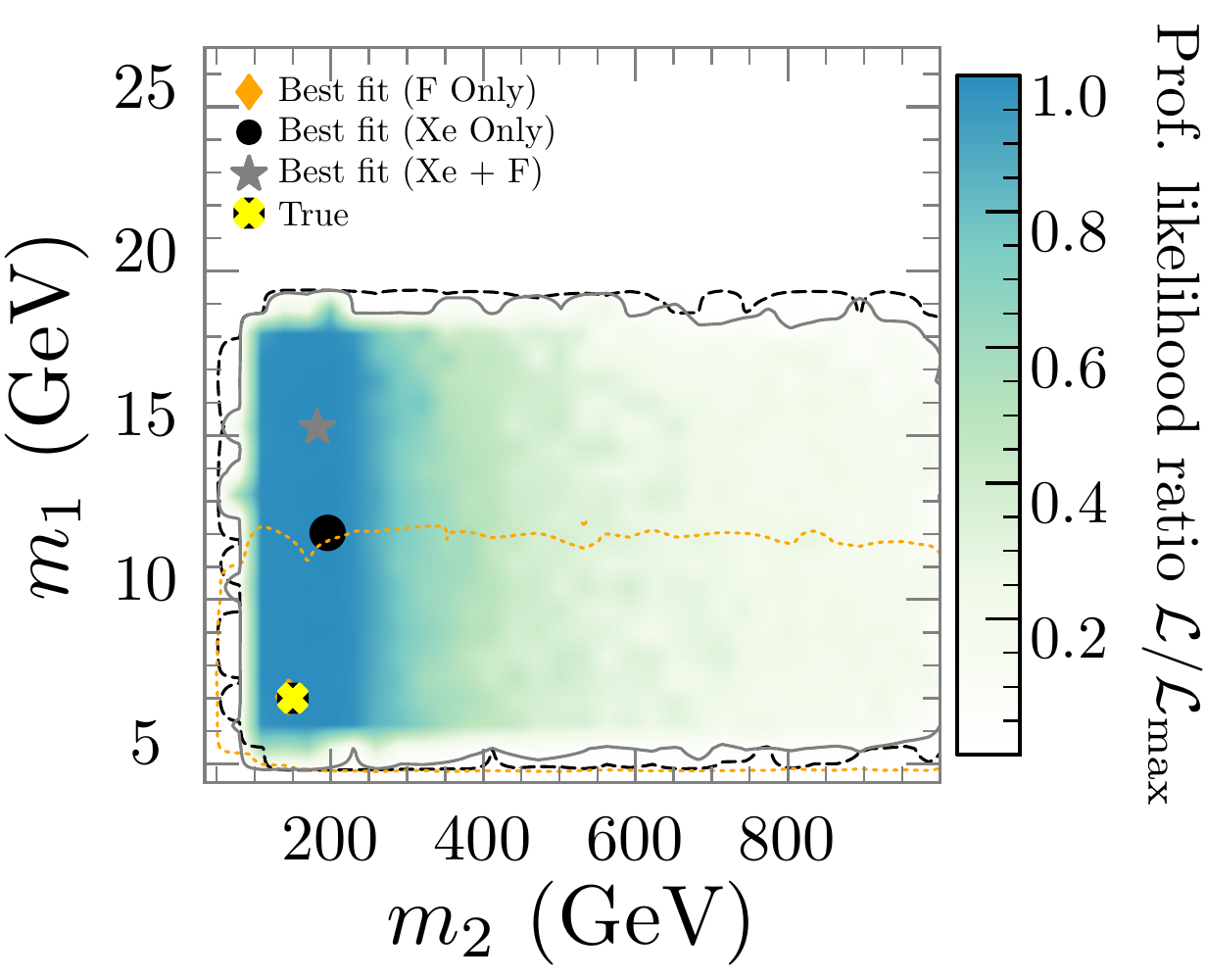}\quad 
	\includegraphics[width=0.42\textwidth]{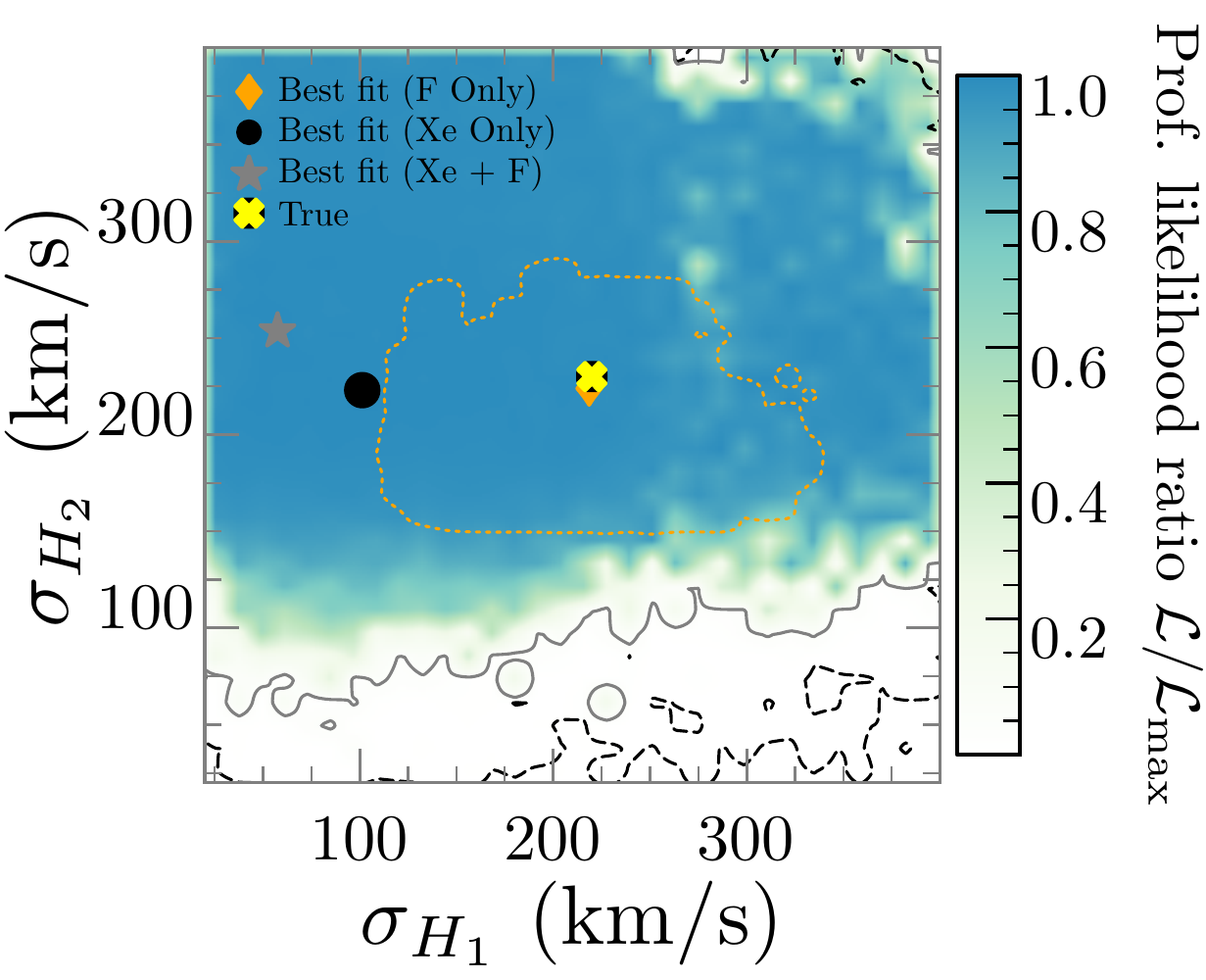}\quad 
	\includegraphics[width=0.42\textwidth]{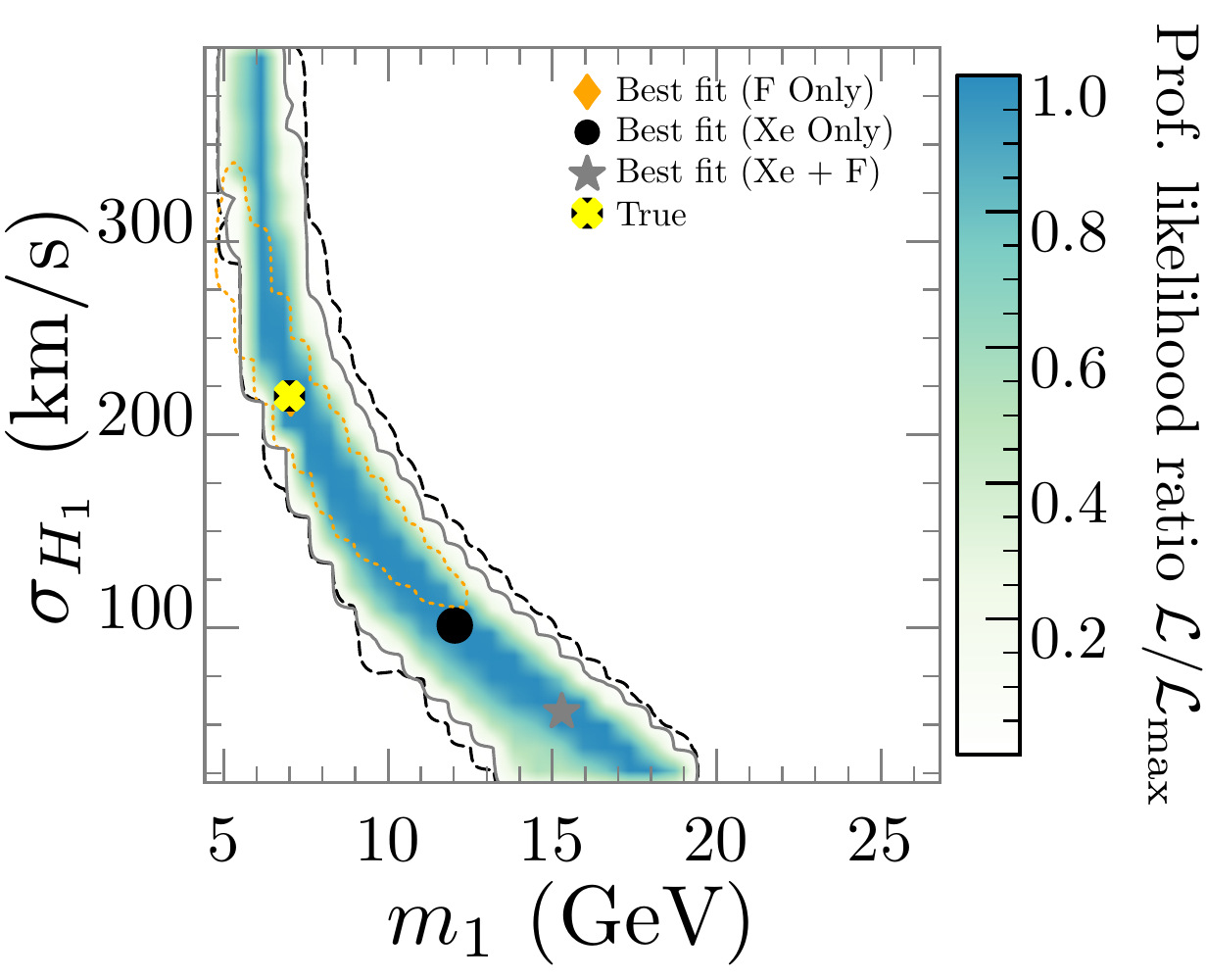}\quad 
	\includegraphics[width=0.42\textwidth]{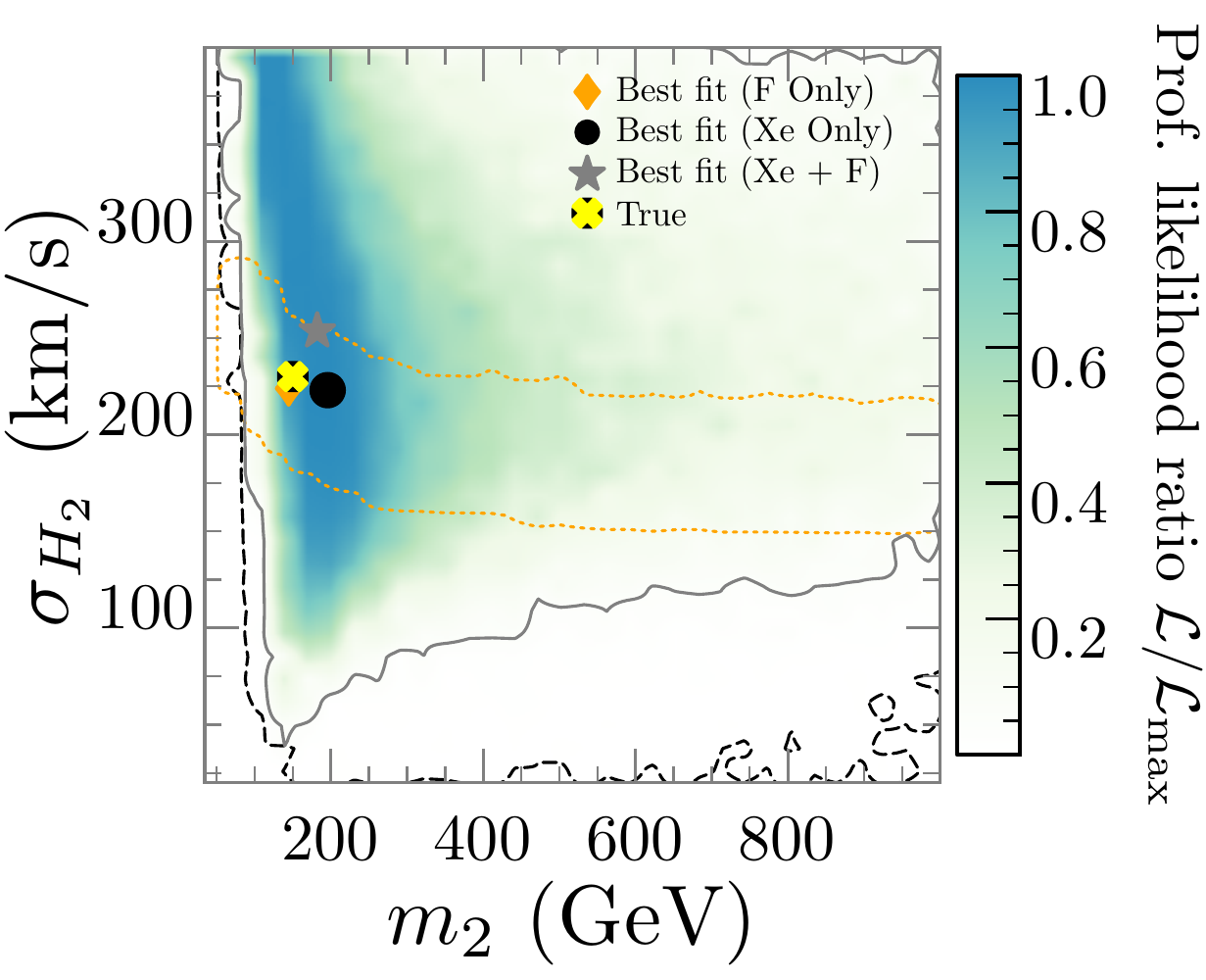}\quad 
	\includegraphics[width=0.42\textwidth]{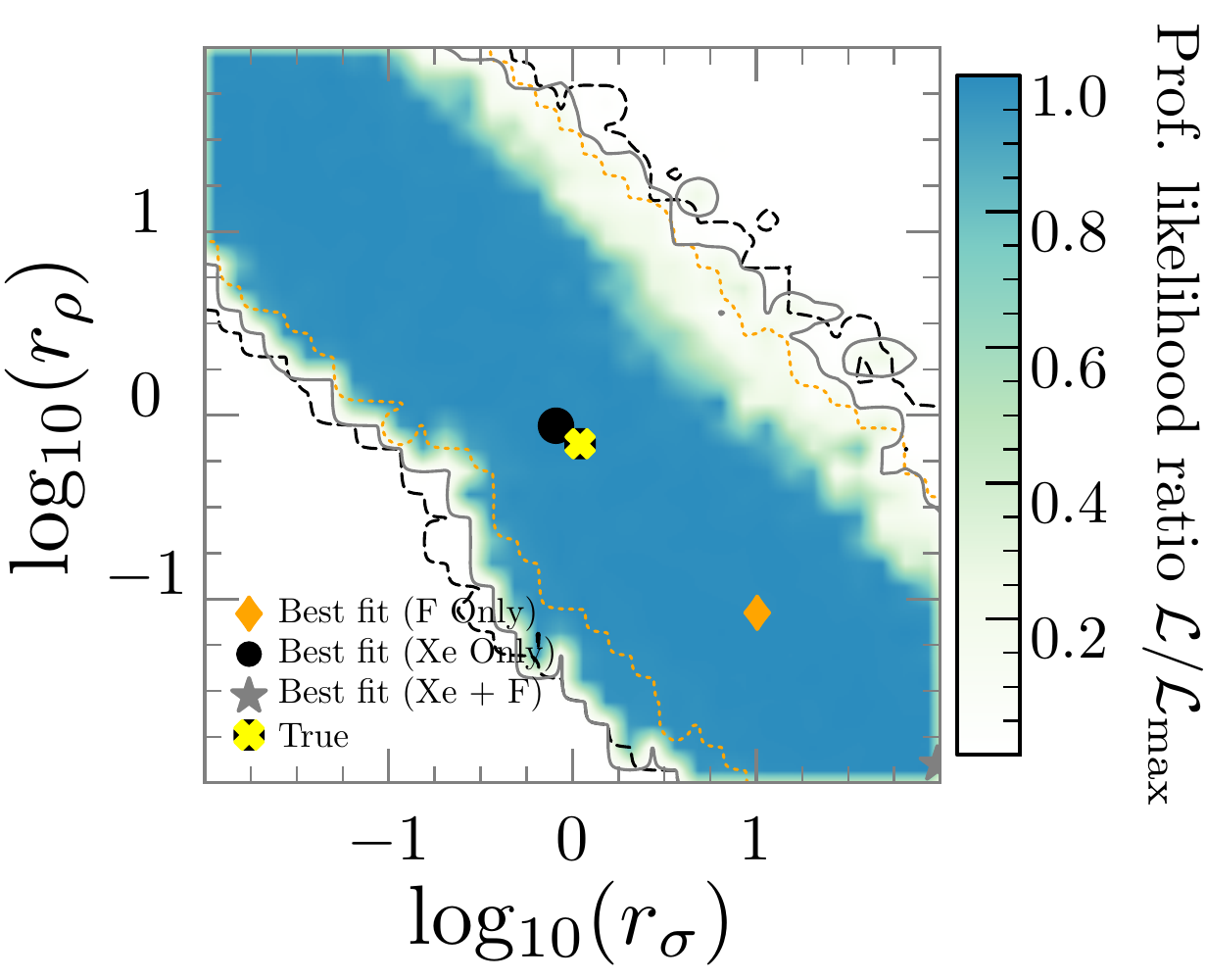}
	\includegraphics[width=0.42\textwidth]{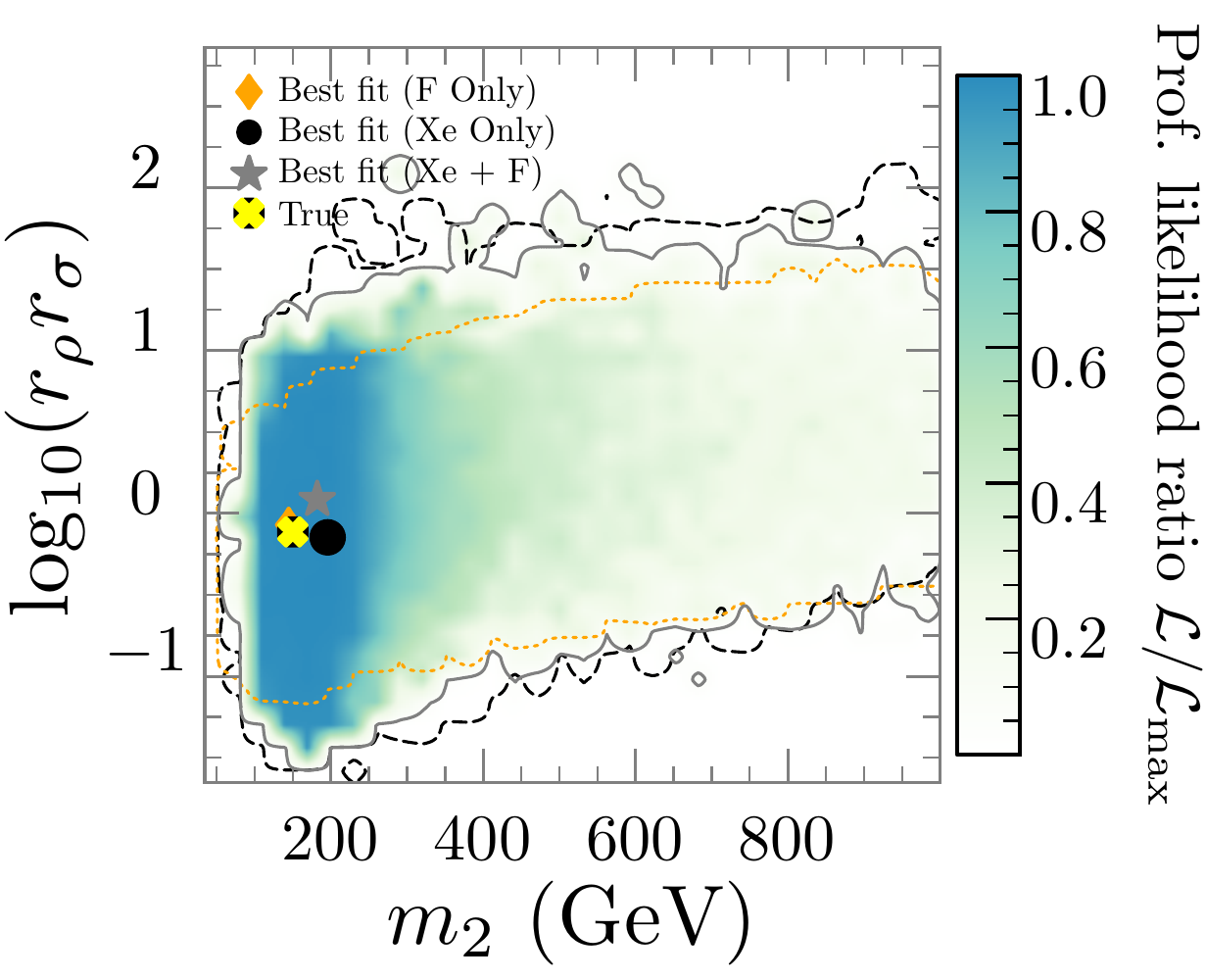}\quad 
	\includegraphics[width=0.42\textwidth]{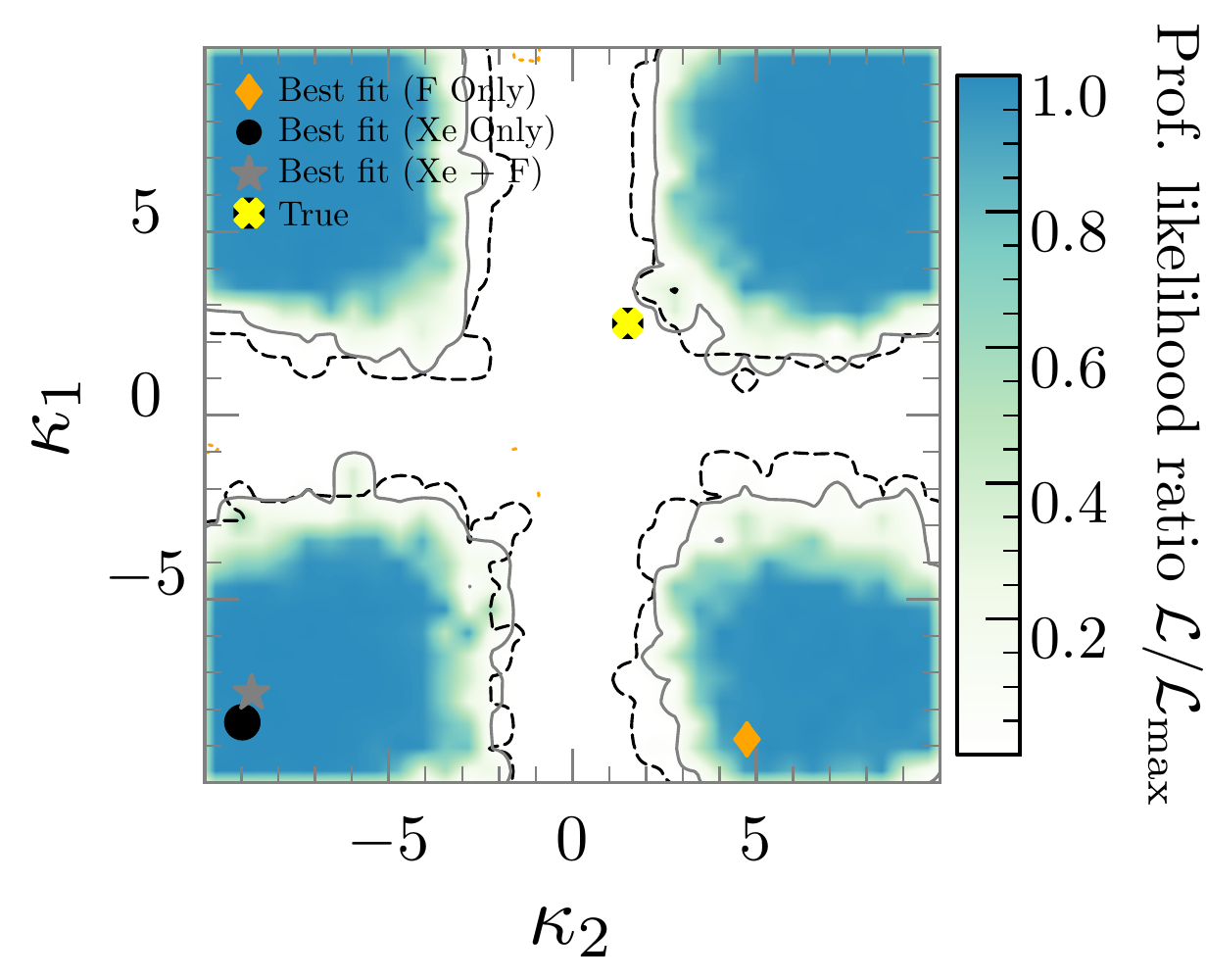}\quad 
	\includegraphics[width=0.42\textwidth]{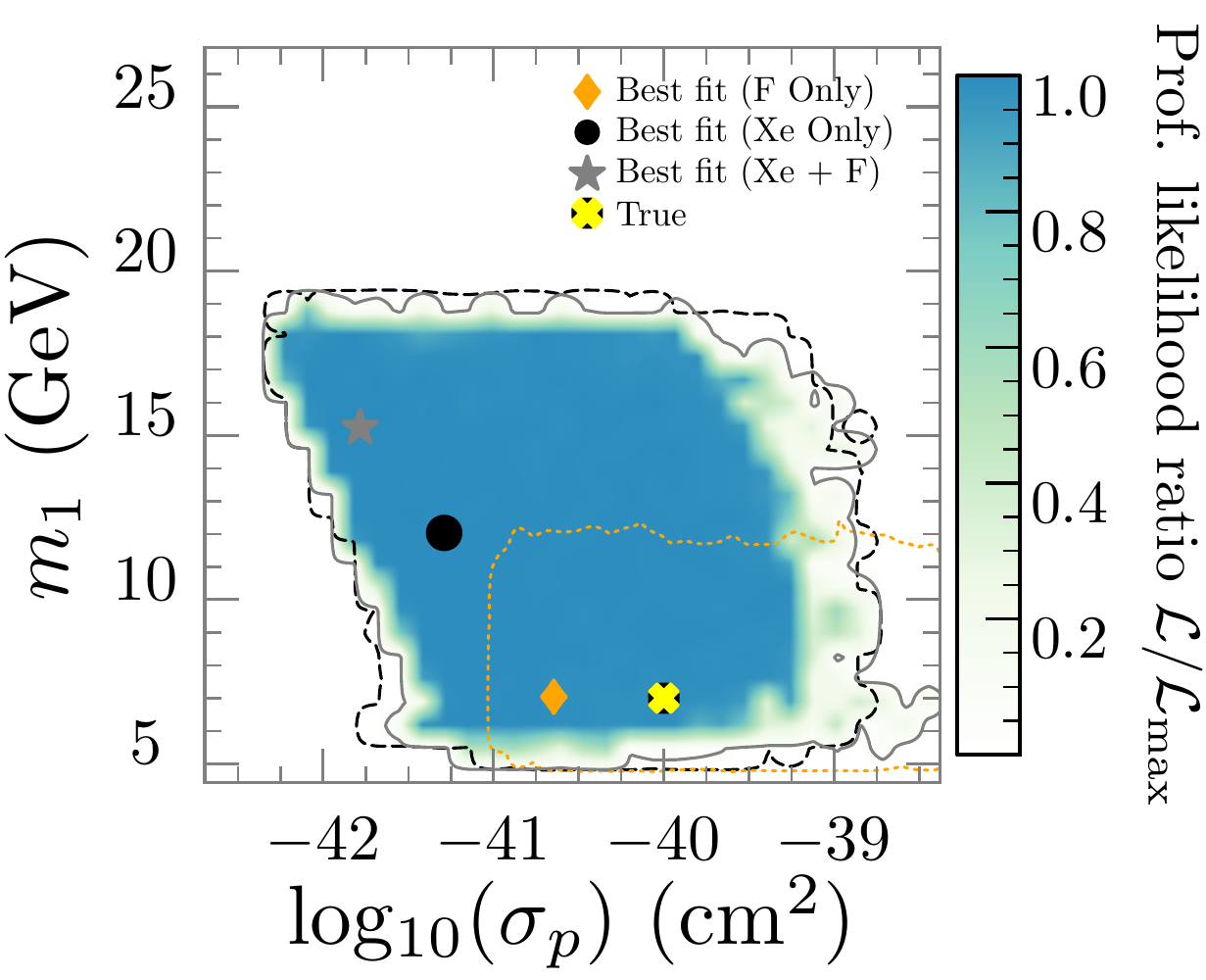}\quad 	
	\caption{SD results of the parameter estimation from the fit to one signal in Xe/F, and two signals in Xe + F. Figure properties are the same as for Fig.~\ref{2exp_fit}. The solid (grey), dashed (black) and dotted (orange) lines show $2\sigma$ C.L contours for the combined Xe+F signal, Xe-only signal and F-only signal respectively.} \label{2exp_fit_SD}
\end{figure}

In Figs.~\ref{2exp_fit} and~\ref{2exp_fit_SD} we show parameter estimates using the PLR method for the combination of two signals Xe+Ge (SI) and Xe+F (SD) respectively. 
Visualisation is done with the \texttt{Pippi} plotting package \cite{Scott:2012qh}. We show the 2$\sigma$ C.L contours (solid line), as well as the best-fit point and normalised profile likelihood density $\L/\L_{\rm max}$ on the colour scale. For comparison, we show the best-fit points and 2$\sigma$ C.L contours (dashed/dotted lines) for the one-signal cases (Xe/Ge for SI and Xe/F for SD) in order to display any improvement in parameter realisation.
We show results in the same planes that were used for the previous section, as well as parameter combinations that showed an interesting degree of uncertainty/degeneracy.  We notice in general that the PLR distribution of the SD results are in general more non-localised and uniform, as opposed to the SI cases where definitive regions of high PLR display a good degree of resolution for certain parameters. We also notice that the combined signal in the SD case generally produces regions that are more extended than the F-only case. This may initially seem counter-intuitive. The reason is that we have an eight-parameter scan with a likelihood which is highly multi-modal and non-trivial. As a result, one should not expect parameter estimates from the combined and one signal cases to give the same best-fit point, but rather significantly different regions of PLR. In fact, in general, we will see that for SD interactions the xenon form factor makes the parameter estimation much harder in the combined case than in the case of just a signal in fluorine, where the form factor suppression is absent. This makes the combined signal somewhat worse than the individual one.


\FloatBarrier
\FloatBarrier

Looking first at the top-left panel of Fig.~\ref{2exp_fit}, one can see excellent realisation of the two DM masses with the best-fit PLR, however the uncertainty in $m_2$ extends over a large region of the $m_2$ parameter space (although it is peaked around the true value). This is because for large $m_2$, $v_{\rm min}$ and therefore $\eta(E_R)$ become independent of $m_2$. Therefore the same signal can be fitted with increasing $r_\rho r_\sigma \propto m_2$ as can be seen from Eq.~\ref{eq:rate_tot}. In the SD case in Fig.~\ref{2exp_fit_SD}, the uncertainty in $m_1$ is increased and the true point is covered by an extended region of high PLR, however the best-fit point for the combined case does not resolve $m_1$ and $m_2$ accurately. The best-fit point for a signal with only fluorine however does resolve the benchmark point. There is no distinct difference between the Xe+Ge and Xe/Ge-only cases for SI other than a slight improvement in the $m_1$ precision, however the F-only region shows less uncertainty in $m_1$ for the SD case. Notice that the light $m_1$ chosen for SD (equal to $7$ GeV) implies that lighter nuclei (like F) are much more sensitive to it.

The top-right panels of Figs.~\ref{2exp_fit} and~\ref{2exp_fit_SD} show the SI/SD results in the plane of $\sigma_{H_{1}}$--$\sigma_{H_2}$. We notice that in the SI case we are less sensitive to $\sigma_{H_2}$, whilst the best-fit PLR is able to resolve the true $\sigma_{H_1}$. This is because for the heavier particle an experiment is able to probe the entire velocity distribution, whereas for the lighter particle we are more sensitive to fact that we need velocities to provide recoil energies above threshold. This is not the case for the SD result however, which shows that whilst the true point lies well within $2\sigma$ contour, the uncertainty has increased in both $\sigma_{H_1}$ and $\sigma_{H_2}$. Interestingly, the F-only 2$\sigma$ contour is much more resolved than the combined Xe+F result, with the best-fit PLR lying on top of the benchmark point.   

Next we show results in the $m_1$--$\sigma_{H_1}$ plane. For the SI case, we see that the extended degeneracy decreases in length from the Xe/Ge-only to the combined signal case. Moreover, the true point is well estimated by all best-fit points. For the SD case, the benchmark point is only resolved by the best-fit PLR for the F-only case. The degenerate region for the combined signal and Xe-only is significantly more extended than the SI case. Of all results shown in Fig.~\ref{2exp_fit_SD}, the SD case seems to have the most dramatic increase in the 2$\sigma$ region going from the F-only signal case to the combined signal and Xe-only case. 

Next, we consider the $m_2$--$\sigma_{H_2}$ plane where we can see the true point lies well within $2\sigma$ for the combined signals in the SI and SD case, and is well resolved by the best-fit PLR. The PLR density in both the SI and SD cases is high for the true $m_2$, however it is not localised in the $\sigma_{H_2}$ direction. The F-only signal in the SD case interestingly still covers the true point, and has a larger degree of precision in $\sigma_{H_2}$.   

The next figure shows an extended degeneracy in the $\log_{10}(r_\sigma)$--$\log_{10}(r_\rho)$ plane for both SI and SD cases, i..e, we have that $r_\rho \propto 1/r_\sigma$. Notice that we do not observe an exactly symmetric plot in $r_\rho$ and $r_\sigma$ since $r_\rho$ also enters the rate in the pre-factor $\propto1/(1+r_\rho)$. The best-fit points for the combined and one signal cases are not able to resolve true values for both the SI and SD scenarios, except for the Xe-only SD case. Given such a large degeneracy of high PLR density however, one can safely ignore the significance of the best fit points on this plot. The true point does however fall within the $2\sigma$ C.L region for all combined and single signal cases. The results of this figure show that in general, the median experiment will have trouble resolving the individual $r_{\rho,\sigma}$.

We next show the result for the combination $\log_{10}(r_\rho r_\sigma)$ vs. $m_2$ in order to explicitly explore the degeneracy that was previously mentioned. In both the SI and SD results we observe similar extended 2$\sigma$ regions for the combined signal, as well as very good accuracy for the best-fit PLR in the product $r_\rho r_\sigma$, much better than for the individual $r_{\rho,\sigma}$ (c.f. previous figure and discussion). In both cases, the PLR density is peaked ($>0.9$) at roughly the correct $m_2$.   

The next panel in each figure shows a fourfold degeneracy in the $\kappa_2$--$\kappa_1$ plane. This is because, as we saw in hypothesis testing section, we are insensitive to the sign of $\kappa$ for large values, as then $A_{\rm eff}^2\propto|\kappa|^2$.  In the SI case, the degeneracy is resolved by the best-fit PLR for the combined signal (also the PLR density is peaked in the correct quadrant) as well as the Ge-only signal. The Xe and Ge-only cases produce 2$\sigma$ contours that are all equally large indicating equal preference for all degenerate regions. This is not the case for the combined signal, where the PLR density favours the correct quadrant. In the SD case, the F-only signal has no sensitivity, and thus its contour covers the whole region ($\kappa_{1,2}\in[-10,10]$). In addition the fourfold degeneracy is not broken by either Xe-only or combined signal best-fit PLR. Furthermore, each quadrant of the degeneracy contains an equally high ($\gtrsim 0.9$) PLR. Therefore we can conclude that the couplings to neutrons and protons are difficult to resolve in the case of SD interactions.

Lastly, we show results in the $\log(\sigma_p^1)$--$m_1$ plane. The combined signal for the SI case contains the true point in a region of high PLR and well within the $2\sigma$ contour. The combined signal also offers a slightly more constrained region than the Xe-only and Ge-only cases. This is not true in the SD case, where the combined signal produces a large region of high PLR density, and the F-only case provides a best-fit that is closer to the benchmark, as well as a more tightly constrained region. 


\section{Conclusions} \label{sec:conc}

We have studied the implications of multi-component DM in DD signals. On general grounds, in the case of similar energy densities, we have argued that it is not possible to know which type of particle dominates the direct detection rate, the heavier or the lighter. For the cases in which one expects similar number densities, and assuming that the interactions of the dark sector are roughly the same, the total rate is suppressed by the heaviest DM particle, which also dominates the (suppressed) rate. Therefore, if this is the case, there is an upper bound on the heaviest DM expected to be able to be detected, and we have estimated it for several nuclei. For this case, indirect detection rates are also suppressed by the heaviest DM particles. In the case of similar masses, the approximations of equal number/energy densities are of course equivalent.

Focusing on two-component DM, we showed that the discerning feature of the two component scenario is a \emph{kink} in the (somewhat low energy) recoil spectrum. Such a feature cannot be present in the one-component scenario, and can thus be used to discriminate between the one/two component hypotheses. We then simulated the sensitivity of an `average' experiment to the one or two-component hypothesis. We adopted several experimental configurations (energy threshold, exposure) for a variety of different nuclear targets to roughly simulate the next generation of direct detection experiments for both SI and SD interactions. Our first results assumed equal cross-sections, energy densities, couplings to neutrons and protons and velocity dispersions. We showed that the mass splitting between the two WIMP components provides the best means for the median experiment to reject the one-component hypothesis in favour of the two-component hypothesis, and there are both lower and upper bounds on it for discrimination to be possible. In general, that the heaviest DM mass should be smaller than a hundred times the lightest one. The mass configurations that maximise the median significance are roughly $m_1=15$, $m_2=200$ (SI) and $m_1=7$, $m_2=150$ (SD). After fixing the mass of the heavier WIMP at $m_2 = 200\, (150)$ GeV for SI (SD), we then incrementally included more parameters into the hypotheses parameter space, starting with the velocity dispersion of the WIMPS $\sigma_{H_{1,\,2}}$ then moving onto the ratios of the local energy density and WIMP-proton cross-section $r_\rho/r_\sigma$. In general, small variations in these parameters around $\sim 1$ do not affect significantly the results. We finished by scanning over the WIMP-neutron/proton coupling ratios $\kappa_{1,\,2}$. In each of these three generalised scenarios we observed that the median significance can increase in localised regions of the parameter space. 

For the regions of parameter space in which there is especially strong hypothesis discrimination, we have estimated the precision and accuracy with which the model parameters can be extracted. We conducted parameter estimates for both SI and SD scenarios, and we considered the case of a combined signal from two experiments and compared the results to the one signal case. We observed that parameter estimation is worsened for SD scenarios, where regions of high PLR density were less localised than in the SI case. The parameters that show the greatest degree of uncertainty are the mass of the heavier WIMP and the individual ratios of energy densities and cross-sections ($r_{\rho,\,\sigma}$), as well as the velocity dispersion of the heavier WIMP, $\sigma_{H_2}$. However, the product $r_{\rho}\,r_\sigma$ was much better resolved. We observed degeneracies in the parameter space of $r_\rho$-$r_\sigma$ as well as $\kappa_1$-$\kappa_2$.  In the SI case, the $\kappa_1$-$\kappa_2$ degeneracies were broken by the combined-signal best-fit PLR. No degeneracies in any parameter space were observed to be broken in the SD case by going from single to dual experimental signals.  
 
We would also like to emphasize that our analysis assumes a simplified idealised scenario, with known background and perfect energy resolution and efficiency. Hence, more sophisticated experimental simulations are needed once a signal is observed. We note that the median sensitivity scales with $\sim\sqrt{\rm exposure}$ (before systematics dominate) and hence our results serve as a guide for more detailed studies of future experimental designs.
 
Although we focused on elastic SI and SD interactions, further studies can be performed with more complicated DM interactions, like inelastic endothermic/exothermic DM~\cite{TuckerSmith:2001hy,TuckerSmith:2004jv,Graham:2010ca,Schwetz:2011xm,Bozorgnia:2013hsa,Frandsen:2014ima}, or with interactions with nuclei mediated by different operators, which can have non-standard velocity/momentum dependencies~\cite{DelNobile:2013sia,Liu:2017kmx,Dobrescu:2006au,Fitzpatrick:2012ib,Anand:2013yka,Gresham:2014vja,Fitzpatrick:2012ix,Bishara:2016hek}. An interesting avenue to pursue would be to study to what extent non-trivial interactions can be degenerate with multi-component DM, i.e., can show a \emph{kink}-like feature.  

To finish we would like to mention that multi-component DM could give rise to specific signals also in indirect detection and at colliders. The smoking gun signal would seem to observe two gamma ray lines at different DM masses. However this could also be produced by a single DM annihilating into a boson (Z or Higgs) and a photon~\cite{Profumo:2009tb}, and therefore other information is necessary to break such degeneracy. However, even the continuum gamma ray spectrum could in principle show features pointing to multi-component DM, like the presence of a \emph{kink}, in a similar way as in DD. For similar densities and interactions, depending on whether the DM particles decay or annihilate, the heaviest or the lightest DM particle would give the strongest signal, as the dependence of the decay width with the mass is very strong ($\propto m^5$ for decaying DM with four-fermion interactions, for example). Therefore the preferred energy ranges and strategies to search for indirect DM signals are model-dependent. 

Regarding collider searches, if the masses of the particles in the dark sector are $\lesssim \mathcal{O}(1)$ TeV, they could be produced at the LHC. In this case, disentangling whether there is one type of DM produced or more is not an easy task~\cite{Giudice:2011ib}, as different DM masses or operators (models) could generate similar missing energy distributions.

Optimistically, one can imagine a situation in which DM is first observed being multi-component in direct detection, the underlying interaction is extracted (say SD), and that is used to predict halo-independently a signal at colliders~\cite{Blennow:2015gta,Herrero-Garcia:2015kga} and from annihilation into neutrinos in the Sun~\cite{Blennow:2015oea,Blennow:2015hzp}.

\vspace{1cm}
{\bf Acknowledgements:} JH-G would like to thank Thomas Schwetz for useful discussions regarding multi-component dark matter. AS would like to thank Ankit Beniwal for useful discussions regarding technical aspects. The work of JH-G, AS and AGW is supported by the Australian Research Council through the Centre of Excellence for Particle Physics at the Terascale CE110001004. MW is supported by the Australian Research Council Future Fellowship FT140100244.
\vspace{1cm}

\appendix
\section{Distribution of the $\mathcal{T}$ statistic when $\Ht$ is true}
\label{appendix:Tstat}

In sec.~\ref{sec::stats} we claim that one can easily show that for Gaussian distributed data, the $\T$ statistic under $\Ht$ is Gaussian with mean given by $\mathcal{T}^{\rm 2DM}_0$ as defined in Eq.~\eqref{T0}, and standard deviation given by $2\sqrt{\mathcal{T}^{\rm 2DM}_0}$. In this appendix we show why this is true, following the method of Ref.~\cite{ciuffoli2013sensitivity}. We start by writing the data $x_i$ as a Gaussian distributed variable
\begin{align}
 x_i= \mu_i(\theta_{\Ht}^{\rm true}) + g\sigma_i\;,
 \end{align}
  where $\mu_i(\theta_{\Ht}^{\rm true})$ is the true data under the two-DM hypothesis $\Ht$, $\sigma_i = \sqrt{ \mu_i(\theta_{\Ht}^{\rm true})}$ and $g$ is a unit Gaussian variable. Then, from the definition of $\T$ from Eq.~\eqref{TestStat}, one finds that
\begin{align}
\mathcal{T^{\rm 2DM}}&= \min_{\theta_{\Ho}} \chi^2(\theta_{\Ho}) -\min_{\theta_{\Ht} }  \chi^2(\theta_{\Ht}) \\
&=\sum_i^N\left(\min_{\theta_{\Ho}} \frac{\left[x_i-\mu_i (\theta_{\Ho})\right]^2}{\sigma_i^2} - \min_{\theta_{\Ht}} \frac{\left[x_i-\mu_i (\theta_{\Ht})\right]^2}{\sigma_i^2}\right) \\
&=\sum_i^N\left(\min_{\theta_{\Ho}} \frac{\left[
	\mu_i(\theta_{\Ht}^{\rm true}) + g\sigma_i-\mu_i (\theta_{\Ho})
	\right]^2}{\sigma_i^2} - \min_{\theta_{\Ht}}\frac{\left[
	\mu_i(\theta_{\Ht}^{\rm true}) + g\sigma_i-\mu_i (\theta_{\Ht})
	\right]^2}{\sigma_i^2}\right)\\
&=\min_{\theta_{\Ho}} \sum_i^N\frac{\left[
	\mu_i(\theta_{\Ht}^{\rm true})-\mu_i (\theta_{\Ho})
	\right]^2}{\sigma_i^2} + \frac{2\left[
	\mu_i(\theta_{\Ht}^{\rm true})- \mu_i (\theta_{\Ho})
	\right]}{\sigma_i}g\;.
\end{align}
The last line here explicitly shows that $\mathcal{T^{\rm 2DM}}$ is a Gaussian random variable with mean given by the first term 
\begin{align}
\min_{\theta_{\Ho}} \sum_i^N\frac{\left[
	\mu_i(\theta_{\Ht}^{\rm true})-\mu_i (\theta_{\Ho})
	\right]^2}{\sigma_i^2} \equiv \mathcal{T}^{\rm 2DM}_0\;,
\end{align} 
and standard deviation given by the second term
\begin{align}
\min_{\theta_{\Ho}} \sum_i^N\frac{2\left[
	\mu_i(\theta_{\Ht}^{\rm true})- \mu_i (\theta_{\Ho})
	\right]}{\sigma_i}g = 2\sqrt{\mathcal{T}^{\rm 2DM}_0}\;.
\end{align}

\section{Bayesian parameter estimation}
\label{app:bayes}
In this appendix we conduct the same parameter estimation study as in sec.~\ref{sec:par_estimation} but in a Bayesian framework. We first give an overview of the methods used followed by the results.

Bayes' theorem
\begin{align}
P(\boldsymbol{\theta}\,|\mathbf{x}) = \frac{\L(\mathbf{x}|\boldsymbol{\theta})\cdot\pi(\boldsymbol{\theta})}{\int d\boldsymbol{\theta}\,\L(\mathbf{x}|\boldsymbol{\theta})\cdot\pi(\boldsymbol{\theta})}\;,
\end{align}
allows one to explicitly solve for the probability of a given set model parameters $\boldsymbol{\theta}$ having observed some data $\mathbf{x}$. This probability function $P(\boldsymbol{\theta}\,|\mathbf{x})$ is called the `posterior' probability, and is a function of the likelihood $\L(\mathbf{x}|\boldsymbol{\theta})$ that the data is observed given the model parameters as well as the `prior probability' function $\pi(\boldsymbol{\theta})$, which parameterises one's prior degree of belief in $\boldsymbol{\theta}$ as well as sets the allowed size of the parameter space. The integral in the denominator of Bayes' theorem is called the `Bayesian evidence' and ensures that the posterior is normalized to 1 with respect to the model parameters. In order to constrain a slice of the full parameter space, one can integrate over unwanted nuisance parameters to give the `marginalised posterior'
\begin{align}
\label{2DpostDef}
P(\theta_1,\theta_2\,|\mathbf{x}) = \int d\theta_3...d\theta_{n}\,P(\boldsymbol{\theta}\,|\mathbf{x})\;.
\end{align}

\begin{figure}[t!]
	\centering
	\includegraphics[width=0.42\textwidth]{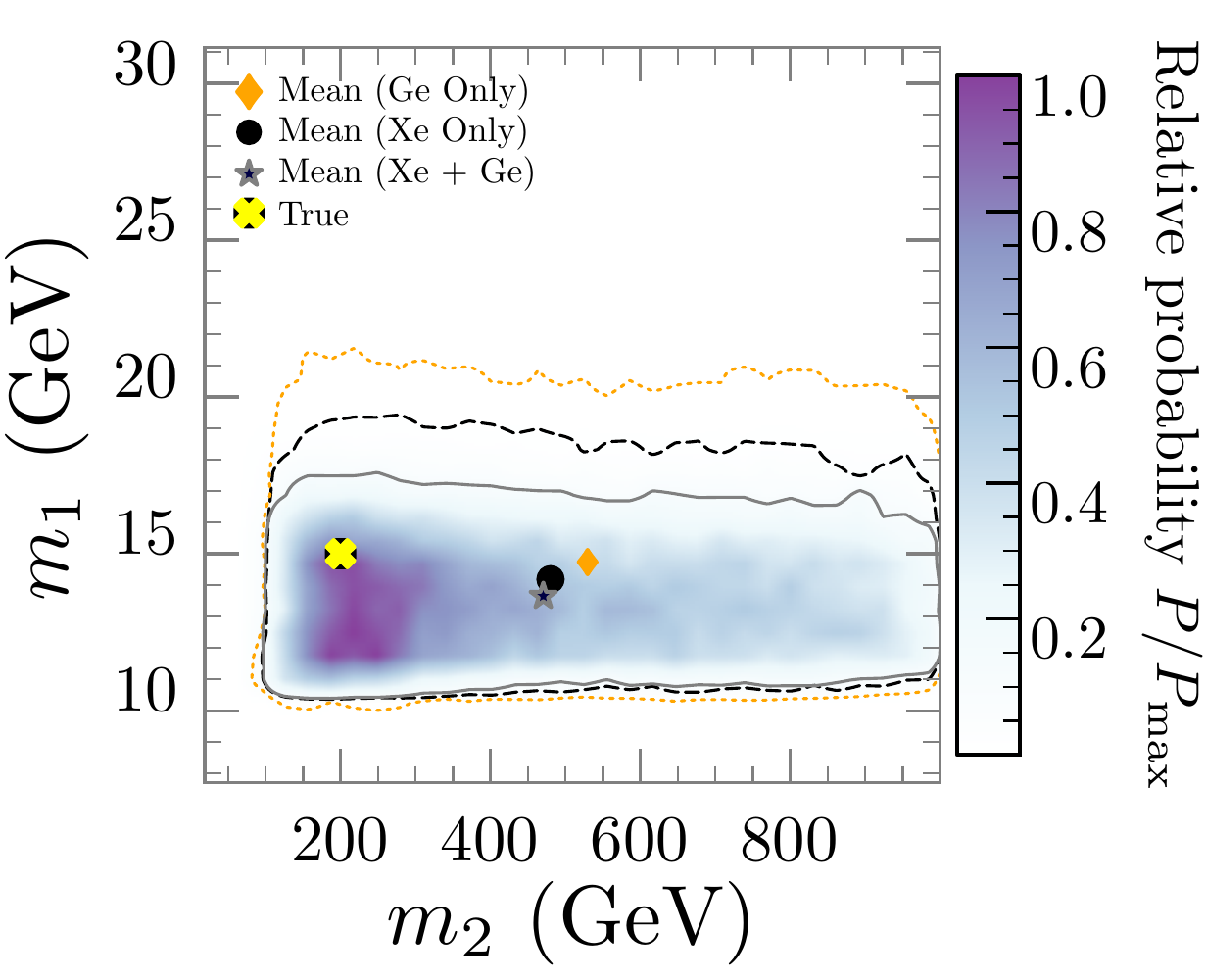}\quad 
	\includegraphics[width=0.42\textwidth]{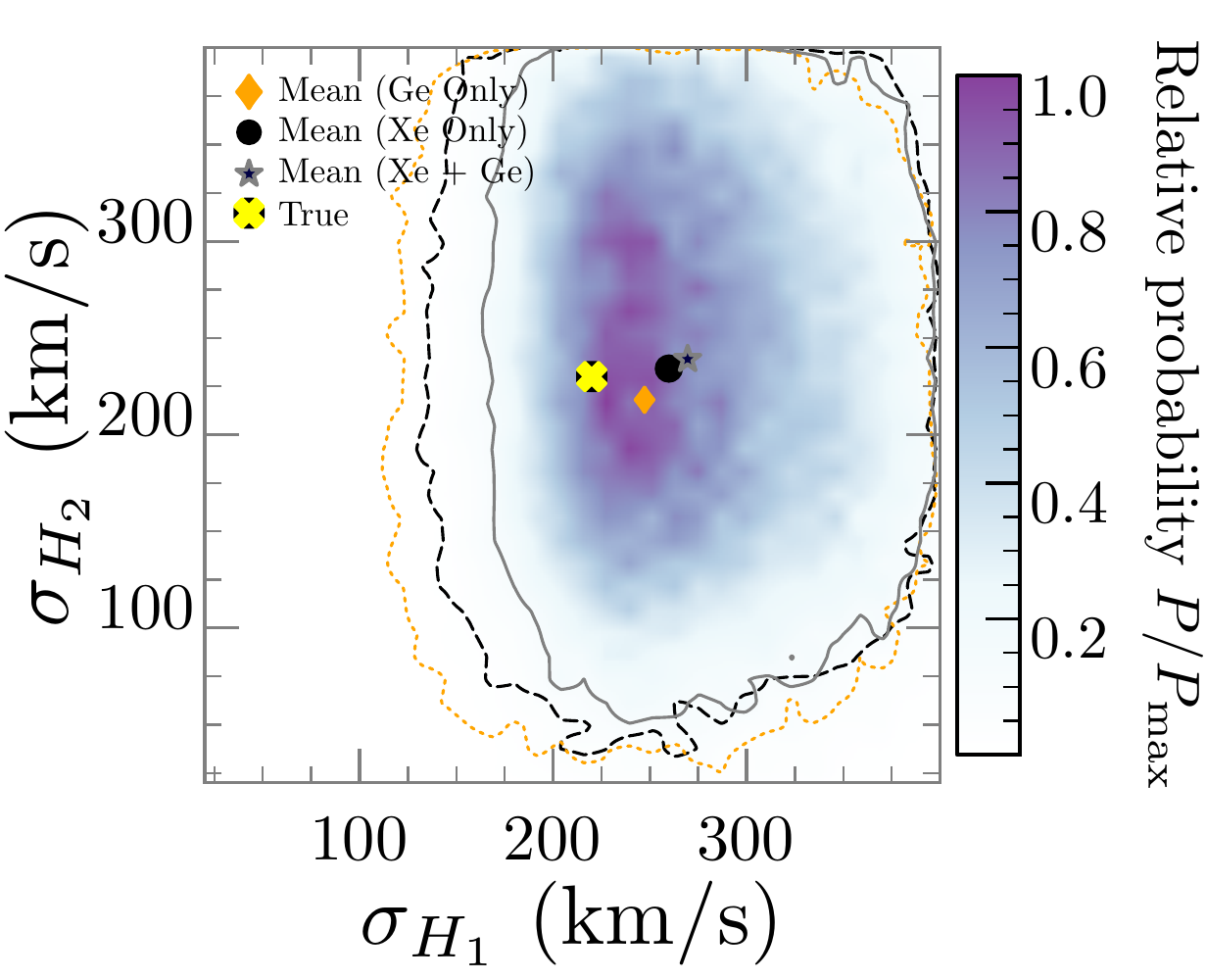}\quad 
	\includegraphics[width=0.42\textwidth]{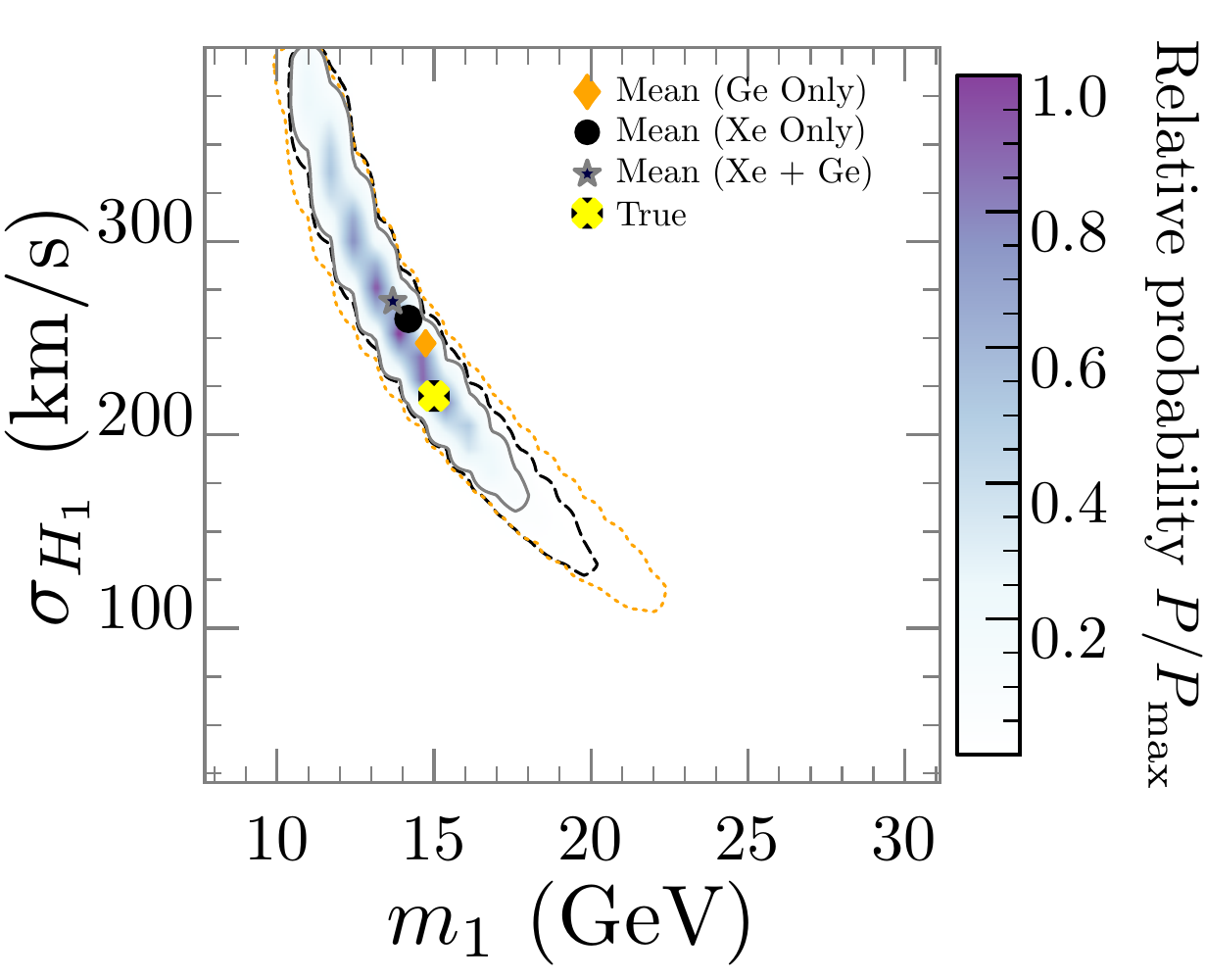}\quad 
	\includegraphics[width=0.42\textwidth]{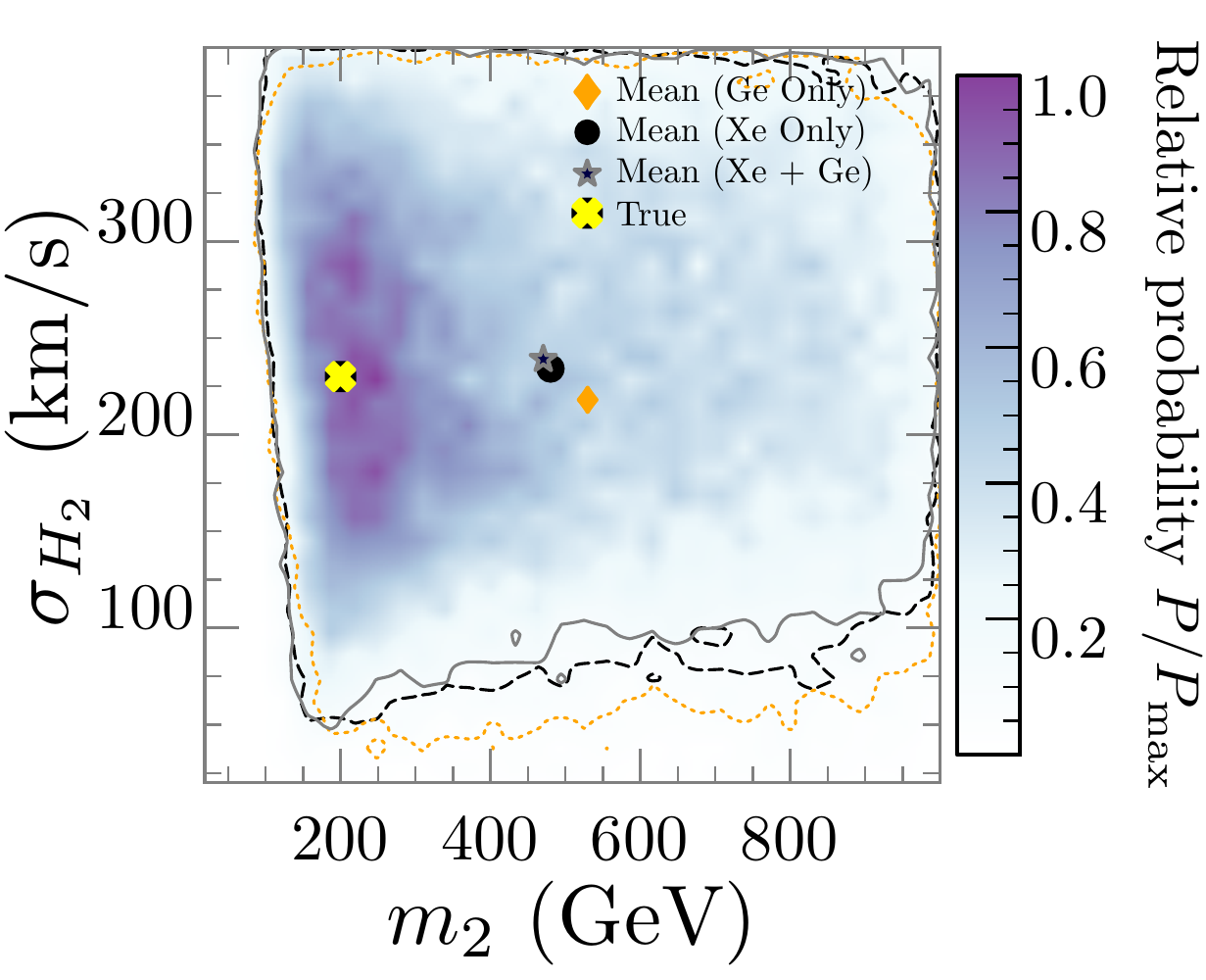}\quad 
	\includegraphics[width=0.42\textwidth]{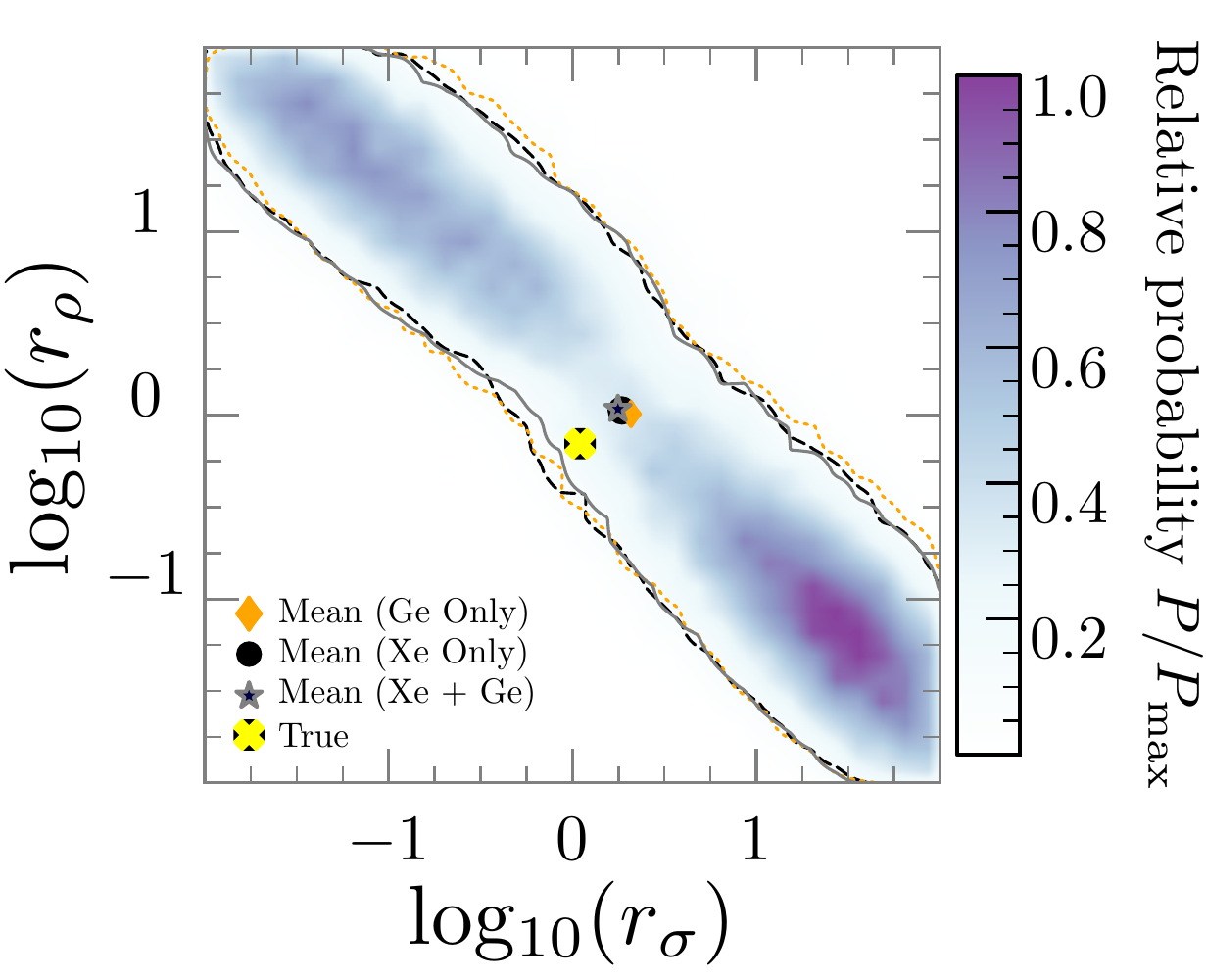}
	\includegraphics[width=0.42\textwidth]{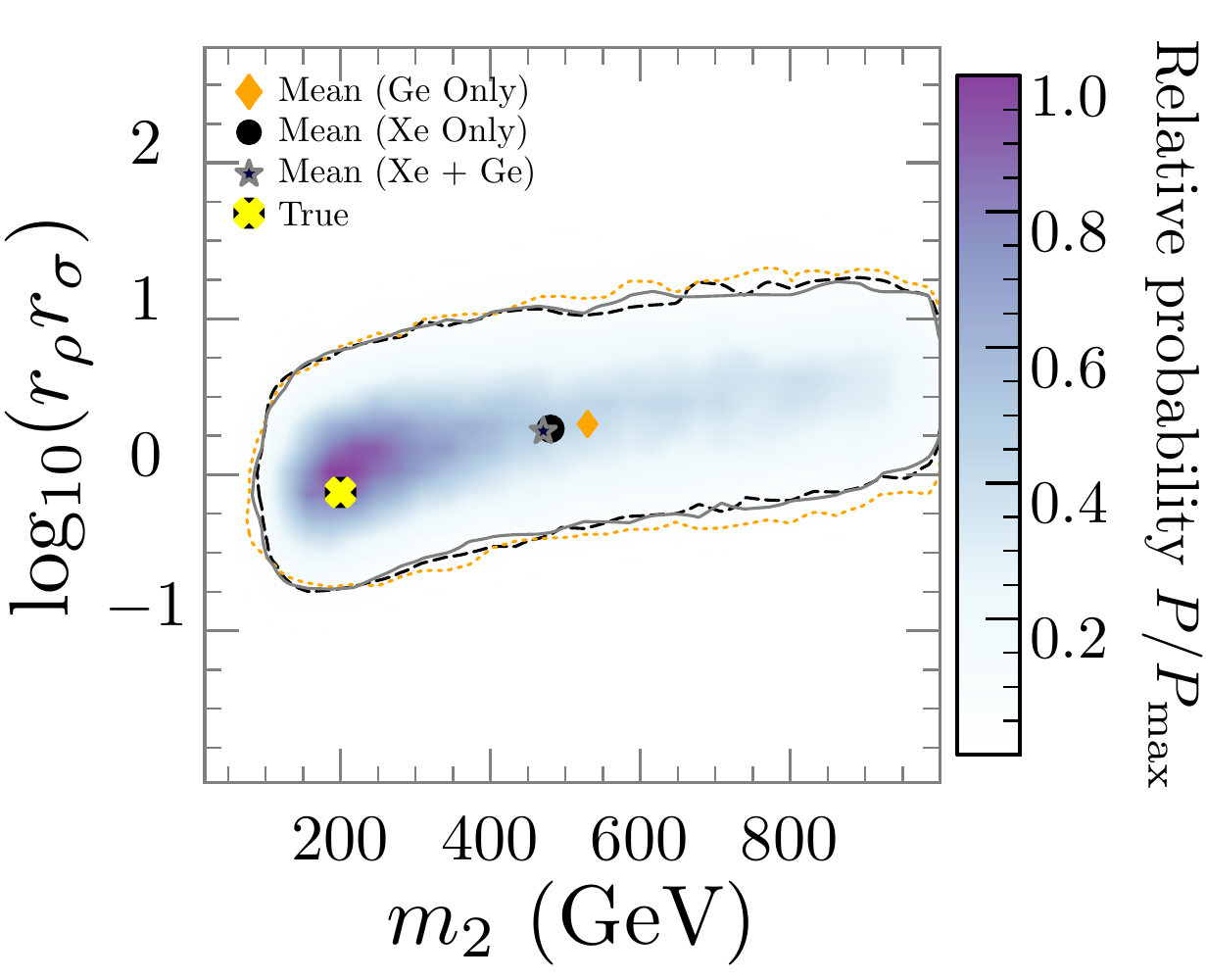}\quad 	\includegraphics[width=0.42\textwidth]{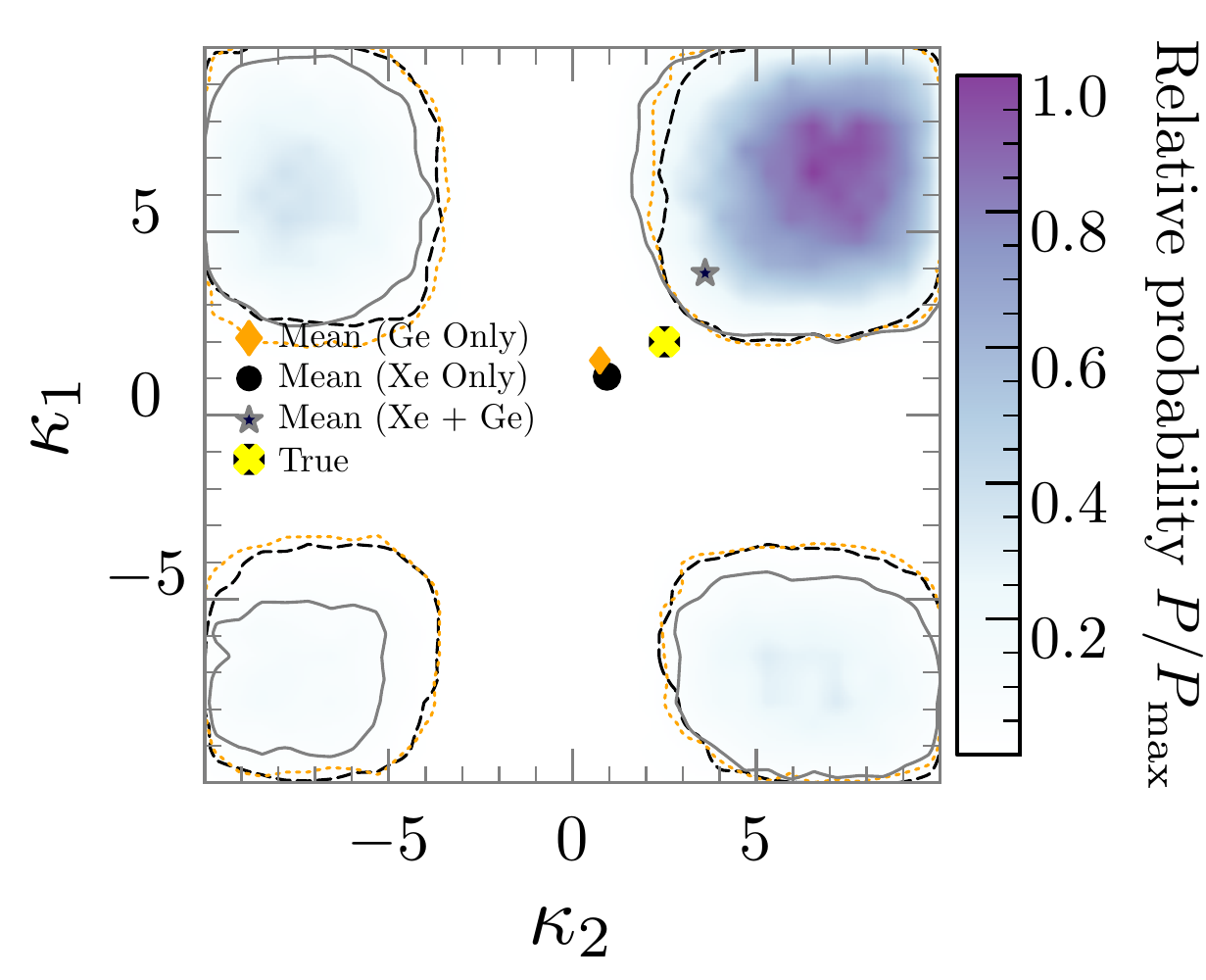}\quad 
	\includegraphics[width=0.42\textwidth]{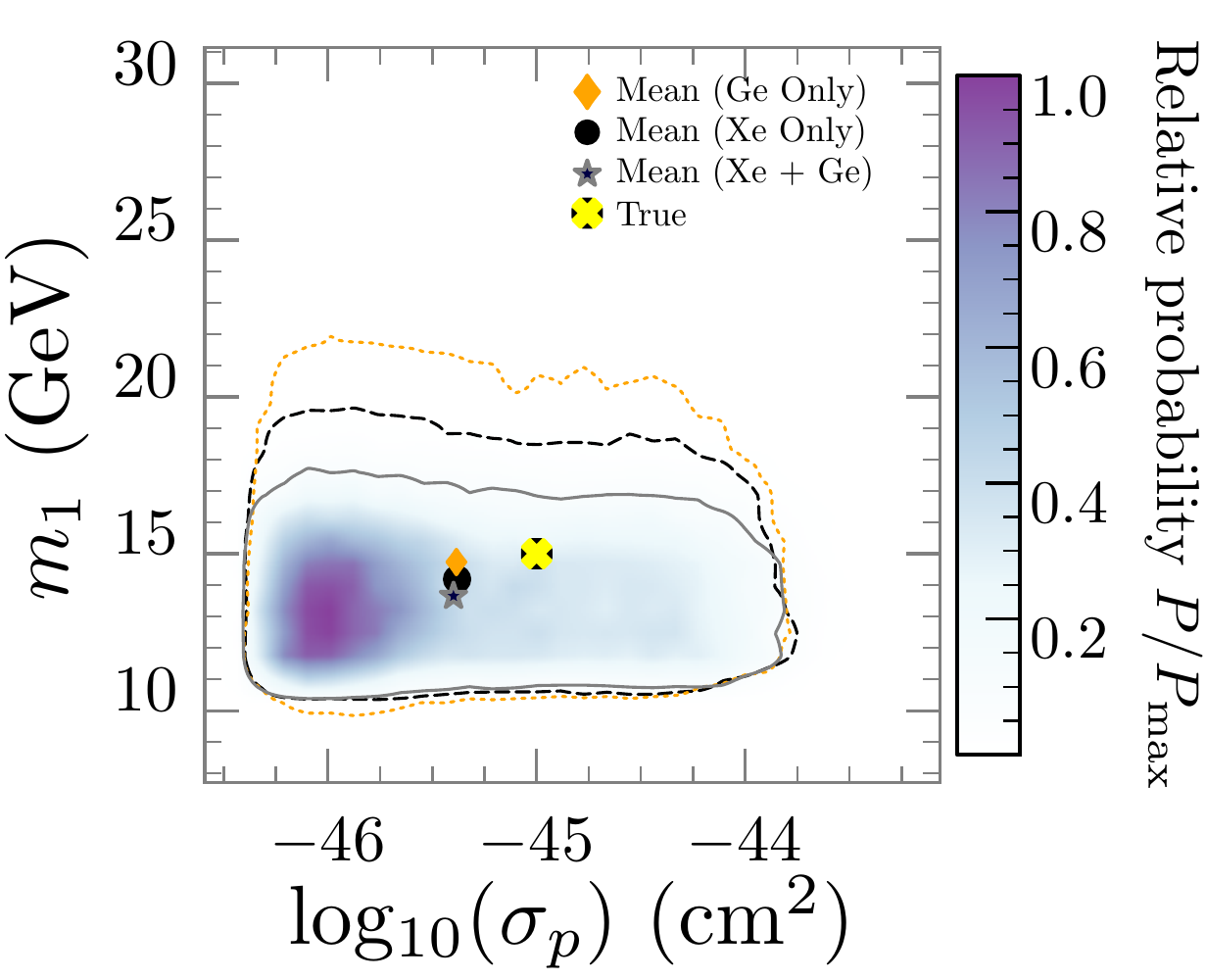}\quad 	
	\caption{Normalised posterior probability from the fit to one and two SI signals in Xe and Ge. Shown are the 2D regions of parameter space as given in Fig.~\ref{2exp_fit}.  
	} \label{2exp_fit_post}
\end{figure}
\FloatBarrier
\begin{figure}[t!]
	\centering
	\includegraphics[width=0.42\textwidth]{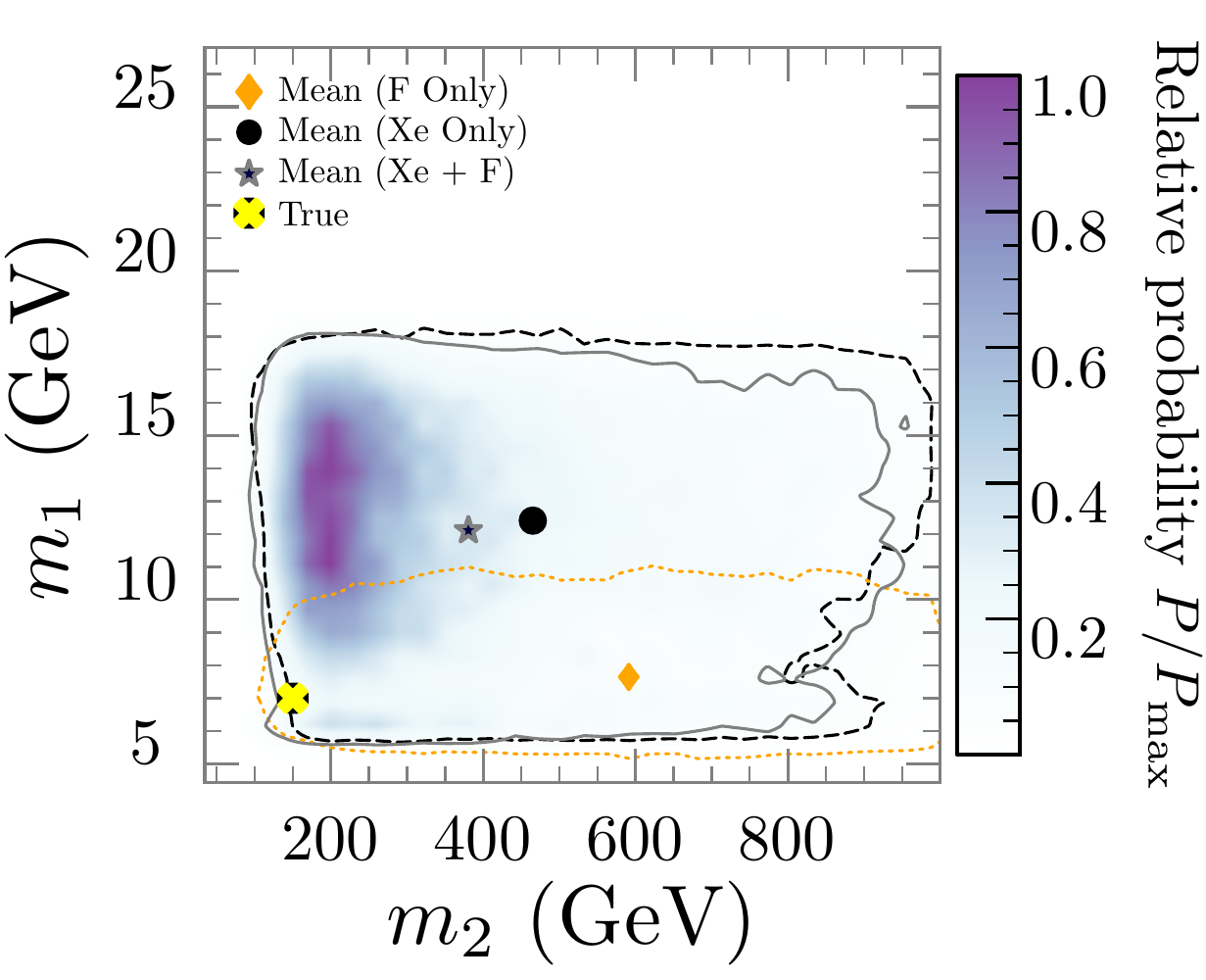}\quad 
	\includegraphics[width=0.42\textwidth]{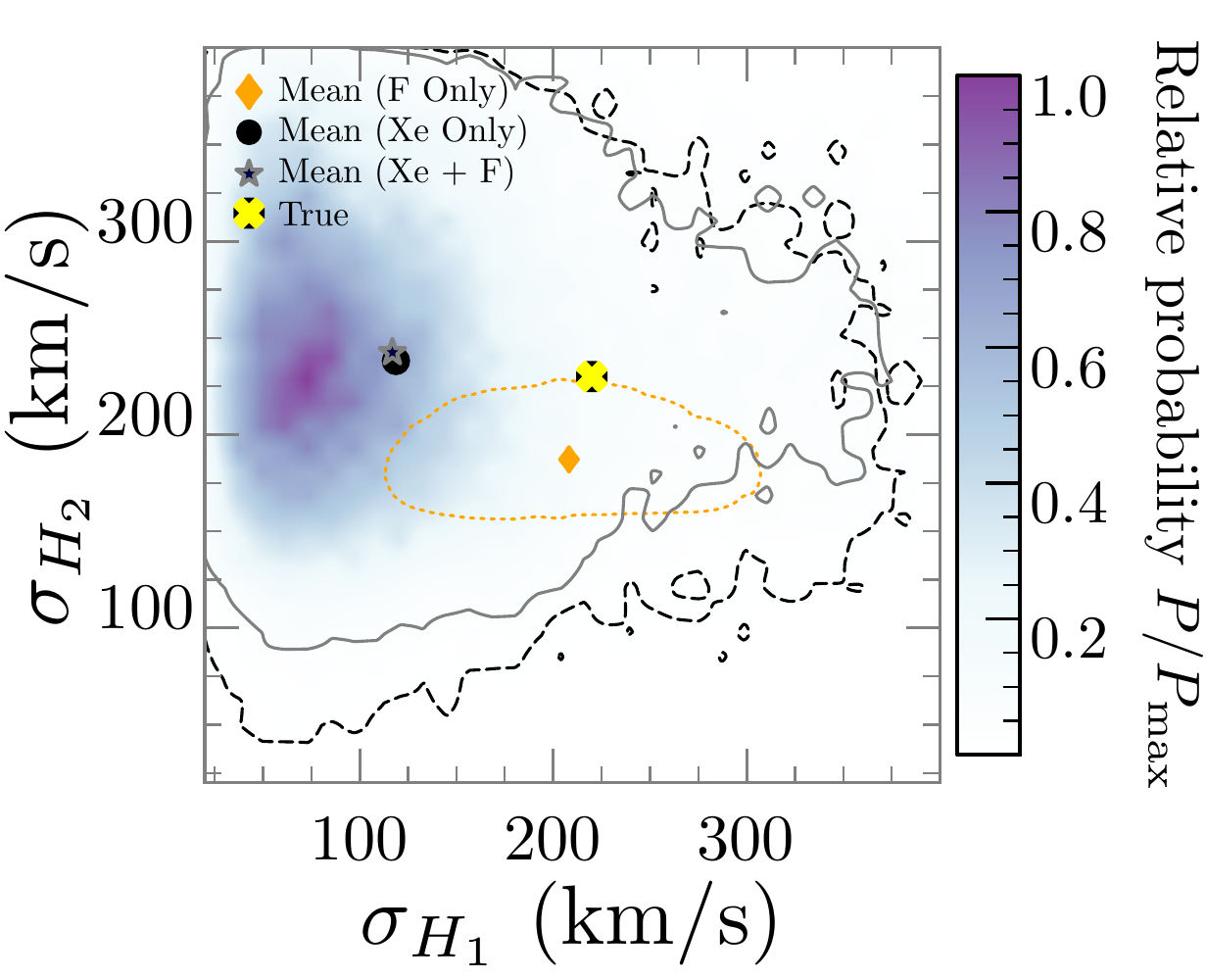}\quad 
	\includegraphics[width=0.42\textwidth]{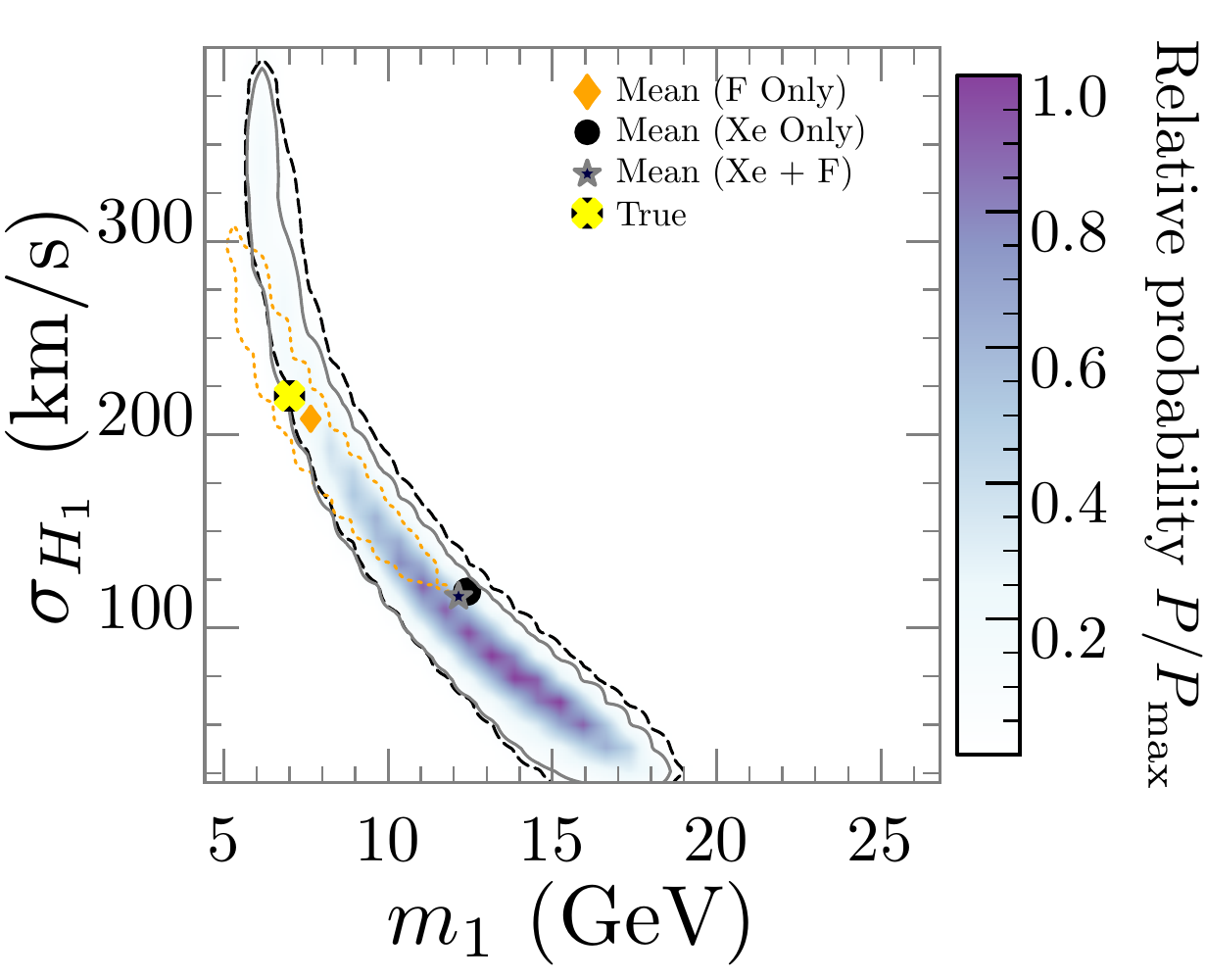}\quad 
	\includegraphics[width=0.42\textwidth]{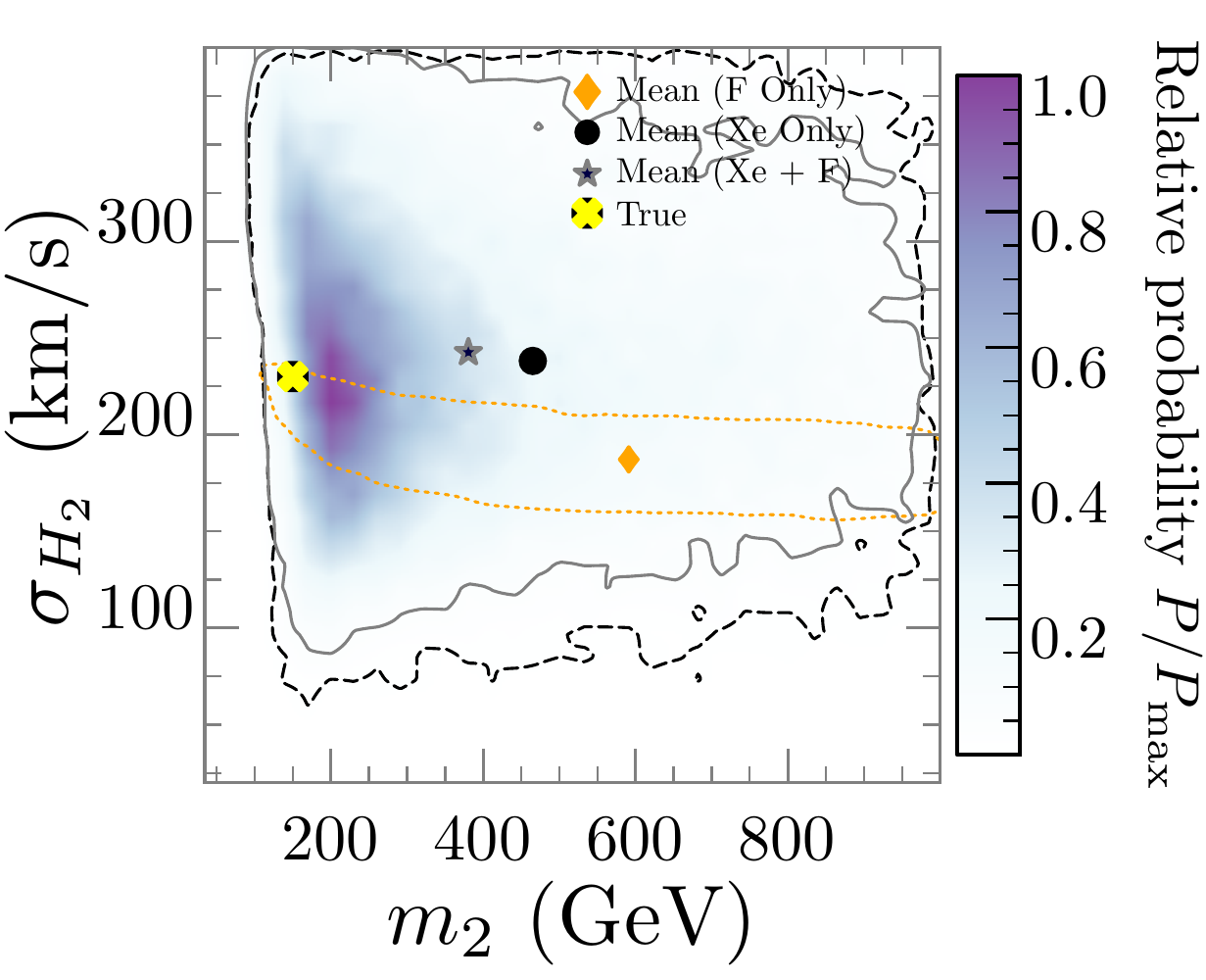}\quad 
	\includegraphics[width=0.42\textwidth]{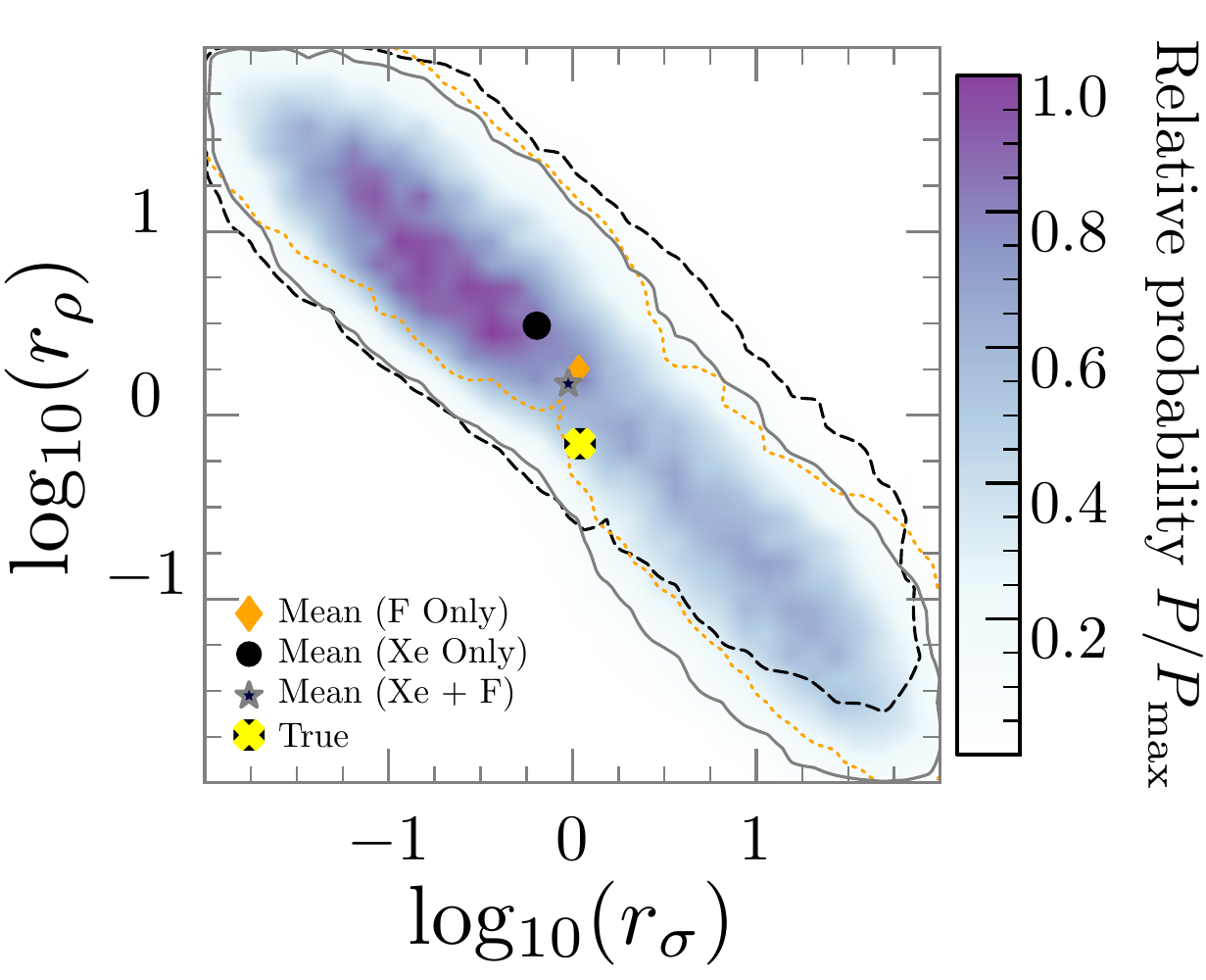}
	\includegraphics[width=0.42\textwidth]{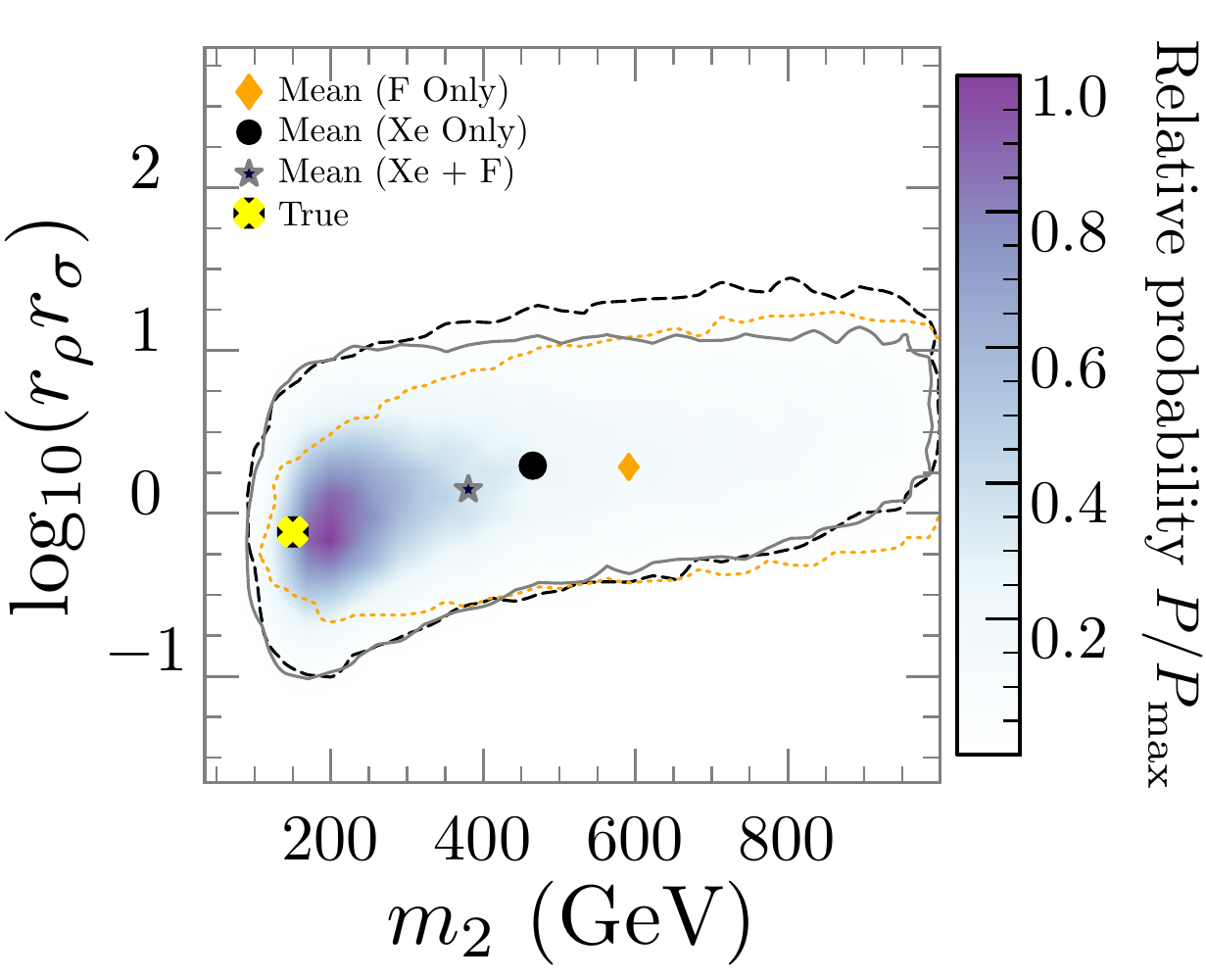}\quad 
	\includegraphics[width=0.42\textwidth]{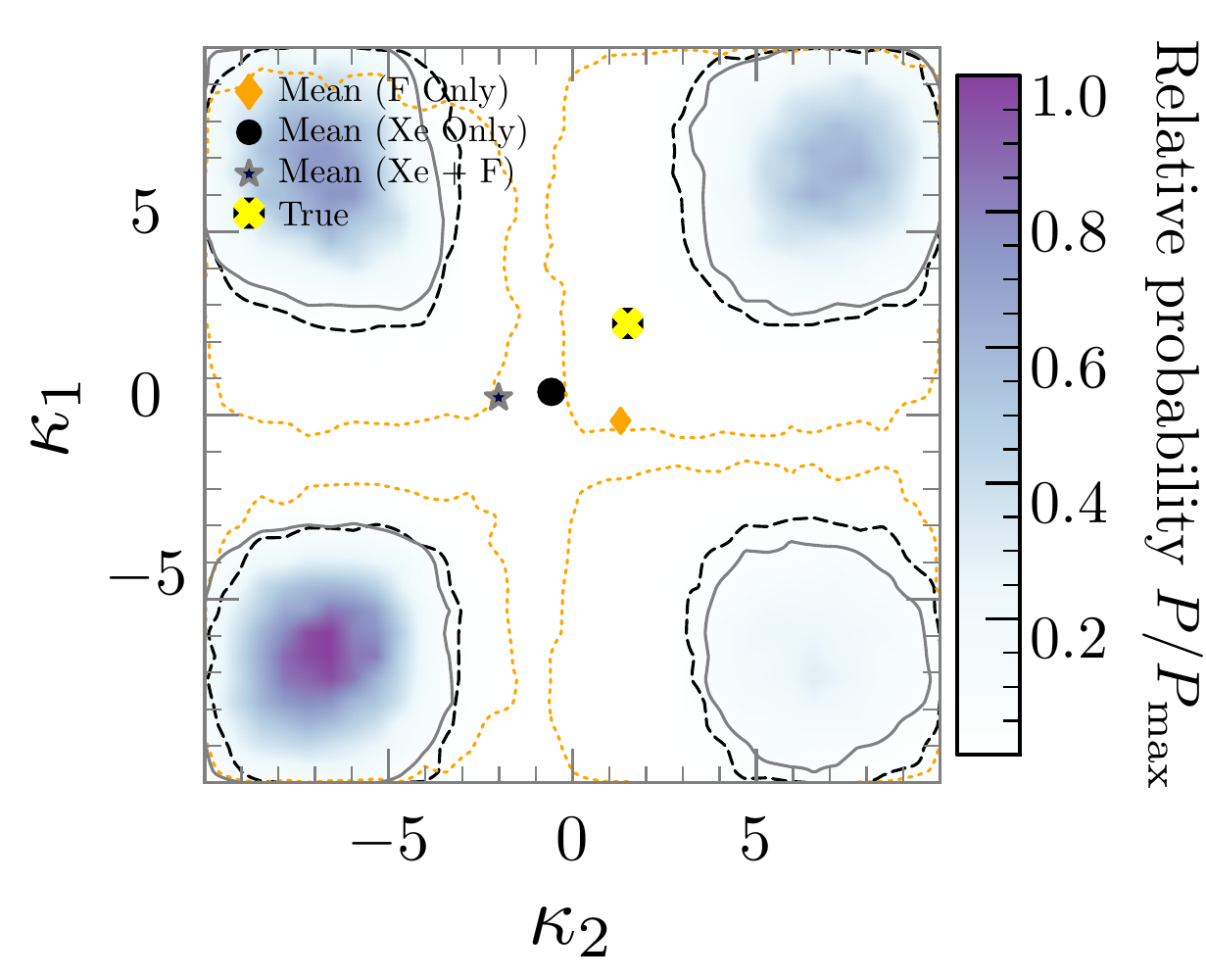}\quad 
	\includegraphics[width=0.42\textwidth]{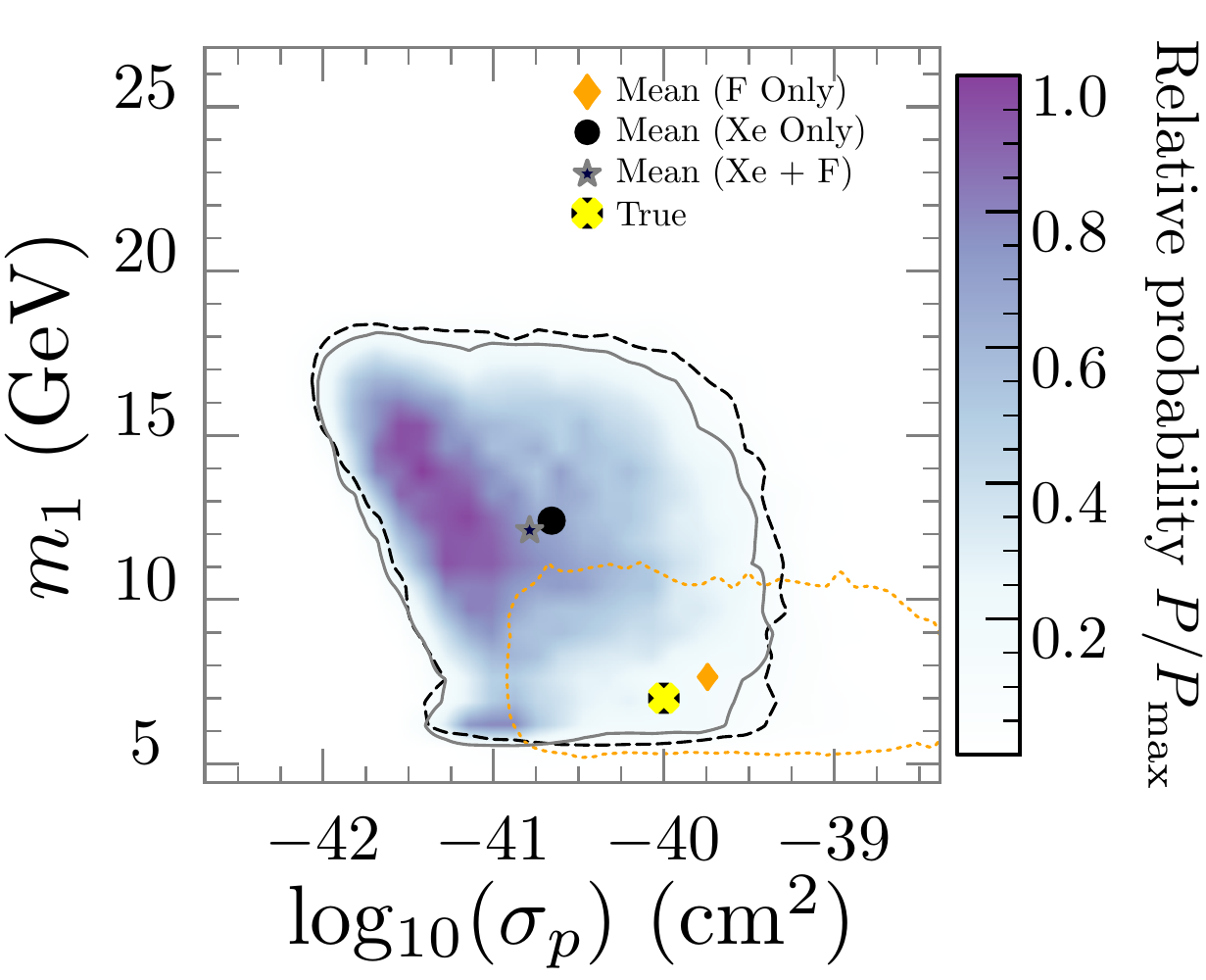}\quad 	
	\caption{Normalised posterior probability from the fit to one and two SD signals in Xe and F. Shown are the 2D regions of parameter space as given in Fig.~\ref{2exp_fit_SD}.} \label{2exp_fit_SD_post}
\end{figure}
\FloatBarrier

The example given in Eq.~\eqref{2DpostDef} would be an example of a 2D posterior. A Bayesian `credible region' (CR)  is the interval on the parameter $\theta_1$ such that 
\begin{align}
1-\alpha = \int d\theta_1\,P(\theta_1,\theta_2\,|\mathbf{x})\;,
\end{align}
where we choose $1-\alpha = 0.9545$ to define a $2\sigma$ CR.

Our results are shown in figs.~\ref{2exp_fit_post} and~\ref{2exp_fit_SD_post}, where the likelihood that enters Bayes' theorem is the Asimov likelihood defined in Eq.~\ref{AmimovLikePS}. We chose a combination of flat and log-flat priors so as to not give any preference to any region of parameter space. Our chosen priors are shown in table~\ref{tab:priors} and the results are not expected to change significantly under translation of the prior intervals.  The results shown are of the normalised posterior probability $P/P_{\rm max}$ with $2\sigma$ CR (solid line) for data obtained from the combined signal Xe + F as well as for comparison the $2\sigma$ CR (dashed line) for the F-only signal.

The posteriors shown in figs.~\ref{2exp_fit_SD} and~\ref{2exp_fit_SD_post} for the most part spatially resemble the PLR results shown in sec.~\ref{sec:par_estimation} (keeping in mind that strictly speaking one can not qualitatively compare the two), however, there are some distinct differences. In general, for both SI and SD results, the posterior probability seems to be able to resolve degeneracies quite well, contrary to the PLR for the SD case. The Bayesian method also seems to resolve certain parameters with less uncertainty, although we caution that the definition of the uncertainty is different in this case. The most notable differences are present in the $\log_{10}(r_\sigma)$--$\log_{10}(r_\rho)$ and $\kappa_2$--$\kappa_1$ planes. In this first case, the extended degenerate region $r_\rho r_\sigma=\text{cte}$ is distinctly separated into two regions of high posterior probability. The true point still does not fall within the $2\sigma$ CR however. In the $\kappa_1$-$\kappa_2$ plane, we observe that in the SI case, the four-fold degeneracy has regions that do not contain the true point, but are smaller than the regions of high PLR. The posterior does however favour the correct quadrant. In the SD case, the benchmark point is contained by the 2$\sigma$ CR for the F-only signal, however the combined signal produces smaller regions with the incorrect quadrant favoured by the posterior density.

\bibliographystyle{my-h-physrev}
\bibliography{multiDM.bib}

\begin{thebibliography}{10}

\bibitem{Goodman:1984dc}
M.~W. Goodman and E.~Witten,
\newblock {\em {Detectability of Certain Dark Matter Candidates}},
\newblock Phys. Rev. {\bf D31}, 3059 (1985).

\bibitem{Adulpravitchai:2011ei}
A.~Adulpravitchai, B.~Batell, and J.~Pradler,
\newblock {\em {Non-Abelian Discrete Dark Matter}},
\newblock Phys. Lett. {\bf B700}, 207 (2011), 1103.3053.

\bibitem{Batell:2009vb}
B.~Batell, M.~Pospelov, and A.~Ritz,
\newblock {\em {Direct Detection of Multi-component Secluded WIMPs}},
\newblock Phys. Rev. {\bf D79}, 115019 (2009), 0903.3396.

\bibitem{Dienes:2012cf}
K.~R. Dienes, J.~Kumar, and B.~Thomas,
\newblock {\em {Direct Detection of Dynamical Dark Matter}},
\newblock Phys. Rev. {\bf D86}, 055016 (2012), 1208.0336.

\bibitem{Profumo:2009tb}
S.~Profumo, K.~Sigurdson, and L.~Ubaldi,
\newblock {\em {Can we discover multi-component WIMP dark matter?}},
\newblock JCAP {\bf 0912}, 016 (2009), 0907.4374.

\bibitem{Bhattacharya:2017fid}
S.~Bhattacharya, P.~Ghosh, T.~N. Maity, and T.~S. Ray,
\newblock {\em {Mitigating Direct Detection Bounds in Non-minimal Higgs Portal
  Scalar Dark Matter Models}},
\newblock (2017), 1706.04699.

\bibitem{Bhattacharya:2016ysw}
S.~Bhattacharya, P.~Poulose, and P.~Ghosh,
\newblock {\em {Multipartite Interacting Scalar Dark Matter in the light of
  updated LUX data}},
\newblock JCAP {\bf 1704}, 043 (2017), 1607.08461.

\bibitem{Chialva:2012rq}
D.~Chialva, P.~S.~B. Dev, and A.~Mazumdar,
\newblock {\em {Multiple dark matter scenarios from ubiquitous stringy
  throats}},
\newblock Phys. Rev. {\bf D87}, 063522 (2013), 1211.0250.

\bibitem{Blennow:2015hzp}
M.~Blennow, S.~Clementz, and J.~Herrero-Garcia,
\newblock {\em {Pinning down inelastic dark matter in the Sun and in direct
  detection}},
\newblock JCAP {\bf 1604}, 004 (2016), 1512.03317.

\bibitem{Bozorgnia:2013hsa}
N.~Bozorgnia, J.~Herrero-Garcia, T.~Schwetz, and J.~Zupan,
\newblock {\em {Halo-independent methods for inelastic dark matter
  scattering}},
\newblock JCAP {\bf 1307}, 049 (2013), 1305.3575.

\bibitem{TuckerSmith:2004jv}
D.~Tucker-Smith and N.~Weiner,
\newblock {\em {The Status of inelastic dark matter}},
\newblock Phys. Rev. {\bf D72}, 063509 (2005), hep-ph/0402065.

\bibitem{Giudice:2011ib}
G.~F. Giudice, B.~Gripaios, and R.~Mahbubani,
\newblock {\em {Counting dark matter particles in LHC events}},
\newblock Phys. Rev. {\bf D85}, 075019 (2012), 1108.1800.

\bibitem{Bertone:2010rv}
G.~Bertone, D.~G. Cerdeno, M.~Fornasa, R.~Ruiz~de Austri, and R.~Trotta,
\newblock {\em {Identification of Dark Matter particles with LHC and direct
  detection data}},
\newblock Phys. Rev. {\bf D82}, 055008 (2010), 1005.4280.

\bibitem{Blennow:2015gta}
M.~Blennow, J.~Herrero-Garcia, T.~Schwetz, and S.~Vogl,
\newblock {\em {Halo-independent tests of dark matter direct detection signals:
  local DM density, LHC, and thermal freeze-out}},
\newblock JCAP {\bf 1508}, 039 (2015), 1505.05710.

\bibitem{Petraki:2013wwa}
K.~Petraki and R.~R. Volkas,
\newblock {\em {Review of asymmetric dark matter}},
\newblock Int. J. Mod. Phys. {\bf A28}, 1330028 (2013), 1305.4939.

\bibitem{PhysRev.104.1466}
R.~H. Helm,
\newblock {\em Inelastic and Elastic Scattering of 187-Mev Electrons from
  Selected Even-Even Nuclei},
\newblock Phys. Rev. {\bf 104}, 1466 (1956).

\bibitem{LEWIN199687}
J.~Lewin and P.~Smith,
\newblock {\em Review of mathematics, numerical factors, and corrections for
  dark matter experiments based on elastic nuclear recoil},
\newblock Astroparticle Physics {\bf 6}, 87  (1996).

\bibitem{Klos:2013rwa}
P.~Klos, J.~Men{\'e}ndez, D.~Gazit, and A.~Schwenk,
\newblock {\em {Large-scale nuclear structure calculations for spin-dependent
  WIMP scattering with chiral effective field theory currents}},
\newblock Phys. Rev. {\bf D88}, 083516 (2013), 1304.7684,
\newblock [Erratum: Phys. Rev.D89,no.2,029901(2014)].

\bibitem{Gelmini:2000dm}
G.~Gelmini and P.~Gondolo,
\newblock {\em {WIMP annual modulation with opposite phase in Late-Infall halo
  models}},
\newblock Phys. Rev. {\bf D64}, 023504 (2001), hep-ph/0012315.

\bibitem{Savage:2006qr}
C.~Savage, K.~Freese, and P.~Gondolo,
\newblock {\em {Annual Modulation of Dark Matter in the Presence of Streams}},
\newblock Phys. Rev. {\bf D74}, 043531 (2006), astro-ph/0607121.

\bibitem{Freese:2012xd}
K.~Freese, M.~Lisanti, and C.~Savage,
\newblock {\em {Colloquium: Annual modulation of dark matter}},
\newblock Rev. Mod. Phys. {\bf 85}, 1561 (2013), 1209.3339.

\bibitem{HerreroGarcia:2012fu}
J.~Herrero-Garcia, T.~Schwetz, and J.~Zupan,
\newblock {\em {Astrophysics independent bounds on the annual modulation of
  dark matter signals}},
\newblock Phys. Rev. Lett. {\bf 109}, 141301 (2012), 1205.0134.

\bibitem{HerreroGarcia:2011aa}
J.~Herrero-Garcia, T.~Schwetz, and J.~Zupan,
\newblock {\em {On the annual modulation signal in dark matter direct
  detection}},
\newblock JCAP {\bf 1203}, 005 (2012), 1112.1627.

\bibitem{Lee:2013wza}
S.~K. Lee, M.~Lisanti, A.~H.~G. Peter, and B.~R. Safdi,
\newblock {\em {Effect of Gravitational Focusing on Annual Modulation in
  Dark-Matter Direct-Detection Experiments}},
\newblock Phys. Rev. Lett. {\bf 112}, 011301 (2014), 1308.1953.

\bibitem{Bozorgnia:2014dqa}
N.~Bozorgnia and T.~Schwetz,
\newblock {\em {Is the effect of the Sun's gravitational potential on dark
  matter particles observable?}},
\newblock JCAP {\bf 1408}, 013 (2014), 1405.2340.

\bibitem{DelNobile:2015nua}
E.~Del~Nobile, G.~B. Gelmini, and S.~J. Witte,
\newblock {\em {Gravitational Focusing and Substructure Effects on the Rate
  Modulation in Direct Dark Matter Searches}},
\newblock JCAP {\bf 1508}, 041 (2015), 1505.07538.

\bibitem{Aprile:2017iyp}
XENON, E.~Aprile {\em et~al.},
\newblock {\em {First Dark Matter Search Results from the XENON1T Experiment}},
\newblock (2017), 1705.06655.

\bibitem{Amole:2016pye}
PICO, C.~Amole {\em et~al.},
\newblock {\em {Improved dark matter search results from PICO-2L Run 2}},
\newblock Phys. Rev. {\bf D93}, 061101 (2016), 1601.03729.

\bibitem{Amole:2017dex}
PICO, C.~Amole {\em et~al.},
\newblock {\em {Dark Matter Search Results from the PICO-60 C$_3$F$_8$ Bubble
  Chamber}},
\newblock Phys. Rev. Lett. {\bf 118}, 251301 (2017), 1702.07666.

\bibitem{Fatemighomi:2016ree}
DEAP-3600, N.~Fatemighomi,
\newblock {DEAP-3600 dark matter experiment},
\newblock in {\em {35th International Symposium on Physics in Collision (PIC
  2015) Coventry, United Kingdom, September 15-19, 2015}}, 2016, 1609.07990.

\bibitem{Calvo:2016hve}
ArDM, J.~Calvo {\em et~al.},
\newblock {\em {Commissioning of the ArDM experiment at the Canfranc
  underground laboratory: first steps towards a tonne-scale liquid argon time
  projection chamber for Dark Matter searches}},
\newblock JCAP {\bf 1703}, 003 (2017), 1612.06375.

\bibitem{Aprile:2014zvw}
XENON1T, E.~Aprile {\em et~al.},
\newblock {\em {Conceptual design and simulation of a water Cherenkov muon veto
  for the XENON1T experiment}},
\newblock JINST {\bf 9}, P11006 (2014), 1406.2374.

\bibitem{Akerib:2015cja}
LZ, D.~S. Akerib {\em et~al.},
\newblock {\em {LUX-ZEPLIN (LZ) Conceptual Design Report}},
\newblock (2015), 1509.02910.

\bibitem{Aalbers:2016jon}
DARWIN, J.~Aalbers {\em et~al.},
\newblock {\em {DARWIN: towards the ultimate dark matter detector}},
\newblock JCAP {\bf 1611}, 017 (2016), 1606.07001.

\bibitem{Undagoitia:2015gya}
T.~Marrod{\'a}n~Undagoitia and L.~Rauch,
\newblock {\em {Dark matter direct-detection experiments}},
\newblock J. Phys. {\bf G43}, 013001 (2016), 1509.08767.

\bibitem{Liu:2017drf}
J.~Liu, X.~Chen, and X.~Ji,
\newblock {\em {Current status of direct dark matter detection experiments}},
\newblock Nature Phys. {\bf 13}, 212 (2017).

\bibitem{Fushimi:2015sew}
PICO-LON, K.~Fushimi {\em et~al.},
\newblock {\em {Dark matter search project PICO-LON}},
\newblock J. Phys. Conf. Ser. {\bf 718}, 042022 (2016), 1512.04645.

\bibitem{Agnese:2016cpb}
SuperCDMS, R.~Agnese {\em et~al.},
\newblock {\em {Projected Sensitivity of the SuperCDMS SNOLAB experiment}},
\newblock Phys. Rev. {\bf D95}, 082002 (2017), 1610.00006.

\bibitem{Feldstein:2014ufa}
B.~Feldstein and F.~Kahlhoefer,
\newblock {\em {Quantifying (dis)agreement between direct detection experiments
  in a halo-independent way}},
\newblock JCAP {\bf 1412}, 052 (2014), 1409.5446.

\bibitem{Kavanagh:2012nr}
B.~J. Kavanagh and A.~M. Green,
\newblock {\em {Improved determination of the WIMP mass from direct detection
  data}},
\newblock Phys. Rev. {\bf D86}, 065027 (2012), 1207.2039.

\bibitem{Herrero-Garcia:2015kga}
J.~Herrero-Garcia,
\newblock {\em {Halo-independent tests of dark matter annual modulation
  signals}},
\newblock JCAP {\bf 1509}, 012 (2015), 1506.03503.

\bibitem{Blennow:2013oma}
M.~Blennow, P.~Coloma, P.~Huber, and T.~Schwetz,
\newblock {\em {Quantifying the sensitivity of oscillation experiments to the
  neutrino mass ordering}},
\newblock JHEP {\bf 03}, 028 (2014), 1311.1822.

\bibitem{Ciuffoli:2017ayi}
E.~Ciuffoli,
\newblock {Statistical Methods for the Neutrino Mass Hierarchy},
\newblock in {\em {Prospects in Neutrino Physics (NuPhys2016) London, London,
  United Kingdom, December 12-14, 2016}}, 2017, 1704.08043.

\bibitem{ciuffoli2013sensitivity}
E.~Ciuffoli, J.~Evslin, and X.~Zhang,
\newblock {\em Sensitivity to the Neutrino Mass Hierarchy},
\newblock arXiv preprint arXiv:1305.5150  (2013).

\bibitem{doi:10.1080/14786440009463897}
K.~P. F.R.S.,
\newblock {\em X. On the criterion that a given system of deviations from the
  probable in the case of a correlated system of variables is such that it can
  be reasonably supposed to have arisen from random sampling},
\newblock Philosophical Magazine {\bf 50}, 157 (1900),
  http://dx.doi.org/10.1080/14786440009463897.

\bibitem{wilks1938large}
S.~S. Wilks,
\newblock {\em The large-sample distribution of the likelihood ratio for
  testing composite hypotheses},
\newblock The Annals of Mathematical Statistics {\bf 9}, 60 (1938).

\bibitem{cowan2011asymptotic}
G.~Cowan, K.~Cranmer, E.~Gross, and O.~Vitells,
\newblock {\em Asymptotic formulae for likelihood-based tests of new physics},
\newblock The European Physical Journal C {\bf 71}, 1554 (2011).

\bibitem{Feng:2011vu}
J.~L. Feng, J.~Kumar, D.~Marfatia, and D.~Sanford,
\newblock {\em {Isospin-Violating Dark Matter}},
\newblock Phys. Lett. {\bf B703}, 124 (2011), 1102.4331.

\bibitem{Yaguna:2016bga}
C.~E. Yaguna,
\newblock {\em {Isospin-violating dark matter in the light of recent data}},
\newblock Phys. Rev. {\bf D95}, 055015 (2017), 1610.08683.

\bibitem{Pato:2010zk}
M.~Pato {\em et~al.},
\newblock {\em {Complementarity of Dark Matter Direct Detection Targets}},
\newblock Phys. Rev. {\bf D83}, 083505 (2011), 1012.3458.

\bibitem{Peter:2013aha}
A.~H.~G. Peter, V.~Gluscevic, A.~M. Green, B.~J. Kavanagh, and S.~K. Lee,
\newblock {\em {WIMP physics with ensembles of direct-detection experiments}},
\newblock Phys. Dark Univ. {\bf 5-6}, 45 (2014), 1310.7039.

\bibitem{feroz2009multinest}
F.~Feroz, M.~Hobson, and M.~Bridges,
\newblock {\em MultiNest: an efficient and robust Bayesian inference tool for
  cosmology and particle physics},
\newblock Monthly Notices of the Royal Astronomical Society {\bf 398}, 1601
  (2009).

\bibitem{Kavanagh:2014rya}
B.~J. Kavanagh, M.~Fornasa, and A.~M. Green,
\newblock {\em {Probing WIMP particle physics and astrophysics with direct
  detection and neutrino telescope data}},
\newblock Phys. Rev. {\bf D91}, 103533 (2015), 1410.8051.

\bibitem{Blennow:2015oea}
M.~Blennow, J.~Herrero-Garcia, and T.~Schwetz,
\newblock {\em {A halo-independent lower bound on the dark matter capture rate
  in the Sun from a direct detection signal}},
\newblock JCAP {\bf 1505}, 036 (2015), 1502.03342.

\bibitem{Ferrer:2015bta}
F.~Ferrer, A.~Ibarra, and S.~Wild,
\newblock {\em {A novel approach to derive halo-independent limits on dark
  matter properties}},
\newblock JCAP {\bf 1509}, 052 (2015), 1506.03386.

\bibitem{Ibarra:2017mzt}
A.~Ibarra and A.~Rappelt,
\newblock {\em {Optimized velocity distributions for direct dark matter
  detection}},
\newblock (2017), 1703.09168.

\bibitem{Scott:2012qh}
P.~Scott,
\newblock {\em {Pippi - painless parsing, post-processing and plotting of
  posterior and likelihood samples}},
\newblock Eur. Phys. J. Plus {\bf 127}, 138 (2012), 1206.2245.

\bibitem{TuckerSmith:2001hy}
D.~Tucker-Smith and N.~Weiner,
\newblock {\em {Inelastic dark matter}},
\newblock Phys. Rev. {\bf D64}, 043502 (2001), hep-ph/0101138.

\bibitem{Graham:2010ca}
P.~W. Graham, R.~Harnik, S.~Rajendran, and P.~Saraswat,
\newblock {\em {Exothermic Dark Matter}},
\newblock Phys. Rev. {\bf D82}, 063512 (2010), 1004.0937.

\bibitem{Schwetz:2011xm}
T.~Schwetz and J.~Zupan,
\newblock {\em {Dark Matter attempts for CoGeNT and DAMA}},
\newblock JCAP {\bf 1108}, 008 (2011), 1106.6241.

\bibitem{Frandsen:2014ima}
M.~T. Frandsen and I.~M. Shoemaker,
\newblock {\em {Up-shot of inelastic down-scattering at CDMS-Si}},
\newblock Phys. Rev. {\bf D89}, 051701 (2014), 1401.0624.

\bibitem{DelNobile:2013sia}
M.~Cirelli, E.~Del~Nobile, and P.~Panci,
\newblock {\em {Tools for model-independent bounds in direct dark matter
  searches}},
\newblock JCAP {\bf 1310}, 019 (2013), 1307.5955.

\bibitem{Liu:2017kmx}
Z.~Liu, Y.~Su, Y.-L.~S. Tsai, B.~Yu, and Q.~Yuan,
\newblock {\em {A combined analysis of PandaX, LUX, and XENON1T experiments
  within the framework of dark matter effective theory}},
\newblock (2017), 1708.04630.

\bibitem{Dobrescu:2006au}
B.~A. Dobrescu and I.~Mocioiu,
\newblock {\em {Spin-dependent macroscopic forces from new particle exchange}},
\newblock JHEP {\bf 11}, 005 (2006), hep-ph/0605342.

\bibitem{Fitzpatrick:2012ib}
A.~L. Fitzpatrick, W.~Haxton, E.~Katz, N.~Lubbers, and Y.~Xu,
\newblock {\em {Model Independent Direct Detection Analyses}},
\newblock (2012), 1211.2818.

\bibitem{Anand:2013yka}
N.~Anand, A.~L. Fitzpatrick, and W.~C. Haxton,
\newblock {\em {Weakly interacting massive particle-nucleus elastic scattering
  response}},
\newblock Phys. Rev. {\bf C89}, 065501 (2014), 1308.6288.

\bibitem{Gresham:2014vja}
M.~I. Gresham and K.~M. Zurek,
\newblock {\em {Effect of nuclear response functions in dark matter direct
  detection}},
\newblock Phys. Rev. {\bf D89}, 123521 (2014), 1401.3739.

\bibitem{Fitzpatrick:2012ix}
A.~L. Fitzpatrick, W.~Haxton, E.~Katz, N.~Lubbers, and Y.~Xu,
\newblock {\em {The Effective Field Theory of Dark Matter Direct Detection}},
\newblock JCAP {\bf 1302}, 004 (2013), 1203.3542.

\bibitem{Bishara:2016hek}
F.~Bishara, J.~Brod, B.~Grinstein, and J.~Zupan,
\newblock {\em {Chiral Effective Theory of Dark Matter Direct Detection}},
\newblock JCAP {\bf 1702}, 009 (2017), 1611.00368.

\end{thebibliography}

\end{document}